\documentclass{article}

\usepackage[margin=1in]{geometry}

\usepackage{graphicx,amsmath,upgreek,bm}
\usepackage{colortbl}
\usepackage{wrapfig}
\usepackage[rgb, table]{xcolor}
\usepackage{booktabs}
\usepackage{makecell}
\usepackage{pifont}
\usepackage{todonotes}
\usepackage[toc,page]{appendix}
\usepackage{multirow}
\usepackage{rotating}
\usepackage{hyperref}
\hypersetup{
   colorlinks=true,
   linkcolor=blue,
   filecolor=magenta,
   urlcolor=blue,
   citecolor=blue,
	}
\usepackage[
  separate-uncertainty = true,
  multi-part-units = repeat
]{siunitx}
\usepackage{caption} 
\usepackage{subcaption}
\usepackage{amsmath}
\usepackage{gensymb}
\usepackage{mathtools}
\usepackage{amssymb}  
\usepackage{appendix}    
\usepackage{booktabs} 
\usepackage[para,online,flushleft]{threeparttable}
\usepackage{textcomp}
\usepackage[noend]{algpseudocode}
\usepackage{stackengine}
\usepackage{arydshln}
\makeatletter
  \renewcommand*\env@matrix[1][*\c@MaxMatrixCols c]{%
    \hskip -\arraycolsep
    \let\@ifnextchar\new@ifnextchar
  \array{#1}}
\makeatother
\newcommand\correspondingauthor{\thanks{Corresponding author.}}

\makeatletter
\def\BState{\State\hskip-\ALG@thistlm}
\makeatother
\definecolor{light-gray}{gray}{0.9}

\title{Transfer Learning for Autonomous Chatter Detection in Machining}

\author{Melih C. Yesilli\\
				Department of Mechanical Engineering\\
				Michigan State University\\
				yesillim@msu.edu
			\and
				Firas A.~Khasawneh\correspondingauthor\\
				Department of Mechanical Engineering\\
				Michigan State University\\
				khasawn3@egr.msu.edu
			\and
				Brian P.~Mann\\
				MEMS Department\\
				Duke University\\
				brianmann@duke.edu
				}
				
\date{}

\begin{document}
\maketitle

\begin{abstract}
Large-amplitude chatter vibrations are one of the most important phenomena in machining processes. It is often detrimental in cutting operations causing a poor surface finish and decreased tool life. Therefore, chatter detection using machine learning has been an active research area over the last decade. Three challenges can be identified in applying machine learning for chatter detection at large in industry: an insufficient understanding of the universality of chatter features across different processes, the need for automating feature extraction, and the existence of limited data for each specific workpiece-machine tool combination. These three challenges can be grouped under the umbrella of transfer learning. This paper studies automating chatter detection by evaluating transfer learning of prominent as well as novel chatter detection methods. We investigate chatter classification accuracy using a variety of features extracted from turning and milling experiments with different cutting configurations. The studied methods include Fast Fourier Transform (FFT), Power Spectral Density (PSD), the Auto-correlation Function (ACF), Wavelet Packet Transform (WPT), and Ensemble Empirical Mode Decomposition (EEMD). We also examine more recent approaches based on Topological Data Analysis (TDA) and similarity measures of time series based on Discrete Time Warping (DTW). We evaluate the transfer learning potential of each approach by training and testing both within and across the turning and milling data sets. Our results show that carefully chosen time-frequency features can lead to high classification accuracies albeit at the cost of requiring manual pre-processing and the tagging of an expert user. On the other hand, we found that the TDA and DTW approaches can provide accuracies and F1 scores on par with the time-frequency methods without the need for manual preprocessing.
\end{abstract}


\section{Introduction}
\label{sec:Intro}

Advancements in sensor technology have enabled researchers to investigate the underlying dynamics of complex systems with improved resolution.  For example, it is now possible to measure and examine the complex vibration patterns of machining processes. The primary purpose of machining processes is to remove or subtract material through cutting and to leave behind a desired three-dimensional object. 
Turning and milling processes are amongst the most common types of manufacturing processes. In turning, the workpiece rotates and a non-rotating cutting tool is fed into the workpiece to achieve cutting. In milling, the opposite is true as the workpiece stays aligned with a stationary reference frame and the cutting tool rotates while being fed into the workpiece. 
One of the most challenging parts of these processes is chatter, which occurs due to self-excited tool vibrations, e.g. see references  \cite{taylor1906,Altintas2004,Quintana2011}.
Chatter is highly undesirable as it can leave behind a poor surface finish and can damage the cutting tool and/or machining center. 
Thus the interest in predicting and detecting chatter has become an ongoing problem of interest.    
%has increased in the last several decades.   
% Main goal is to identify chatter characteristics from experimental signals collected via sensors in machining processes.

In machining, natural frequencies of the system shift when cutting configuration parameters such as overhang distance is changed.
The chatter frequency, where chatter takes place in the frequency domain, also changes. 
Since training a classifier on a data set obtained from each new configuration is cumbersome, we are interested in how a trained classifier on one cutting process can transfer knowledge to the different cutting processes. 
This general idea is known as transfer learning in the literature, and within the context of machining, it has the potential to provide a methodology for pooling data from different manufacturing settings to more robustly detect chatter. 

%Outside of machining, some example applications where transfer learning has used are natural language processing \cite{blitzer2006,houlsby2019,ruder2019}, WiFi localization \cite{pan2007,pan2008,zheng2008}, speech recognition \cite{Kunze2017}, and software defect detection \cite{Nam2013,Nam2018}. 

This study analyzes the performance of transfer learning in each of Wavelet Packet Transform (WPT), Ensemble Empirical Mode Decomposition (EEMD), traditional signal processing tools, Dynamic Time Warping (DTW), and Topological Data Analysis (TDA)-based approaches. 
Classifiers were trained and tested from data gathered from milling and turning experiments.  With the exception of DTW, four different classification algorithms were used for all methods: support Vector Machine (SVM), Logistic Regression (LR), Random Forest classifier (RF), and Gradient Boosting (GB). 
K-Nearest Neighbor was used for measuring the performance of the similarity measure technique DTW.
We categorized the methods into three groups: frequency-based methods (WPT/EEMD, FFT/PSA/ACF (FPA)), similarity measure approach (DTW), and the TDA-based approach.
Our results show that frequency-based methods give the highest accuracy in 13 out of 20 transfer learning combinations between the turning and milling data sets.
When the results are compared with respect to the F1-score, it is seen that this number goes up to 16 out of 20 combinations.
However, the results obtained from FPA and WPT, the methods that give the highest accuracy in most of the cases of transfer learning, provide the best accuracy with a larger standard deviation compared to the DTW and TDA-based approaches.
On the other hand, DTW provides the best accuracy when we apply transfer learning across different cutting processes. 
The TDA-based approach provides the highest accuracy and F1-score in only two and three out of 20 combinations of transfer learning, respectively.

The work presented in this paper is organized as follows. Section 2 explains the data collection for both machining processes---milling and turning. Section 3 provides a brief description of transfer learning and its categorization. Section 4 describes the salient aspects of each feature extraction method. Section 5 presents results and also provides some concluding remarks.

\subsection{Feature Extraction Approaches for Machining}

There are several techniques to detect chatter from time series data.
Wavelet Packet Transform (WPT) and Ensemble Empirical Mode Decomposition (EEMD) are the most common methods used in literature \cite{Choi2003,Yao2010,Yesilli2020,Chen2018,Ji2018,Cao2013,Qian2015,Zhang2016,Chen2017,Li2010, Gonzalez2006}.
Chen and Zheng used a Support Vector Machine (SVM) classifier with a Recursive Feature Elimination (RFE) algorithm to detect chatter in an end milling operation \cite{Chen2017}. 
Li et al. proposed an EEMD-based feature extraction method to predict chatter in a  boring process \cite{Li2010}. 
Each of these methods relies upon feature extraction from the decomposition of signals.
To elaborate, experimental data has been decomposed into wavelet packets or intrinsic mode functions (IMFs). The informative parts of the decomposition were then selected based on frequency domain analysis; features were obtained from the IMFs and reconstructed signals from informative wavelet packets. 
However, these two widely used methods have some drawbacks.  In particular, they require manual pre-processing - a step that is difficult to automate.  These methods require the analyst to inspect the frequency domain and energy ratio plots (for WPT) of the time series data to identify the informative decomposition correctly.

Apart from the WPT and EEMD methods, several authors have applied a more traditional feature extraction method based on the Fast Fourier Transform (FFT), Power Spectral Density (PSD), and/or the Autocorrelation Function (ACF). 
Yesilli et al. studied these methods to identify chatter in measured time series from a turning process \cite{Yesilli2020a}. 
Coordinates of peaks in the FFT, PSD, and ACF plots were used as features for supervised learning classification.
Determining the restriction parameters to select peaks correctly is reported as one of the primary challenges or shortcomings of this approach. 
Gradi\v{s}ek et al. used entropy and coarse-grained information rate to develop an automatic chatter detection approach for grinding \cite{Gradivsek2003}. Aslan and Altintas developed an on-line chatter detection approach that uses the Fourier spectrum of the spindle drive motor commands obtained from a CNC milling machine \cite{Aslan2018}.
Li et al. used multiscale permutation entropy, multiscale power spectral entropy, and the Laplacian score to select features for online chatter detection in the milling process, and these features were used in gradient tree boosting~\cite{Li2020a}. Albertelli et al. introduced a chatter detection algorithm based on cyclostationary theory, and it is tested on the data collected from spindle encoder in milling process~\cite{Albertelli2019}. In another recent study, Caliskan et al. introduced an energy-based chatter detection approach where a nonlinear energy operator is used to search for chatter frequency between two tooth passing frequency~\cite{Caliskan2018}. Wan et al. utilized Adaboost with SVM classifiers and used frequency and time domain features in addition to the ones obtained with stacked denoising autoencoder to predict chatter in milling process~\cite{Wan2021}.

The challenges associated with existing chatter detection methods have continued to motivate the need to develop new approaches.   
For example, the similarity measure, Dynamic Time Warping (DTW) algorithm has been proposed to identify chatter in machining \cite{Yesilli2019b}. 
DTW does not require feature extraction and time series data is only used to compute pairwise distance matrix. 
K-Nearest Neighbor (KNN) is used as a classifier to test the performance of the proposed approach.
In addition, persistent homology from Topological Data Analysis has been used in chatter detection \cite{Khasawneh2018,Khasawneh2016,Yesilli2019a,Yesilli2019}, and several authors have explored the use of deep learning to detect chatter~\cite{Tansel1991,Lamraoui2013,Fu2015,Tran2020}.

Feature extraction has been applied to several other machining operations. Wu et al. used sensor fusion and EEMD to extract features from signals obtained from different sensors to predict the remaining useful life of machining tools~\cite{Wu2018}. Plaza et al. compared time direct analysis and power spectral density to WPT and singular spectrum analysis (SSA) for surface finish monitoring~\cite{Plaza2019}. Another study focuses on building a big data processing scheme for features extraction in electrical discharge machining~\cite{Chen2019a}. Cheng et al. extracted features in the time and frequency domain and proposed a generalized multiclass support vector machine (GenSVM) for monitoring health degradation in machining tools~\cite{Cheng2019}.
\subsection{Transfer Learning Approaches for Machining}

Several studies focus on chatter detection using deep learning and transfer learning. Cherukuri et al. use synthetic data to train an artificial neural network (ANN) to predict chatter \cite{Cherukuri2019}. Postel et al. used a pre-trained Deep Neural Network to predict stability in milling operation \cite{Postel2020}. A synthetic data set is used to train the network and then fine-tuning is performed using the small size of an experimental data set. 
Unver and Sener used a numerical simulation of milling operation to train AleXNet structure for Convolutional Neural Networks, and they test the same network on experimental milling data to detect chatter~\cite{Unver2021}.
In addition to chatter detection, the majority of prior works that apply transfer learning focus on fault detection and tool/machine conditioning instead of chatter detection.  
Further, these works utilize deep learning algorithms that require a large number of observations~\cite{Li2020} and do not provide insight into the signals' most informative features for chatter detection. 
For instance, Wu et al. used 1D Convolutional Neural Networks (CNN) for fault detection in bearings and gears \cite{Wu2020}. They applied two different transfer learning approaches: 1) training and testing a classifier on samples from different working conditions and 2) training on simulation data and testing on experimental data. 
Li and Liang developed a CNN-based approach to diagnosing severe tool wear, tool breakage, and spindle failure during machining processes \cite{Li2020}. They used two different CNC machines to train and test a classifier in an experiment that took six months to collect the data needed to train the CNN. 
% We could only find a study that applies regression analysis using transfer learning.
Kim et al.~used Support Vector Regressor to predict the machining power, and they transferred knowledge from machining power models of steel and aluminum to predict the power model of titanium \cite{Kim2021}. 
Mamledesai et al.~utilized CNN and transfer learning to monitor tool conditions to help the machinist decide whether to keep using the same tool or replace it \cite{Mamledesai2020}.
Marei et al. used Convolution Neural Network-based transfer learning to predict tool wear of the carbide cutting tool flank \cite{Marei2021}. 
Another study that includes transfer learning and deep learning is focused on the estimation of force in the milling process using simulation data and experimental data as a source and target domain, respectively \cite{Wang2021}. 
Wang et al. use the pre-trained network VGG19 to identify machining fault types in rolling bearings. 
They modified the final fully connected layer to reduce the number of network parameters and implement the transfer learning between non-manufacturing data and manufacturing data \cite{Wang2020}. 
Kim et al. proposed another approach that converts cutting force signals into images using a multi-layer recurrence plot (MRP) to estimate the machining quality in laser-assisted micro-milling operation \cite{Kim2021a}. They used a pre-trained ResNet-18 CNN structure and tested it on the images generated from cutting signals.

Traditional machine learning approaches are also adopted in transfer learning approaches for machining applications. 
For instance, Gao et al. implemented extreme vector machines and transfer learning to build a prediction model for remaining tool life~\cite{Gao2021}. 
Yesilli et al. combined traditional signal decomposition tools and machine learning algorithms such as support vector machines, random forest classifier, and gradient boosting to detect chatter in experimental turning signals~\cite{Yesilli2020}. 
Fast Fourier Transform, Auto-correlation Function, and Power Spectral Density are also combined with similar machine learning algorithms to identify unstable time series obtained from turning experiments~\cite{Yesilli2020a}. 
Shen et al. combined the TrAdaBoost transfer learning algorithm ~\cite{Dai2007} and singular value decomposition-based feature extraction to identify different fault types in a bearing data set~\cite{Shen2015}. 
The trAdaBoost algorithm is also used in tool tip dynamics prediction~\cite{Chen2019}. 

\subsection{Research Contribution}

%In this paper we do not apply transfer learning within a deep learning framework. 
In this work, we present the first study on using state-of-the-art feature extraction tools to transfer chatter knowledge across turning and milling operations using experimental data. 
The main goal is to automate chatter detection for different cutting conditions and operations and to reduce the amount of data and time needed to train a classifier.
% To our knowledge, this study is the first to apply transfer learning between different cutting operations. 
Once a classifier is trained using a given data set, the gained information can be utilized for different operations without needing large and completely new training data sets from the target process.

This work is different from the prior work of a subset of the authors of this paper on transfer learning both in focus and in the application domain.  
Specifically, in \cite{Yesilli2020} Yesilli et al.~used Wavelet Packet Transform (WPT) and Ensemble Empirical Mode Decomposition (EEMD) to detect chatter in time series data collected only from turning experiments. 
They utilized transfer learning to transfer knowledge in turning operations that were performed using different overhang distances~\cite{Yesilli2020}. 
The goal of that study was to compare the transfer learning capabilities of WPT and EEMD and to examine corresponding feature extraction methods. 
In addition, Yesilli and Khasawneh used only turning data sets to test the transfer learning performance of traditional feature extraction tools such as Fast Fourier Transform (FFT), Autocorrelation Function (AF), and Power Spectral Density (PSD) in \cite{Yesilli2020a}.

Therefore, the work presented here is distinct from prior works, and it provides an approach to leverage existing data sets from one machining operation to make decisions about another operation. 
In contrast to prior works on transfer learning for chatter detection, this manuscript studies a large number of feature extraction methods including WPT, EEMD, FFT, AF, PSD as well as two other methods that have only recently been applied in the context of machining: Topological Data Analysis (TDA) methods and similarity-based methods using Dynamic Time Warping (DTW).

\section{Experimental Procedure}
\label{sec:E_P}

\begin{figure}[hbt!]
\begin{center}
\centering\includegraphics[width=0.7\columnwidth,keepaspectratio]{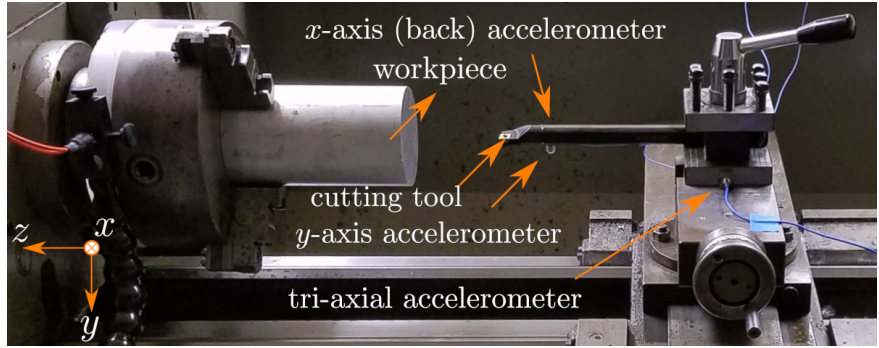}
\caption{Picture of the turning experimental setup showing the cutting tool, workpiece, and accelerometer sensors.}
\label{fig:Exp_Setup}
\end{center}
\end{figure}

\begin{figure}[!htb]
    \centering
    \begin{minipage}{.3\textwidth}
        \centering
        \includegraphics[width=0.8\columnwidth, keepaspectratio]{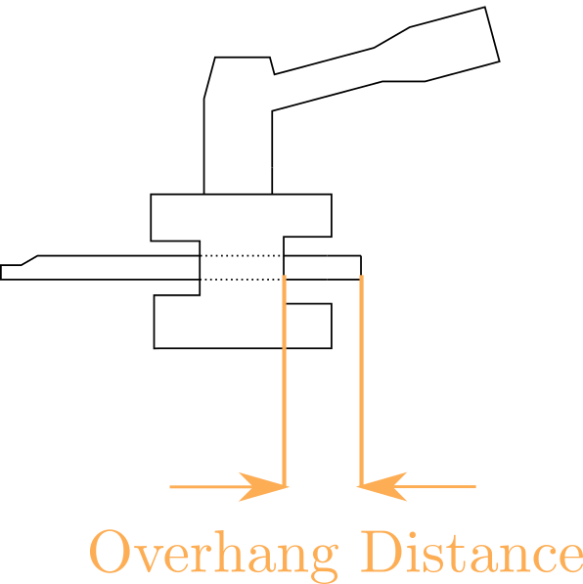}
        \caption{Illustration of the turning process shown to highlight the tool overhang distance.}
        \label{fig:od}
    \end{minipage}%
    \begin{minipage}{0.7\textwidth}
        \centering
        \includegraphics[width=1\columnwidth, keepaspectratio]{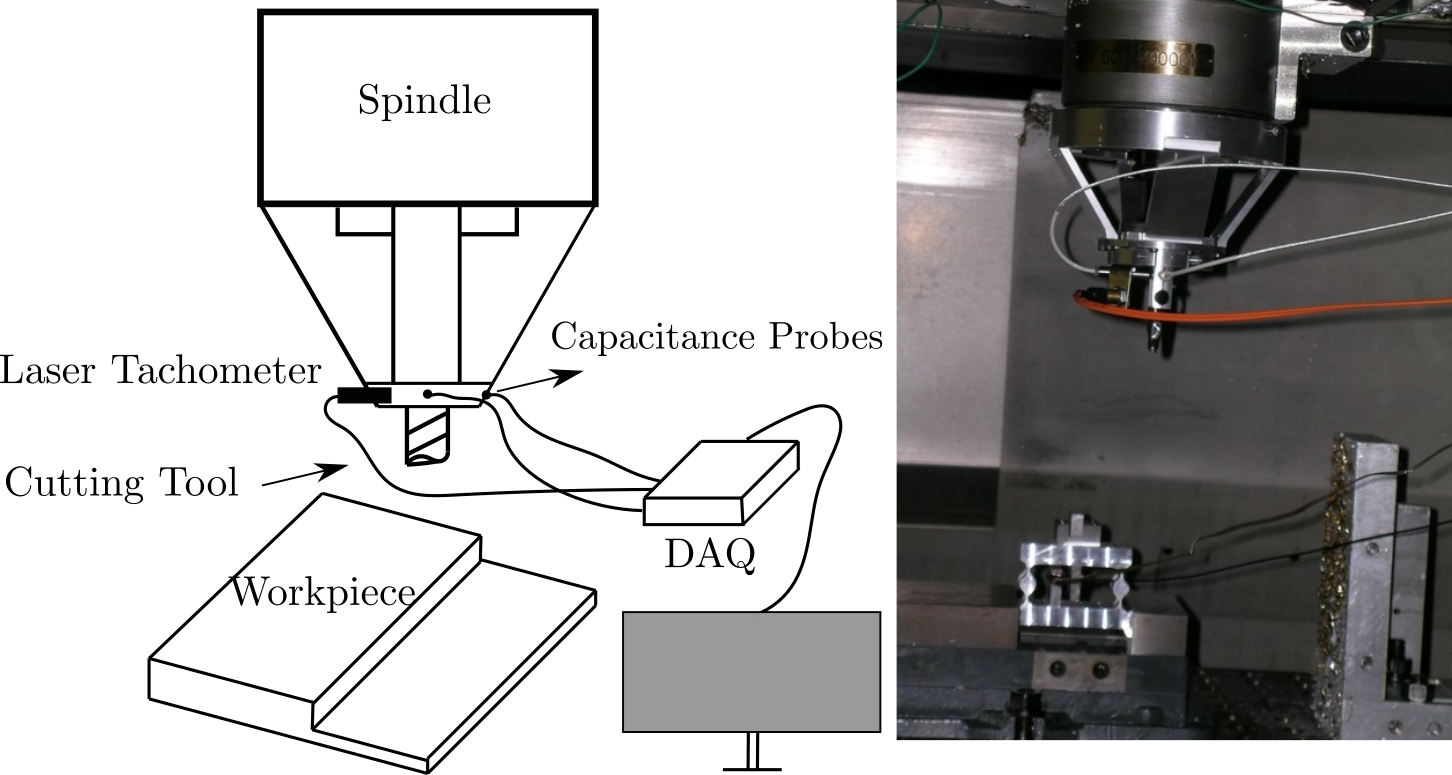}
        \caption{Experimental setup of milling cutting experiments.(left) Illustration of the setup (right) picture of the cutting tool and the workpiece.}
        \label{fig:milling_experimental_setup}
    \end{minipage}
\end{figure}

This section describes the experimental setups and the data collected from each experiment.  
In the turning experiment, three accelerometers were attached to the setup, see locations in  Fig.~\ref{fig:Exp_Setup}. The workpiece was fixed to the spindle and the cutting tool was positioned at the tip of a boring bar.
The distance between the heel of the boring bar and the back of the tool holder is called the overhang distance, see the illustration of  Fig.~\ref{fig:od}.  
The overhang distance is an important characteristic for the turning experiment since the stiffness of the boring bar can be altered by simply changing this distance. 
In this experiment, data is collected for four different overhang distances with varying rotational speeds and cutting depths; the overhang distances used in the experiments were 5.08, 6.35, 8.89, and 11.43 cm. 
The range of the rotational speeds and the depth of cuts for turning and milling experiment is provided in Tab.~\ref{tab:cutting_parameters}. 
We only used the $x$-axis data from the triaxial accelerometer since it had the best signal-to-noise ratio.
The experimental data was oversampled at 160 kHz since we did not use an in-line analog filter.
We then used a low-pass filter (Butterworth) with an order of 100, and we downsampled the data to 10 kHz. Note that the downsampling does not lead to a loss of useful information since the accelerometers used in the experiment had a 10 kHz bandwidth. 

\begin{table}[h]
\centering
\caption{Cutting parameters for the turning and milling experiments.}
%\resizebox{0.6\columnwidth}{!} {
\begin{tabular}{c|c c c}
\label{tab:cutting_parameters}
\makecell{Cutting operation}  & \makecell{Rotational Speeds (rpm)} & Depth of Cut (mm) & \makecell{Feed Rate}\\
\hline
Turning - 5.08 cm &320, 425, 570, 770 & 0.0254 - 1.27 &  0.00508 cm/rev (0.002 inch/rev)\\
Turning - 6.35 cm & 570, 770&  0.0508 - 0.381&  0.00508 cm/rev (0.002 inch/rev)\\
Turning - 8.89 cm & 570, 770, 1030 & 0.0381 - 0.762  &  0.00508 cm/rev (0.002 inch/rev)\\
Turning - 11.43 cm & 570, 770, 1030 &  0.127 - 1.016&  0.00508 cm/rev (0.002 inch/rev)\\
Milling & 11206-32161 & 0.381 - 3.556 & 0.191 mm/tooth/rev\\
\hline
\end{tabular}
%}
\end{table} 

Each time series was tagged based on a time and frequency domain analysis. 
The label options were stable (chatter-free), intermediate (mild chatter), and unstable (chatter). Low-amplitude signals in the time and frequency domain were tagged as stable. 
If the time series had a low amplitude in the time domain and a large amplitude in the frequency domain, it was tagged as mild chatter. 
The signals with large amplitude in both domains were labeled as unstable. 
The remaining signals were assigned as unknown since they do not fit into these criteria.  

%\begin{figure}[hbt!]
%\begin{center}
%\centerline{\includegraphics[width=0.25\columnwidth,keepaspectratio]{Stickout_Explanation}}
%\caption{Illustration of the turning process shown to highlight the tool overhang distance.}
%\label{fig:od}
%\end{center}
%\end{figure}
%
%\begin{figure}[hbt!]
%\begin{center}
%\centerline{\includegraphics[width=0.45\columnwidth,keepaspectratio]{experimental_setup_milling}}
%\caption{Experimental setup of milling cutting experiments.}
%\label{fig:milling_experimental_setup}
%\end{center}
%\end{figure}

\begin{figure}[hbt!]
\begin{center}
\centerline{\includegraphics[width=1\columnwidth,keepaspectratio]{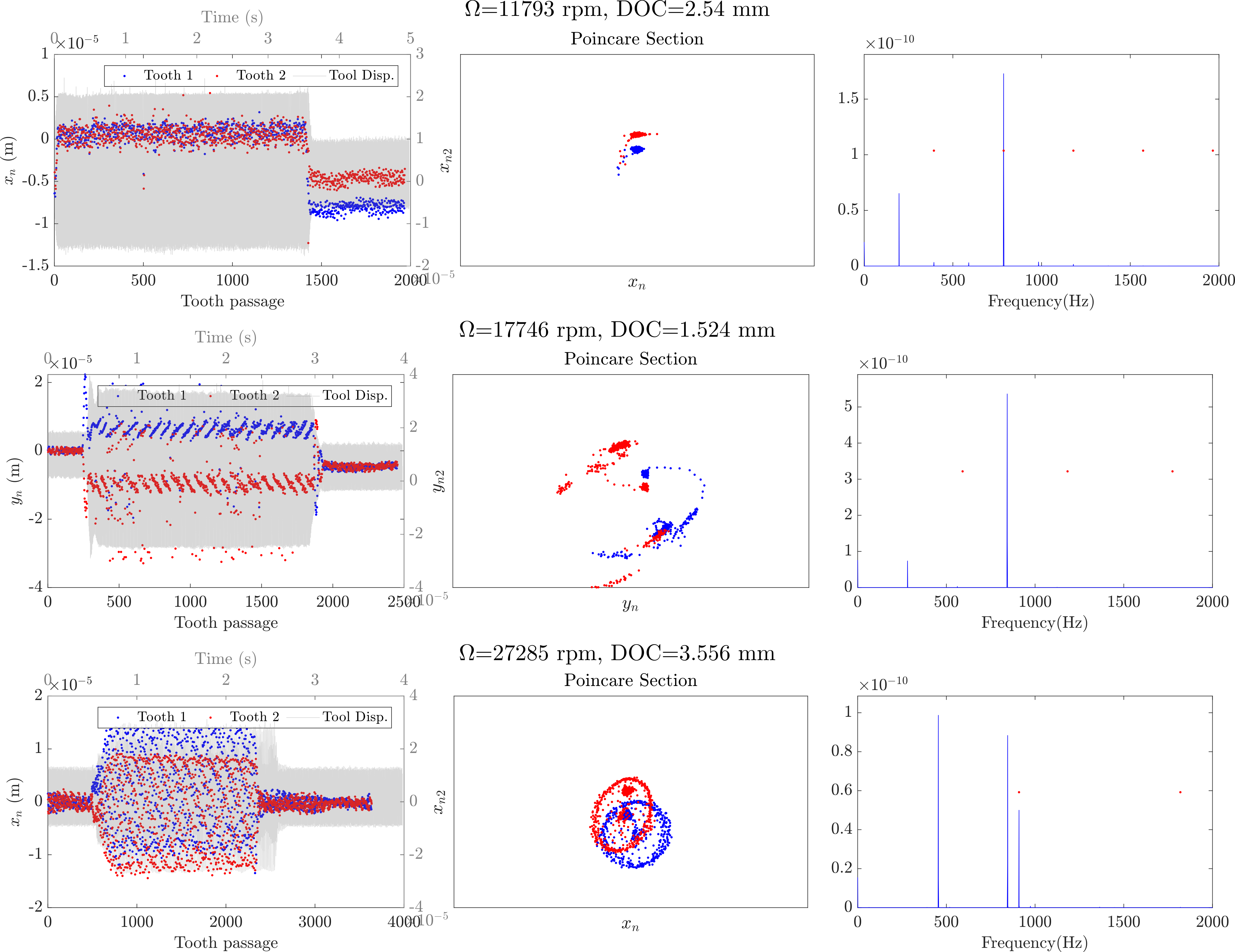}}
\caption{The first column represents tool displacements for two teeth and the second column provides Poincare sections for two time series obtained with three different rotational speeds and depth of cuts in milling experiments. The third column shows the PSD plots of the three-time series whose Poincare sections are shown in the first column. Red dots represent tooth passage frequency. The time series shown in the first row is stable with cutting conditions $\Omega=11793$ rpm and depth of cut of $2.54$ mm, while the one in
the second row is unstable with cutting conditions $\Omega=17746$ rpm and depth of cut of $1.524$ mm.
The third row represents an unstable milling signal with cutting conditions $\Omega=27285$ rpm and depth of cut of $3.556$ mm.}
\label{fig:poincare}
\end{center}

\end{figure}

An illustration showing the milling experimental system is shown in Fig.~\ref{fig:milling_experimental_setup}.
An Ingersol machining center with a Fischer 40000 rpm and 40kW spindle was used to conduct experiments on an aluminum workpiece (7050-T7451). The type of milling conducted in these experiments is down milling. The depth of cut is 2.03 mm and radial immersion is kept constant at 5\%.
Lion precision capacitive probes were used to collect the tool displacements along the $x$ and $y$ axes \cite{Mann2005}. 
The data were sampled at 25 kHz. 
As in the turning experiments, a low pass filter was used and the data was downsampled to 12.5 kHz.  In addition, a laser tachometer was used to independently verify the spindle rotational speed from the machine setting. 
The cutting tool was a 19.05 mm end mill with two teeth and a 106 mm overhang distance. 
Data tagging was performed using power spectral density (PSD) plots and Poincar\'{e} sections. 
Tool displacement plots in the $x$ direction, along with the corresponding Poincar\'{e} sections, are shown in Fig.~\ref{fig:poincare}.

% \begin{figure}
% \begin{center}
% \centerline{\includegraphics[width=0.6\columnwidth,keepaspectratio]{PSD_Plots}}
% \caption{PSD plots of the two time series whose poincare sections shown in Fig.~\ref{fig:poincare}. Red dots represent tooth passage frequency. (First row) stable time series, (second row) unstable time series.}
% \label{fig:PSD}
% \end{center}
% \end{figure}

The first milling example shown in Fig.~\ref{fig:poincare} is a stable cut ($\Omega=19488$~rpm, $1.524$~mm cutting depth) whereas the second example shows a Hopf bifurcation example ($\Omega=27285$~rpm, $3.556$~mm cutting depth). The first column of Fig.~\ref{fig:poincare} presents the tool displacements for two teeth and the second column provides the Poincare sections of these time series. $x_{n}$ and $x_{n2}$ represent the time-delayed coordinates. In this work, we use constant time delay parameter chosen as 6.
The third column of Fig.~\ref{fig:poincare} shows the power spectrum or a PSD plot that helped us better see the frequency content of the time series \cite{Mann2004}. 
If the spectral peaks were synchronous with the tooth passage frequency, that meant the corresponding time series was stable.  However, if the spectral peaks were not aligned with the tooth passage frequency, the cutting test was unstable, as shown in the second row and third column of Fig.~\ref{fig:poincare}.  

\begin{figure}[!htb]
    \centering
    \begin{minipage}{.4\textwidth}
        \centering
        \includegraphics[width=0.9\columnwidth, keepaspectratio]{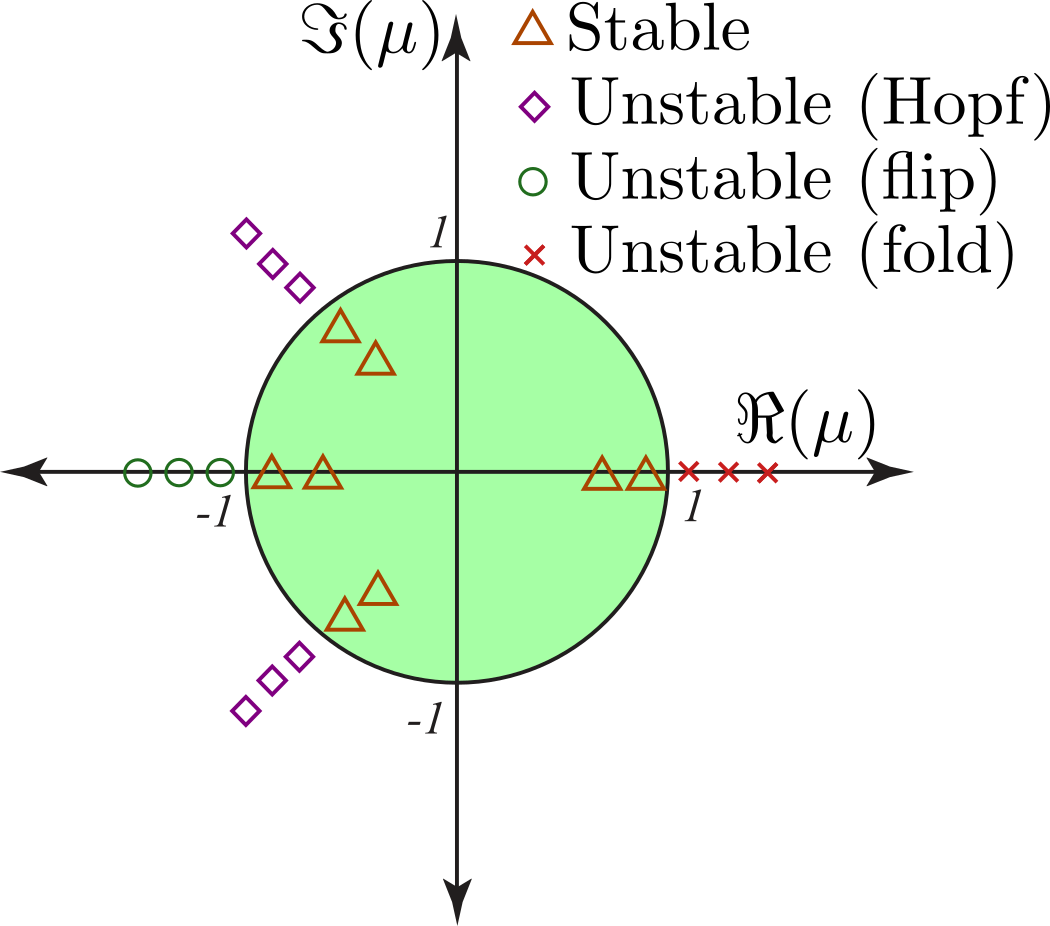}
        \caption{Illustration for the stability analysis using eigenvalues of the dynamic map obtained using spectral element approach~\cite{Khasawneh2011}.}
        \label{fig:stab_circ}
    \end{minipage}%
    \begin{minipage}{0.6\textwidth}
        \centerline{\includegraphics[width=1\columnwidth,keepaspectratio]{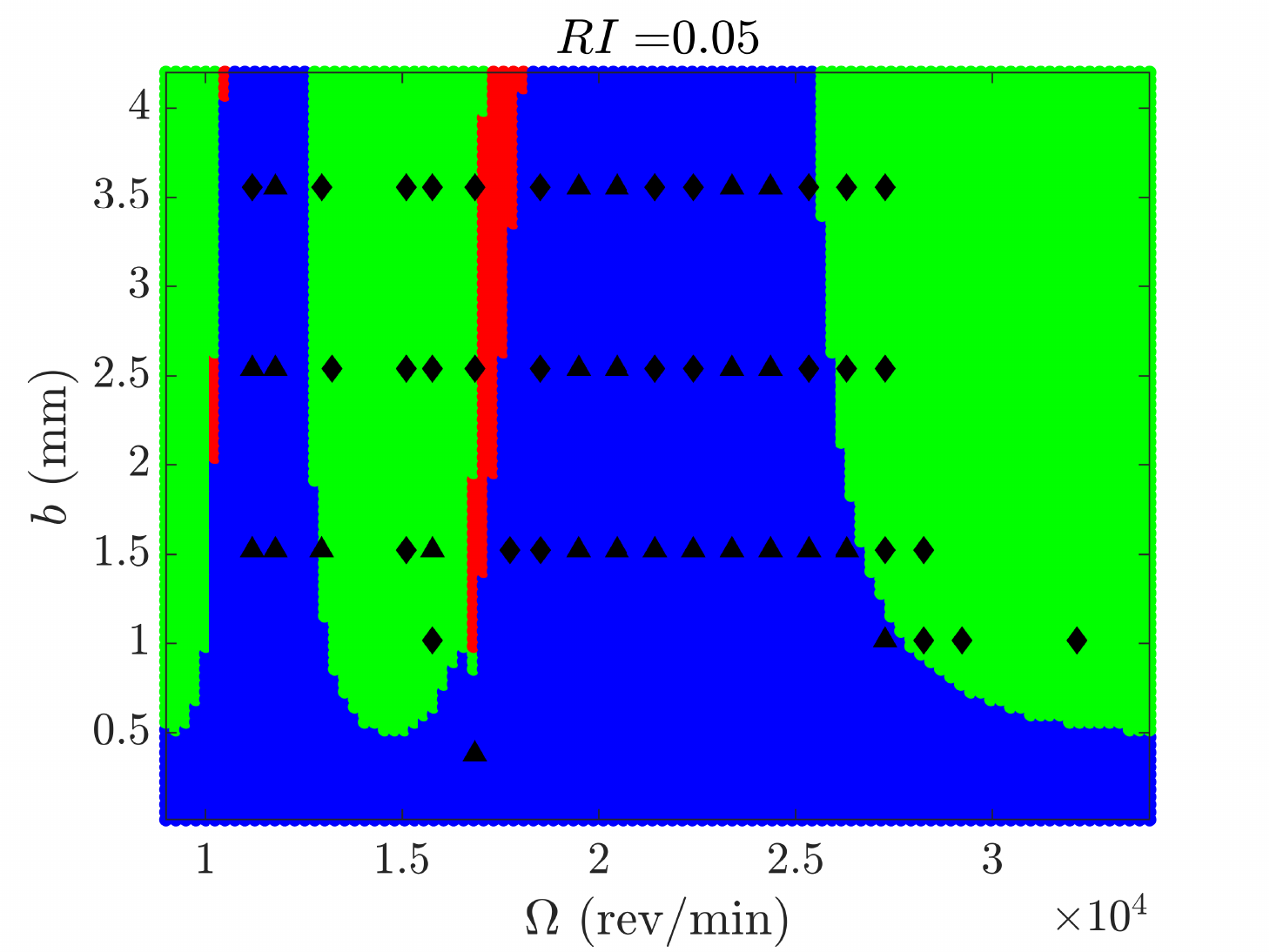}}
        \caption{The stability of time series obtained using the analytical model provided in Ref.~\cite{Yesilli2019} with different depth of cut ($b$) and spindle speed ($\Omega$) on $100\times100$ grid. The green color corresponds to time series with Hopf bifurcation (unstable), while the blue color represents the stable time series. The red color shows the time series with flip bifurcation. Experimental data, whose stability is defined based on the Poincare section and PSD plots, is shown with diamond (unstable) and triangle (stable) symbols.}
        \label{fig:Stability_Diagram}
    \end{minipage}
\end{figure}

Stability predictions were made for the milling system using the measured modal parameters and the spectral element approach \cite{Khasawneh2011}; the resulting stability diagram is shown in Fig.~\ref{fig:Stability_Diagram}.  The spectral element method uses eigenvalues of a dynamic map to determine the stability of the process \cite{Mann2004,Yesilli2019}. 
If the magnitude of the eigenvalue is smaller than 1, the process is stable.  If the eigenvalue has only a positive real part that is larger than 1, then the process is unstable.
Eigenvalues with only negative real part less than -1 represent the flip bifurcations, and Hopf bifurcation occurs when an eigenvalue has amplitude larger than 1. The illustration for the stability analysis using the real and imaginary parts of the eigenvalues is also provided in Fig.~\ref{fig:stab_circ}.
Based on this stability criteria, 10000 time series for a $100\times 100$ grid of points in the rpm vs. cutting depth parameter space were used to produce the stability diagram shown in Fig.~\ref{fig:Stability_Diagram}. 
The stability of the experimental data set is decided based on the Poincare section and the PSD plots. Then we included experimental data set in the stability diagram shown with black diamonds (unstable) and triangle (stable) markers in Fig.~\ref{fig:Stability_Diagram}.
It is seen that the labels obtained using the eigenvalues and the ones obtained using frequency domain analysis and Poincare section may not match. We use the labels obtained from the analysis shown in Fig.~\ref{fig:poincare} to perform classification.

% \begin{figure}
% \begin{center}
% \centerline{\includegraphics[width=0.6\columnwidth,keepaspectratio]{figures/Stability_Diagram_2Labels.pdf}}
% \caption{Time series on $100\times100$ grid and their stability shown with colors. Green color corresponds to time series with Hopf bifurcation (unstable), while blue color represents the stable time series. Red color shows the time series with flip bifurcation. Experimental data, whose stability is defined based on Poincare section and PSD plots, is shown with diamond (unstable) and triangle (stable) symbols.}
% \label{fig:Stability_Diagram}
% \end{center}
% \end{figure}

\section{Transfer Learning}
\label{sec:TL}

\begin{figure}[hbt!]
\begin{center}
\centerline{\includegraphics[width=0.75\columnwidth,keepaspectratio]{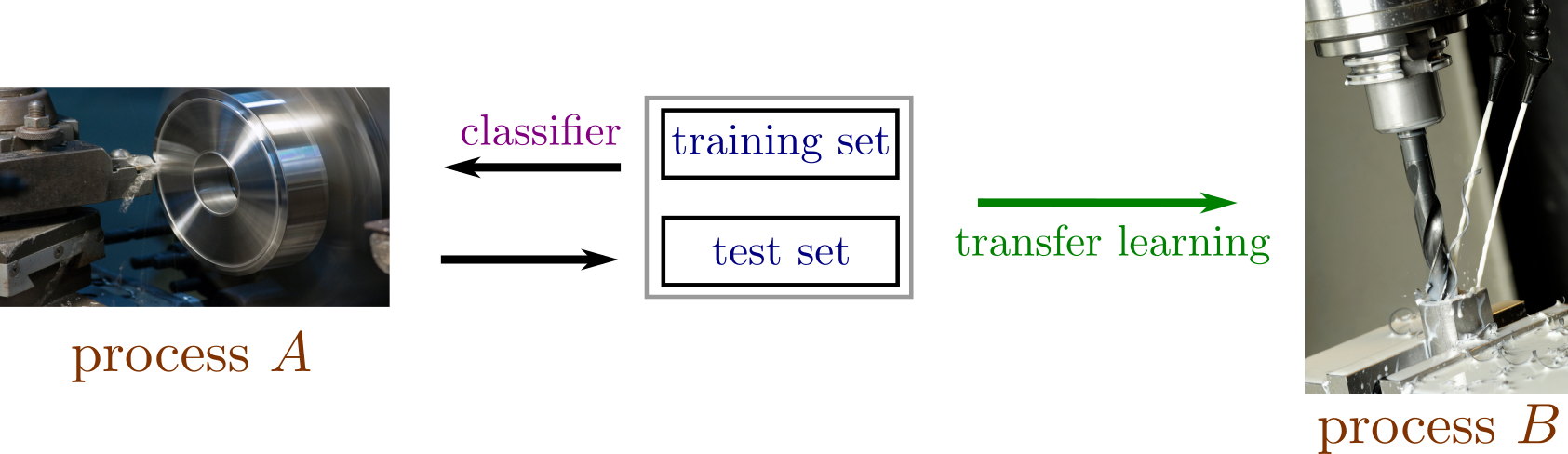}}
\caption{An example of transfer learning where training for chatter detection is performed using a turning process (the source), and the gained knowledge is imported via transfer learning to a milling operation (the target).}
\label{fig:TL}
\end{center}
\end{figure}

In traditional machine learning, a classifier is trained and tested on a data set originating from the same source.
However, real-life applications, such as chatter or fault detection in machining, can experience a shift in the parameters between the time the classifier was trained and the time the system is put into operation. 
This means that the data collected from these applications may no longer have the same feature space as the training set.  
Therefore, traditional machine learning can require data collection for each parameter combination thus leading to increased cost and low automation potential. 
As another motivating example, some experiments are expensive to set up and perform. 
This includes chatter studies which result in long downtime for production machines and personnel during the data collection phase. 
Besides the cost, some sensor data may be collected during machining one-off products, and therefore may be considered of limited use in traditional machine learning settings. 
Therefore, it is extremely beneficial to leverage extracted features related to similar phenomena across different settings and operations. 
In this case, \textit{Transfer Learning} presents a useful machine learning framework that allows training and testing on data sets from different sources. 
As an example, Fig.~\ref{fig:TL} shows a transfer learning application where a chatter classifier was trained using a turning process, and the gained information is then transferred for detecting chatter in a milling operation.

Transfer learning is categorized according to the similarity between tasks and the domain of each source and target.
The source is the system we use to train a classifier, while the target is the system where the classifier is tested. 
There are two main terms in the definition of transfer learning, and these are domain and task. 
A domain can be described as the combination of a feature space $\mathcal{F}$ and the marginal probability of the feature space $P(F)$, while the task contains a label space $\mathcal{L}$ and the conditional probability ($P(l|f)$)~\cite{Pan2010}.
$\mathcal{F}$ represents the space of feature vectors, $\mathbf{x_{i}}$, and $F$ is the an instance set such that $F = \{\mathbf{f} \mid \mathbf{f_{i}} \in \mathcal{F}, i=1,\ldots, n\}$~\cite{Zhuang2020}.
For a given domain, $\mathcal{D} = (\mathcal{F},P(F))$, a task is defined as $\mathcal{T} = (\mathcal{L},P(l|f))$. $P(l|f)$ is also considered as a predictive function $f$ which estimates the label for a given feature space.

\begin{figure}[hbt!]
\begin{center}
\centerline{\includegraphics[width=0.5\columnwidth,keepaspectratio]{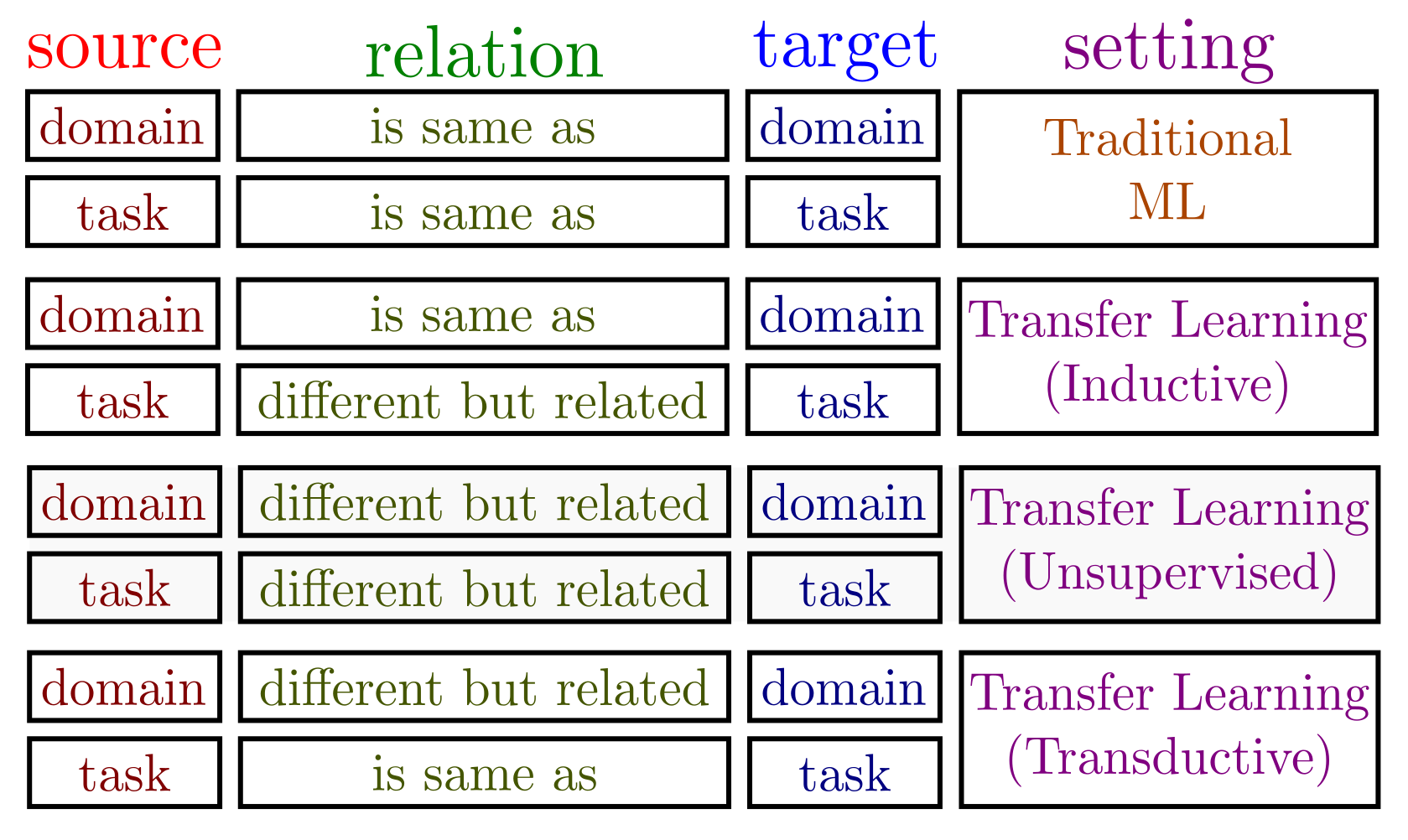}}
\caption{Categorization of transfer learning.}
\label{fig:Categorization}
\end{center}
\end{figure}

Based on the differences between domains and tasks of the source and the target, several transfer learning settings can be obtained (see Fig.~\ref{fig:Categorization}). We refer the interested reader to~\cite{Pan2010,Weiss2016,Lu2015} for more details on transfer learning. 
In this study, our machine learning framework is included under \textit{inductive} transfer learning category because we use the same sets of features for the source and the target. The main purpose of inductive transfer learning is to improve the performance of target prediction function $f_{T}$ using the information in domain and task of the source $\mathcal{D}_{S}$ and $\mathcal{T}_{S}$, respectively~\cite{Pan2010}. There are several approaches to transfer learning. These include instance-transfer, feature representation transfer, parameter-transfer, and relational knowledge transfer~\cite{Zhuang2020}. In this work, we transfer the knowledge of parameters by using the same trained classifier in the testing phase. We used the same set of features for training and testing.
However, the distribution of the features is different in each domain since the source and the target are represented by two different machining processes: turning and milling.
More details about the application of \textit{inductive} transfer learning are available in Sec.~\ref{sec:Results}.

\section{Methods}
\label{sec:Methods}

\begin{figure}
\begin{center}
\centerline{\includegraphics[width=0.8\textwidth,keepaspectratio]{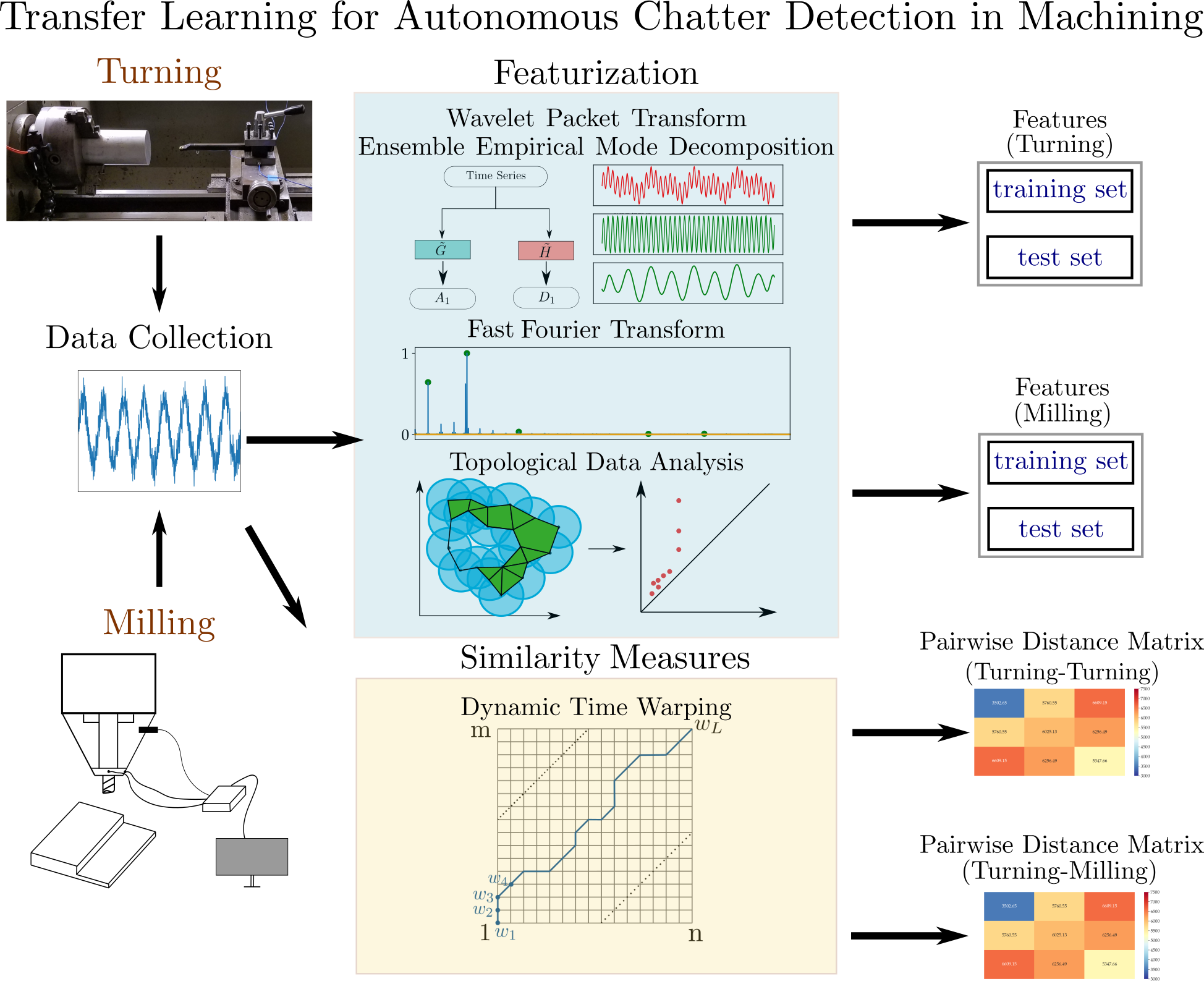}}
\caption{Outline of the general procedure and the featurization methods used in this paper.}
\label{fig:block_diagram}
\end{center}
\end{figure}

\begin{figure}
\begin{center}
\centerline{\includegraphics[width=1\textwidth,keepaspectratio]{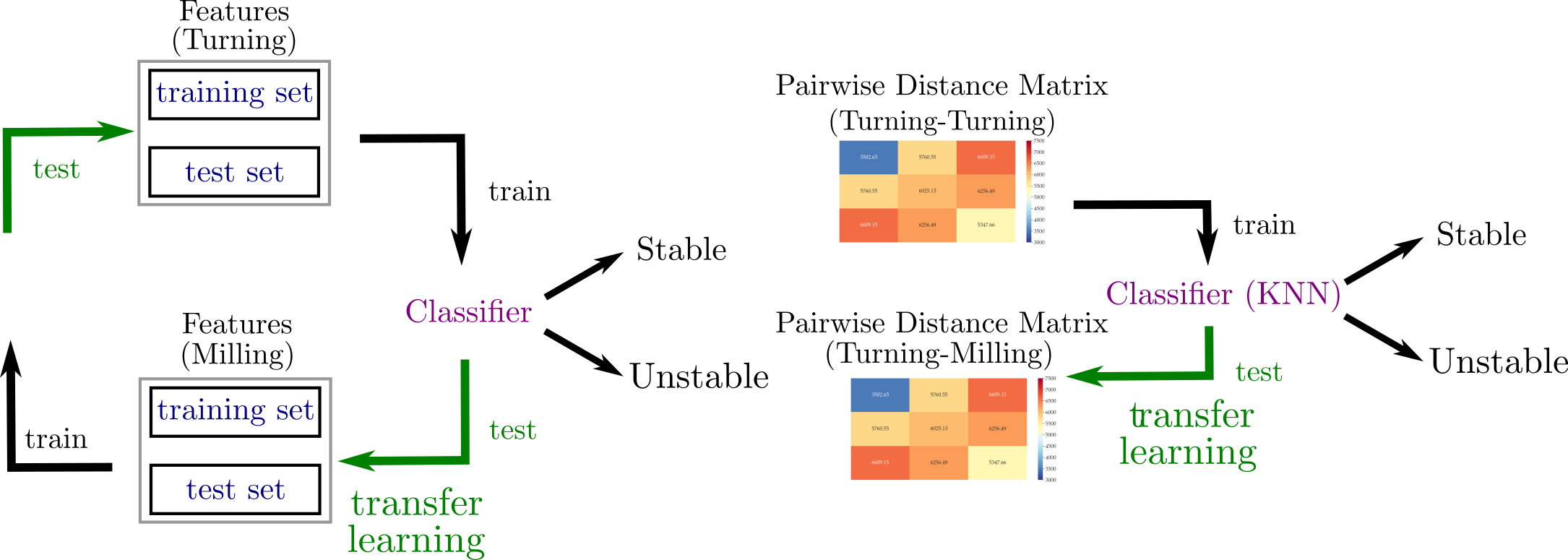}}
\caption{Transfer learning approach used in this paper for feature extraction and similarity measure-based approaches.}
\label{fig:block_diagram2}
\end{center}
\end{figure}

% This section provides a description of each feature extraction approach.  
This section provides a brief description of each feature extraction method that we used. In addition Fig.~\ref{fig:block_diagram} provides a block diagram that explains the procedure we followed in this study. Specifically, the leftmost block shows the experimental setup and the data collection process. This is followed by the middle block which lists the featurization methods used, as well as the similarity-based approach using DTW. The rightmost block shows the pairwise distance matrices and feature matrices obtained from the similarity-based approach and the feature extraction approaches, respectively. Figure~\ref{fig:block_diagram2} provides a cartoon of the transfer learning framework whereby classifiers trained on the turning data are used to detect chatter in milling and vice versa. Each of the following subsections covers one of the methods described in the middle block as follows: Section \ref{sec:wpt} summarizes WPT, Section \ref{sec:eemd} discusses EEMD, Section \ref{sec:traditional} describes FFT, PSD, and ACF approaches, Section \ref{sec:dtw} explains DTW, while Section \ref{sec:tda} deals with topological features from TDA.

%%

% ------------------------------------- WPT -------------------------------------------------%
\subsection{Wavelet Packet Transform (WPT)} \label{sec:wpt}
\begin{figure}
\begin{center}
\centerline{\includegraphics[width=0.5\columnwidth,keepaspectratio]{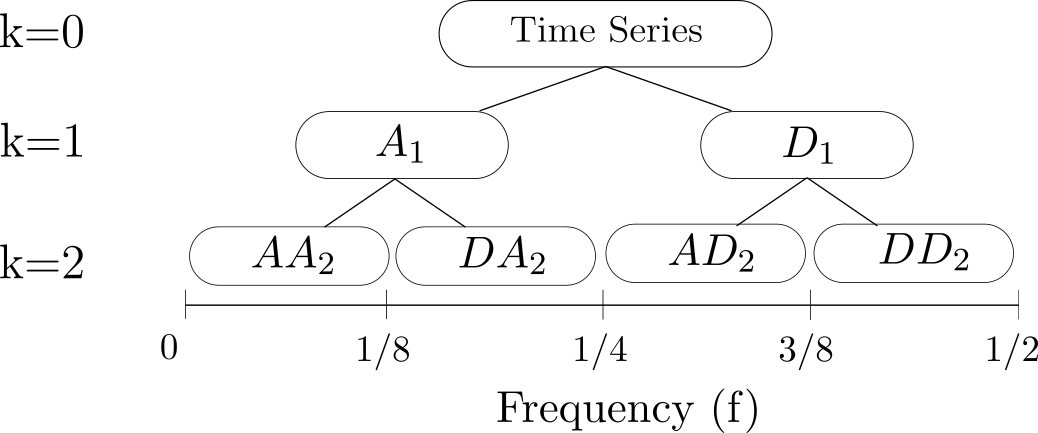}}
\caption{Illustration of two level WPT.}
\label{fig:wavelet_tree}
\end{center}
\end{figure}

This section describes the salient details for the Wavelet Packet Transform (WPT) method. One can refer to Ref.~\cite{Yesilli2020} for more information. The WPT method decomposes a signal into approximation and detail coefficients at each level of the transform.  Figure~\ref{fig:wavelet_tree} provides the decomposition of a time series into two levels of WPT and shows the corresponding frequency content for each wavelet packet. 
Detail and approximation coefficients are obtained after applying the high-pass and low-pass filters, respectively.
They are denoted as $D_{i}$ and $A_{i}$ as shown in Fig.~\ref{fig:wavelet_tree}. 
At each level of transform, we added additional letters $A$ or $D$ to the left side of the previous notation, and the indices change with respect to the level of the transform. 
For example, in the second level of transform, the approximation coefficient $A_{1}$ passes through the high pass filter and becomes $DA_{2}$ (see Fig.~\ref{fig:wavelet_tree}).
In addition, the number of wavelet packets at level $k$ of the transform is $2^{k}$. 

\begin{figure}[hbt!]
\begin{center}
\centerline{\includegraphics[width=0.7\columnwidth,keepaspectratio]{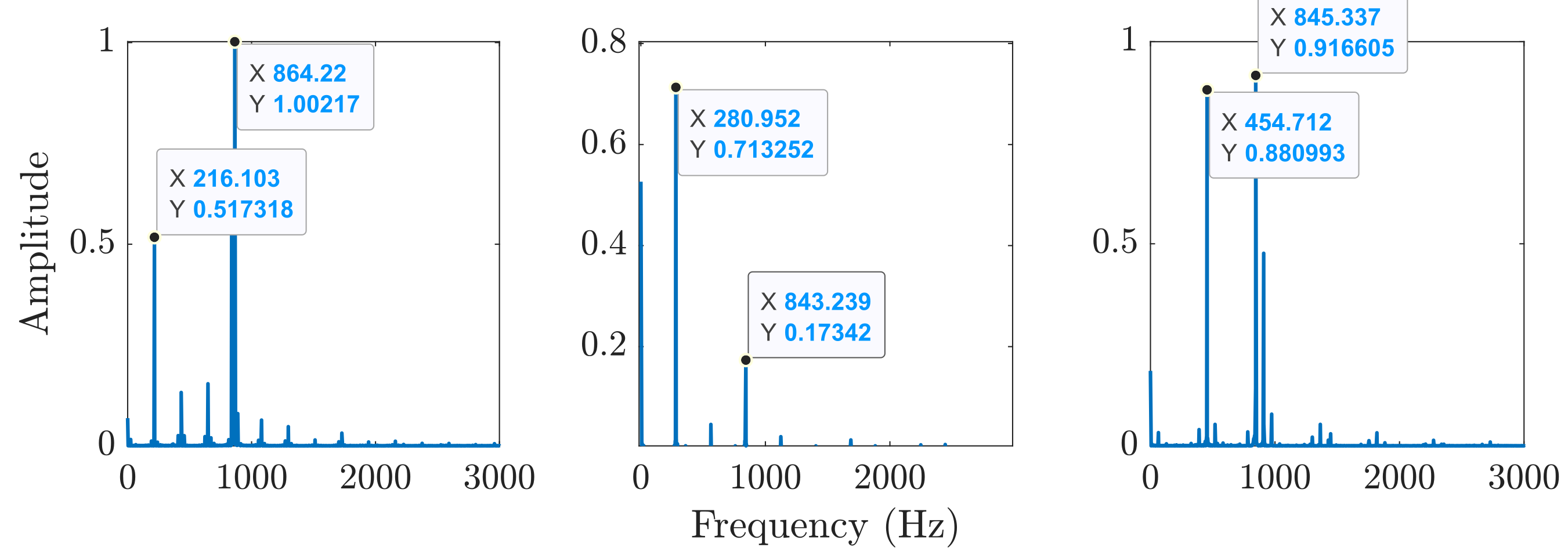}}
\caption{Spectrum of three different time series from milling experiment: (left) 13227 rpm, 2.54 mm depth of cut (doc), unstable,(mid) 16861 rpm, 1.905 mm doc, stable, (right) 27285 rpm, 1.905 doc, unstable.}
\label{fig:FFT_plots}
\end{center}
\end{figure}

\begin{figure}[hbt!]
\begin{center}
\centerline{\includegraphics[width=0.6\columnwidth,keepaspectratio]{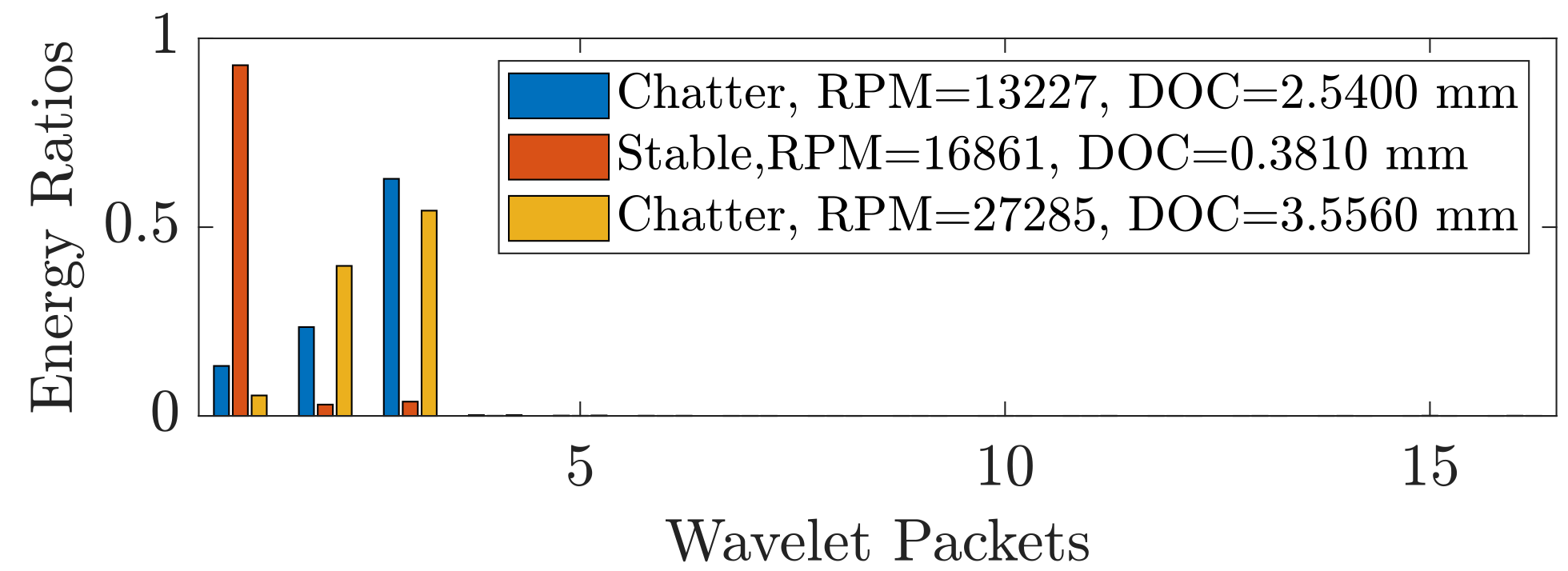}}
\caption{Energy ratio of the wavelet packets obtained from decomposition of the three time series whose spectrum is provided in Fig.~\ref{fig:FFT_plots}. (Blue bars) Milling - Unstable - RPM=13227 - DOC = 2.54mm. (Red bars) Milling - Stable - RPM=16861 - DOC = 0.38 mm. (Orange bars) Milling - Unstable - RPM=27285 - DOC = 3.556 mm.}
\label{fig:Energy_ratio}
\end{center}
\end{figure}

\begin{figure*}[hbt!]
\begin{center}
\centerline{\includegraphics[width=1\columnwidth,keepaspectratio]{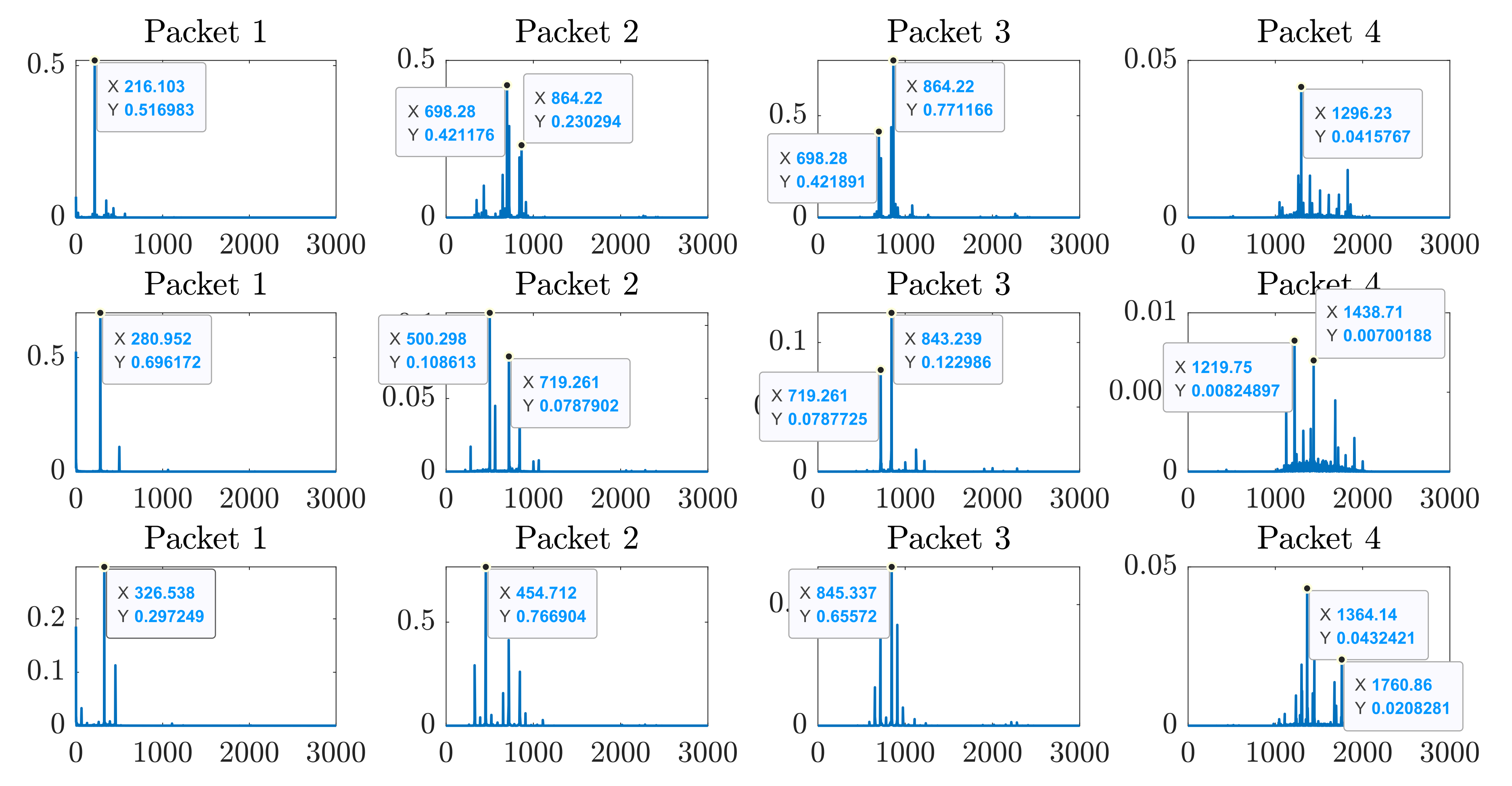}}
\caption{Spectrum of reconstructed signals from first four wavelet packets of three different time series whose spectrum shown in Fig.~\ref{fig:FFT_plots}. (First row) Milling, 13300 rpm, 2.54 mm depth of cut (doc), unstable, (second row) milling, 17300 rpm, 0.3810 mm doc, stable, and (last row) milling, 28000 rpm, 3.5560 mm doc, unstable.}
\label{fig:FFT_recon_plots}
\end{center}
\end{figure*}

\textbf{Milling data set}:
The procedure followed in this study is the same as the one described in Ref.~\cite{Yesilli2020}.
For both the milling and turning data sets, we applied level 4 WPT to downsampled time series.
The first step was to define the chatter frequency by checking the spectrum of the downsampled data. 
Figure~\ref{fig:FFT_plots} provides FFT plots of three different time series from the milling data set. 
It can be seen that the chatter frequency is around 850 Hz which is close to the resonant frequency of 728.7 Hz; this leads us to look for wavelet packets that also have a frequency content near this frequency.
Time series were decomposed into wavelet packets and the energy ratio of each wavelet packet was computed. 
The energy ratio plots and the Fast Fourier Transform (FFT) of the signals, reconstructed from the packets, have been provided in  Fig.~\ref{fig:Energy_ratio} and~\ref{fig:FFT_recon_plots}. 
Fig.~\ref{fig:Energy_ratio} show that most of the energy belongs to the third wavelet packet for the unstable time series.
It is also seen that the spectrum of the third wavelet packet, the unstable time series, has a frequency content of around 1000 Hz. 
Thus the third wavelet packet can be selected as the informative packet for feature extraction.  


\textbf{Turning data set:} We used the same informative wavelet packets defined in Ref.~\cite{Yesilli2020} to extract features. The informative wavelet packet numbers are listed in Tab.~\ref{tab:informative_wavelets} for four overhang distances.

\begin{table}[h]
\centering
\caption{Informative wavelet packet numbers retrieved from Ref.~\cite{Yesilli2020}.}
\resizebox{0.5\columnwidth}{!} {
\begin{tabular}{c|c c}
\label{tab:informative_wavelets}
\makecell{Overhang \\distance (cm)} & \makecell{Chatter frequency\\ range (Hz)} & \makecell{Informative \\ wavelet packet}\\
\hline
5.08 & 900-1000 & Level 4: 3\\
6.35 & 1200-1300 & Level 4: 4\\
8.89 & 1600-1700 & Level 4: 6\\
11.43 & 2900-3000 & Level 4: 10\\
\hline
\end{tabular}}
\end{table} 

After defining the informative wavelet packets for both data sets, we reconstructed the signals from the informative wavelet packets and computed time and frequency domain features, as described in Ref.~\cite{Yesilli2020}.

% ------------------------------------- EEMD -------------------------------------------------%
\subsection{Ensemble Empirical Mode Decomposition (EEMD)} \label{sec:eemd}
Ensemble Empirical Mode Decomposition (EEMD) is a  modified version of Empirical Mode Decomposition (EMD). 
EMD is a nonlinear transform and it is suitable for non-stationary signals \cite{Huang1998}.
The main difference between  EMD and WPT is that the decomposition in EEMD does not correspond to a certain frequency region, while wavelet packets in WPT have a frequency content only in a certain frequency range. 
The decomposition of a signal $s(t)$ is obtained via a sifting process and the steps are as follows:
\begin{enumerate}
\item Find all the local minimum and maximums and obtain the upper and lower envelope of $s(t)$
\item  Take mean ($m_{i}(t)$) of the upper and lower envelope
\item Subtract the mean from $s(t)$ and the remaining is called $h_{1i}(t)$.
\end{enumerate}

 $h_{1i}(t)$ is the first guess of the first IMF. However, it needs to satisfy the following two conditions, see \cite{Huang1998}, to become an IMF:
 
\begin{itemize}
\item The mean value of the upper and lower envelope should be zero at any point.
\item The number of zero crossings and the number of extrema should be the same or the difference between them should be at most one. 
\end{itemize}

The sifting process was thus repeated until $h_{1i}(t)$ had the aforementioned properties, and $h_{1i}(t)$ was treated as new data in each iteration. The standard deviation between the consecutive estimates of an IMF was used as a  stoppage criterion and it was generally set to a value between $0.2$ and $0.3$ \cite{Huang1998}.
After satisfying the criterion, the IMF $c_{i}(t)$ was obtained and the residue was  computed using 

\begin{equation*}
r_{i}(t) = r_{i-1}(t)-c_{i}(t).
\end{equation*}

Note the original signal was the first residue $r_{0}(t)$. The sifting process was repeated until we had a monotonic function as a residue and thus could not extract an IMF from that function anymore.  Thus the original signal could be written in terms of the IMFs and the last residue as 
\begin{equation*}
s(t) = \sum_{i=1}^{n}c_{i}+r_{n}.
\end{equation*}
One of the main drawbacks of EMD is the mode mixing problem, which is that an IMF can contain signals with totally different frequency content or the same frequency content appears in different IMFs. To solve this problem, Ensemble Empirical Mode Decomposition (EEMD) was proposed by  Wu and Huang \cite{ZHAOHUA2009}. 
The following steps were used to apply an EEMD algorithm
\begin{enumerate}
\item White noise is added to original data 
\item Obtain IMFs of the signal using EMD
\item Repeat (1) and (2) with different white noise each time. The number of repetitions is called the ensemble size.
\item Compute the mean of corresponding IMFs and residues in the ensemble.
\end{enumerate}
\begin{figure}[hbt!]
\begin{center}
\centerline{\includegraphics[width=0.7\columnwidth,keepaspectratio]{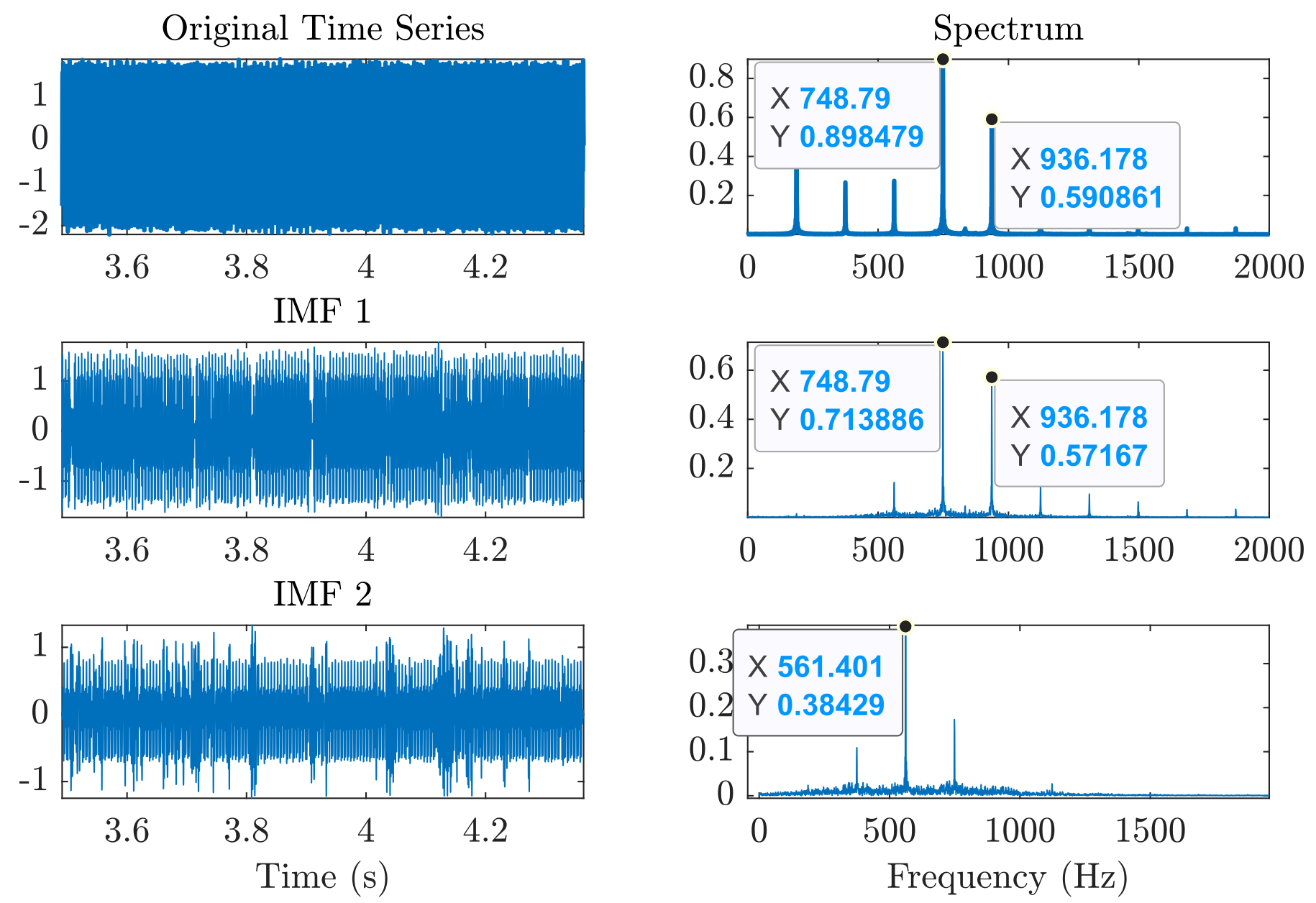}}
\caption{Intrinsic mode functions and their spectrum for the time series with 11210 rpm and 3.556 mm depth of cut from milling experiments.}
\label{fig:EEMD_informative_IMF}
\end{center}
\end{figure}

The experimental signal was decomposed into IMFs and the informative IMFs were selected to generate a feature matrix.  The spectrum from the original signal and the IMFs was then compared to determine the overlap between them.
The IMF with the largest overlap was selected as the informative IMF.
Fig.~\ref{fig:EEMD_informative_IMF} provides an example for the selection of the informative IMF. 
The original time series has frequency content around 1000 Hz and the first two IMF are the candidates to be informative IMF.
Since the spectrum of the first IMF overlaps with the original signal's spectrum, it is selected as an informative IMF.
Ideally, the spectrum of all signals and their decomposition should be checked to determine the informative IMF.  However, this is a manually intensive and time-consuming process. We thus repeated this process only for a couple of time series and chose an informative IMF to be used for all time series.
Then, the selected informative IMF was used to compute the features given in Ref.~\cite{Yesilli2020}, and a feature matrix was generated as an input for a supervised classification algorithm.

% ------------------------------------- FFT/PSD/ACF---------------------------------------%
\subsection{Fast Fourier Transform (FFT), Power Spectral Density (PSD) and Auto-correlation Function (ACF)} \label{sec:traditional}
This method computes the Fast Fourier Transform (FFT), Power Spectral Density (PSD), and Auto-correlation (ACF) for each downsampled data set. 
The next step was to find the significant peaks in these plots and use their $x$ and $y$ coordinates as a feature in the classification algorithm. 
Since built-in functions in computing software for peak finding can result in incorrect peaks, we used two restriction parameters for peak selection that enabled us to find the true peaks.
These parameters are minimum peak height (MPH) and minimum peak distance (MPD). 
The definition for minimum peak height is provided in Ref.~\cite{Yesilli2020a} as
\begin{equation*}
MPH= y_{min} + \alpha (y_{max}-y_{min}),
\end{equation*}
where $\alpha \in$ [0,1], $y_{min}$ and $y_{max}$ correspond to $5^{th}$ and $95^{th}$ percentile of the amplitude of FFT/PSD/ACF plots. The $\alpha$ parameter is defined with respect to the peak amplitudes. Since auto-correlation function has negative amplitudes, the choice for $\alpha$ is chosen separately, while we use the same $\alpha$ value for FFT and PSD plots.
In this implementation, $\alpha$ was 0.1 and 0.5 for FFT/PSD and ACF plots, respectively.

\begin{figure}
\begin{center}
\centerline{\includegraphics[width=0.5\columnwidth,keepaspectratio]{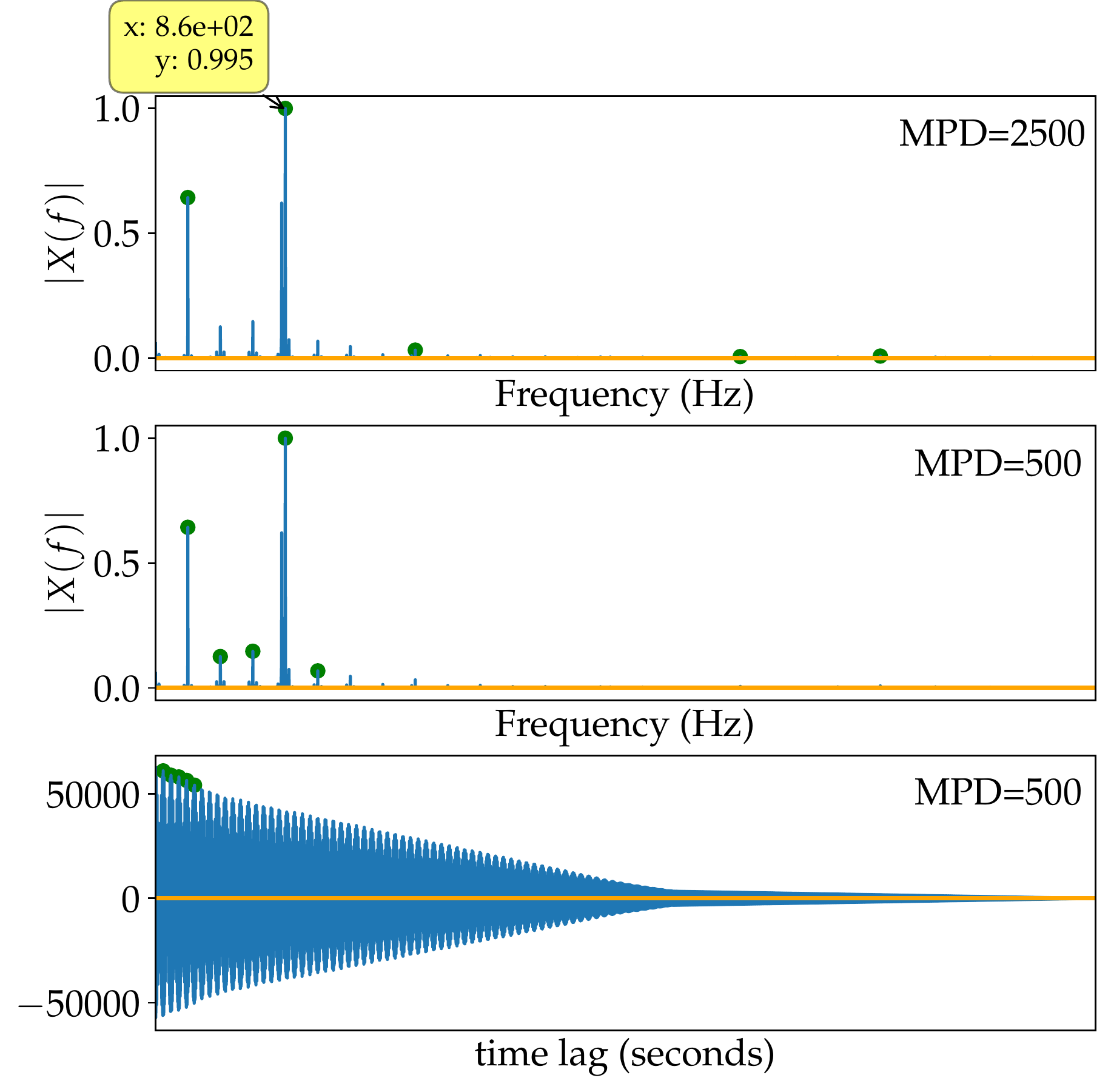}}
\caption{Effect of different MPD values on selected peaks in FFT and ACF plots of time series with RPM=13300 and DOC=2.54 mm from milling experiments. (Top) FFT plot and selected peaks with MPD=2500 (top) and MPD=500 (middle). Auto-correlation function with MPD=500 (bottom). Orange lines represent the MPH.}
\label{fig:sample_peaks_MPD}
\end{center}
\end{figure}

The second parameter, MPD, was defined by visual inspection on FFT/PSD/ACF plots of several time series.
An example is provided in Fig.~\ref{fig:sample_peaks_MPD}. 
This figure shows the effect of the chosen MPD value on the detection of the peaks in the FFT and ACF plots. 
The first two plots provide the spectrum of a time series and the peaks found by a peak detection algorithm with two MPD values.
It is seen that a smaller MPD value brings the selected peaks closer to each other and results in the detection of the true peaks. 
Therefore, MPD was chosen for FFT and PSD plots as 500.
The same value was also used in the ACF function.
After defining the two constraints for peak detection, MPD and MPH, we decided how many peaks to use to generate feature matrices.
In this implementation, we used coordinates of the first two peaks as features and they were given to supervised classification algorithms. 

% ------------------------------------- DTW ---------------------------------------------%
\subsection{Dynamic Time Warping (DTW)} \label{sec:dtw}
\begin{figure}[hbt!]
\begin{center}
\centerline{\includegraphics[width=0.8\columnwidth,keepaspectratio]{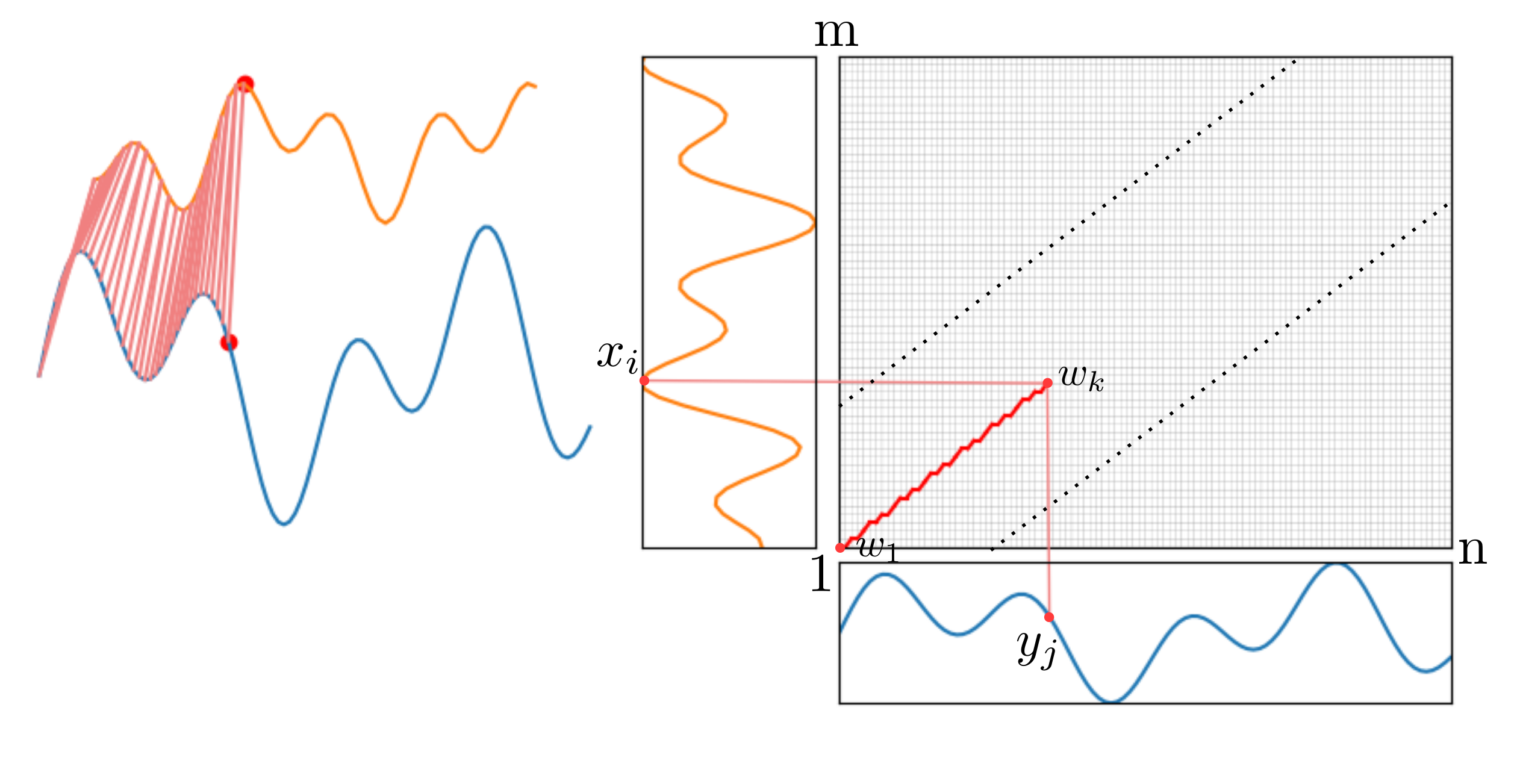}}
\caption{Illustration of a warping path between two time series with length $m$ and $n$.}
\label{fig:warping_path}
\end{center}
\end{figure}

Dynamic Time Warping (DTW) is often used to measure the similarity between two time series whose lengths are different. For example, 
assume that we have two time series $X$ and $Y$ such that
\begin{gather*}
X = x_{1},x_{2},\ldots,x_{m}\\
Y = y_{1},y_{2},\ldots,y_{n},
\end{gather*}
where $m$ and $n$ are the length of the time series. Berndt and Clifford \cite{berndt1994} define a mapping between the time series called warping path. 
Figure~\ref{fig:warping_path} illustrates an example of a warping path.
Warping is composed of nodes that represent the matching between elements of time series.
For instance, $w_{k}$ in Fig.~\ref{fig:warping_path} corresponds to matching between $x_{i}$ and $y_{j}$.
Warping paths are defined with respect to four different constraints and these are monotonicity, continuity, boundary condition, and slope constraint. For more details about these constraints, one can refer to Ref.~\cite{Sakoe1978}.
Several warping paths can be generated based on the constraints, however, the DTW algorithm chooses the one that gives the minimum distance such that
\begin{equation}
\label{eq:DTW}
D_{TW}(X,Y) = min\Big(\sum_{k=1}^{L}d(w_{k})\Big),
\end{equation}
where $d(w_{k})$ represents the distance between two matched elements of the time series, and $L$ is the length of the warping path.  
DTW algorithms look for the optimal path that gives us the smaller sum of the distances between the matching elements of the time series (see Eq.~\eqref{eq:DTW} ). Since there are numerous options for warping paths in the grid shown in Fig.~\ref{fig:warping_path}, restriction parameters are needed to reduce the number of possible warping paths. These parameters include monotonicity, continuity, adjustment window condition, slope constraint, and boundary conditions~\cite{Sakoe1978}.
\begin{itemize}
    \item Monotonicity: $x_{i}$ and $y_{j}$ represents the matching elements in time series. Monotonicity condition says that $i$ and $j$ can not decrease such that $j(k)\geq j(k-1)$, and $j(k)\geq j(k-1)$, where k is the index in Eq.~\eqref{eq:DTW}.
    \item Continuity: $i$ and $j$ can not be increased by more than 1.
    \item Boundary condition: The start point of the warping path should where $i=1$ and $j=1$, and the ending point of the path should be where $i=m$ and $j=n$.
    \item Adjustment window condition: The grid shown in Fig.~\ref{fig:warping_path} is restricted with the two dashed lines to decrease the area where we look for optimal warping path.
    \item Slope constraint: This condition is introduced to avoid moving significantly in one direction. Slope constraint says that the algorithm can not move more than $a$ steps in horizontal or vertical direction without having at least $b$ steps in diagonal direction~\cite{Sakoe1978}. The ratio between $b$ and $a$ is introduced as P=b/a, and it is generally chosen as 1.
\end{itemize}

\begin{figure}[hbt!]
\begin{center}
\centerline{\includegraphics[width=1\columnwidth,keepaspectratio]{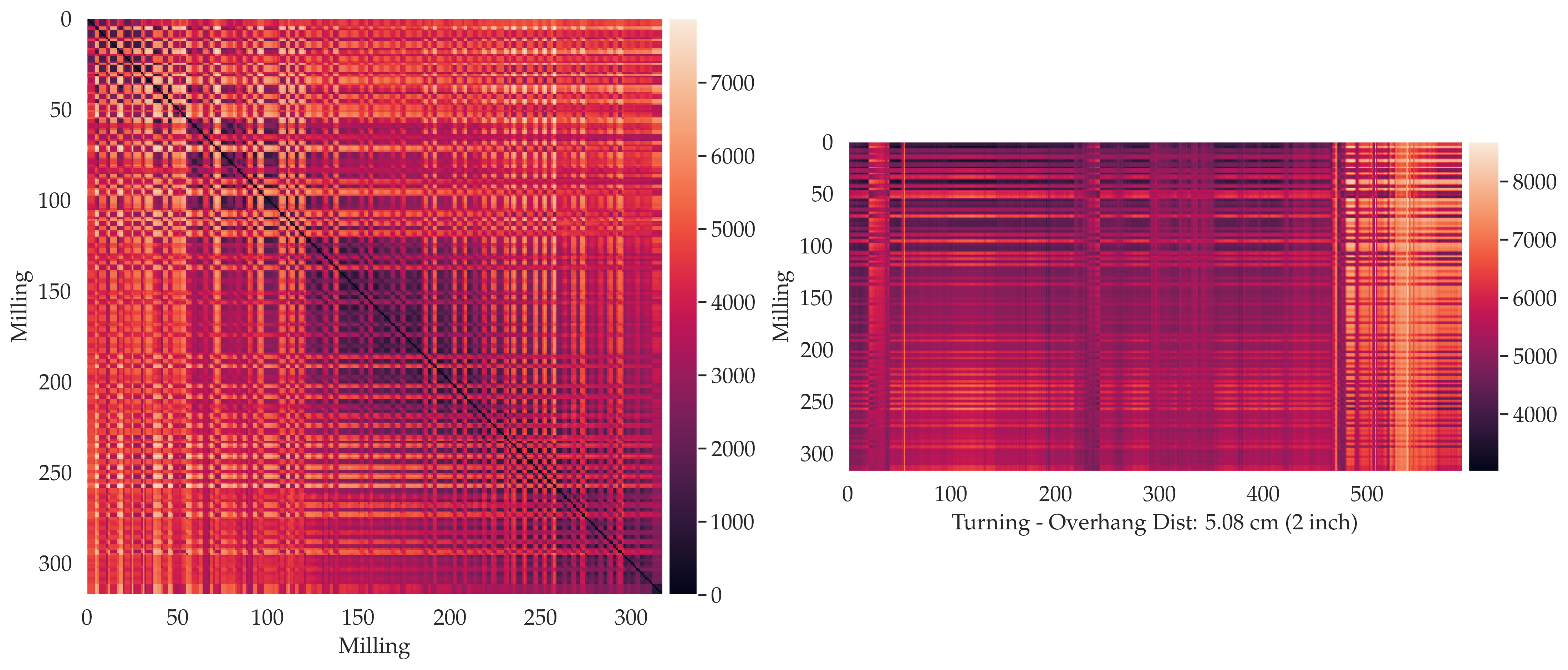}}
\caption{Heatmaps for the pairwise distances between milling time series (left), and for the distances between milling data set and 5.08 cm case of turning data set (right). The numbers on $x$ and $y$ axes represent the number of the time series in the data set.}
\label{fig:heatmaps}
\end{center}
\end{figure}

In this implementation, we used  \href{https://pypi.org/project/cdtw}{cDTW package} and the Manhattan distance to compute pairwise distances between the time series.
The pairwise distances between all time series in the data sets were computed.  The pairwise distances between the time series belong to different overhang cases in the turning data and between turning data cases and milling data. Two examples of heat maps for pairwise distances are given in Fig.~\ref{fig:heatmaps}.
Depending on the application of transfer learning, the corresponding distance matrix can be chosen for the training set and test set. 
For instance, if we train our classifier on $5.08$ cm case in turning data and test it on the milling data, we use the pairwise distances between time series of $5.08$ cm case for training and the distance matrix between $5.08$ cm case and milling data for testing the classifier.
The K-Nearest Neighbor (KNN) algorithm was used to predict the labels of the time series.

% ------------------------------------- TDA ---------------------------------------------%
\subsection{Topological Data Analysis (TDA)} \label{sec:tda}
Topological Data Analysis investigates the shape of the data. 
In this implementation, we mainly focused on the persistent homology tool of TDA and how to extract features using persistent homology.
We briefly explain the persistent homology in this section. 
One can refer to Ref.~\cite{Ghrist2007,Carlsson2009,edelsbrunner2010,Oudot2015,munkres2018,Munch2017} for further details.
Persistence homology extracts the information from the embedded signals in Euclidean space. 
Taken's embedding theorem is used to embed the data higher dimensional space \cite{Takens1981} and 1-D persistent homology is used to generate feature matrices in this study.

The embedded data is called the point cloud and persistent homology investigates connected components, loops, and voids in this point cloud.
The information extracted from the point cloud is represented on a plot called the persistence diagram.
Persistence diagrams can be computed with respect to the shape of interest. 
For example, the information related to connected components in the point cloud is provided in a 0-D persistence diagram. 
In addition to this, 1-D and 2-D persistence diagrams are computed for loops and voids, respectively.

\begin{figure}[hbt!]
\begin{center}
\centerline{\includegraphics[width=1\columnwidth,keepaspectratio]{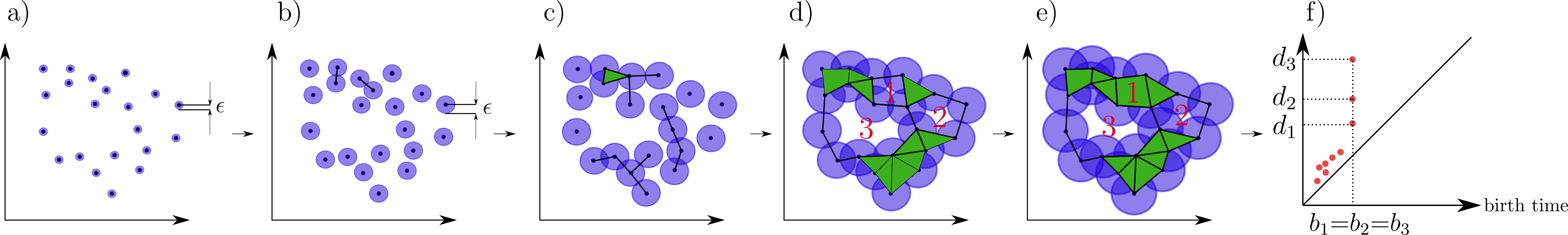}}
\caption{Illustration for explaining the generation of a persistence diagram. a) Balls with radius $\epsilon$ are centered on each point. b) The intersections of two balls create edges. c) The intersection of three balls creates a triangle shown with green color. d) Further increase in radius of balls lead to the generation of cycles shown with 1,2 and 3, and the corresponding radius is assigned as birth time. e) When a circle is filled with balls (see cycle 1), the corresponding radius is said to be the death time of the cycle. f) Birth and death times of the cycles are summarized in a persistence diagram.})
\label{fig:Rips_Complex}
\end{center}
\end{figure}

Figure~\ref{fig:Rips_Complex} provides an illustration on how to generate persistence diagrams.
We start by placing balls with a radius $\epsilon$ centered on each point in the point cloud.  The radius of the balls is then expanded and some of the balls start to intersect with each other. 
The intersection of two balls creates an edge between two data points as shown in Fig.~\ref{fig:Rips_Complex}b. 
As we continue to increase $\epsilon$, three balls intersect with each other and they create a triangle (see Fig.~\ref{fig:Rips_Complex}c).
Larger values of $\epsilon$ result in the intersection of more balls and they generate cyle.
The first time when a cycle appears is named birth time and three cycles appeared in Fig.~\ref{fig:Rips_Complex}d.
Further increase in the radius can lead to the generation of more triangles inside the cycle and a cycle may be filled with triangles. 
The time when a cycle is filled is called the death time. In Fig.~\ref{fig:Rips_Complex}e), the first cycle is filled and its death time is denoted with $d_{1}$ in the persistence diagram (see Fig.~\ref{fig:Rips_Complex}f).
An increase in $\epsilon$ will also lead to the filling of second and third cycles, and their death time is provided in the persistence diagram.
It is shown that the third cycle is the most persistent one since it has more lifetime ($d_{i}-b_{i}$) compared to others.

\textbf{Persistent Homology:}
Persistent homology in 0-D, 1-D and 2-D is denoted as $H_{0}(K)$, $H_{1}(K)$ and $H_{2}(K)$, where $K$ represents the simplicial complex.
As we increase the radius of the balls around the points on the point cloud, we create $n$ dimensional simplicies.
A single data point is called a 0-simplex. while an edge and a triangle are 1-simplex and 2-simplex, respectively.
Since we continuously change the $\epsilon$ (see Fig.~\ref{fig:Rips_Complex}), each $\epsilon$ will result in a different simplicial complex. 
These simplicial complexes are approximated with Rips and C\'ech complex, which are the most commonly used complexes.
In this implementation, we use \href{https://ripser.scikit-tda.org/}{Ripser} package in Python to compute the persistence diagrams, and it employs Rips complex \cite{Ctralie2018}.
The definition of the Rips complex is given as
\begin{equation*}
R_{\epsilon}(K,d) = \{ \sigma \subset K | \underset{(x,y)\in \sigma}{\mathrm{max}} d(x,y)< \epsilon\},
\end{equation*}
where $\sigma$ is the simplicial complex and $d$ is the distance between its vertices.
For each $\epsilon$ value, there is corresponding Rips complex such that
\begin{equation}
R_{1}\subseteq R_{2} \subseteq \ldots \subseteq R_{m}. 
\end{equation}
A cycle can be born in one of those Rips complex $R_{i}$ and it can disappear in another one ($R_{i+k}$). 
The corresponding birth and death time for this cycle is $\epsilon_{i}$ and $\epsilon_{i+k}$, respectively.
The resulting birth and death time for each cycle is plotted on a persistence diagram, and we vectorize the persistence diagrams in this implementation to extract features from the time series. 
We employ four methods and these are Carlsson coordinates, persistence images, persistence landscapes, and template functions.
\subsubsection{Computation of persistence diagrams}
Before explaining the feature extraction methods of TDA, we would like to briefly discuss how to compute the persistence diagrams from time series data.
The first step is to embed the data to higher dimensional space using Taken's embedding \cite{Takens1981}.
We need two parameters to embed the data and these are embedding dimension and delay parameters.
Embedding dimension is defined for each time series using False Nearest Neighbor (FNN) algorithm \cite{Abarbanel1994, Kennel1992}.
The embedding delay parameter is found using the method that combines Least Median Square (LMS) \cite{Rousseeuw1984} and FFT. 
Embedded data is then sent to the Ripser package in Python to compute persistence diagrams. 
One can refer to Ref.~\cite{Yesilli2019a} for more detailed information about embedding and the methods used to find the embedding parameters.
% ------------------------------------- Carlsson Coordinates -----------------------------------%

\subsubsection{Carlsson Coordinates}
Carlsson Coordinates are the features that are computed using the coordinates on the persistence diagram.
We use five different features where four of them are defined in Ref.~\cite{Adcock2016} and one of them is obtained from Ref.~\cite{Khasawneh2018}.
These features are not dependent on the number of points in persistence diagram or the order of the points.
The definition of the five features is given as 
\begin{equation*}
\begin{array}{rl}
f_1(D) & =  \sum{b_i(d_i-b_i)}, \\
f_2(D) & = \sum{(d_{\rm max} - d_i)(d_i - b_i)},\\
f_3(D) & = \sum{b_i^2 (d_i - b_i)^4},  \\
f_4(D) & = \sum{(d_{\rm max} - d_i)^2 (d_i-b_i)^4},\\
f_5(D) & = \max\{ (d_i - b_i),
\end{array}
\end{equation*}
where $D$ represents the persistence diagram, $b_{i}$ is the birth time and $d_{i}$ is the death time of the points in the persistence diagram.
These features are computed for each persistence diagram, and we generate feature matrices using $\sum\limits_{i=1}^5{\tbinom{5}{i}}$ combinations of the features.
Then, we performed classification using supervised learning classification algorithms.

% ------------------------------------- Persistence Images  -----------------------------------%
\subsubsection{Persistence Images}
\begin{figure}[hbt!]
\begin{center}
\centerline{\includegraphics[width=0.6\columnwidth,keepaspectratio]{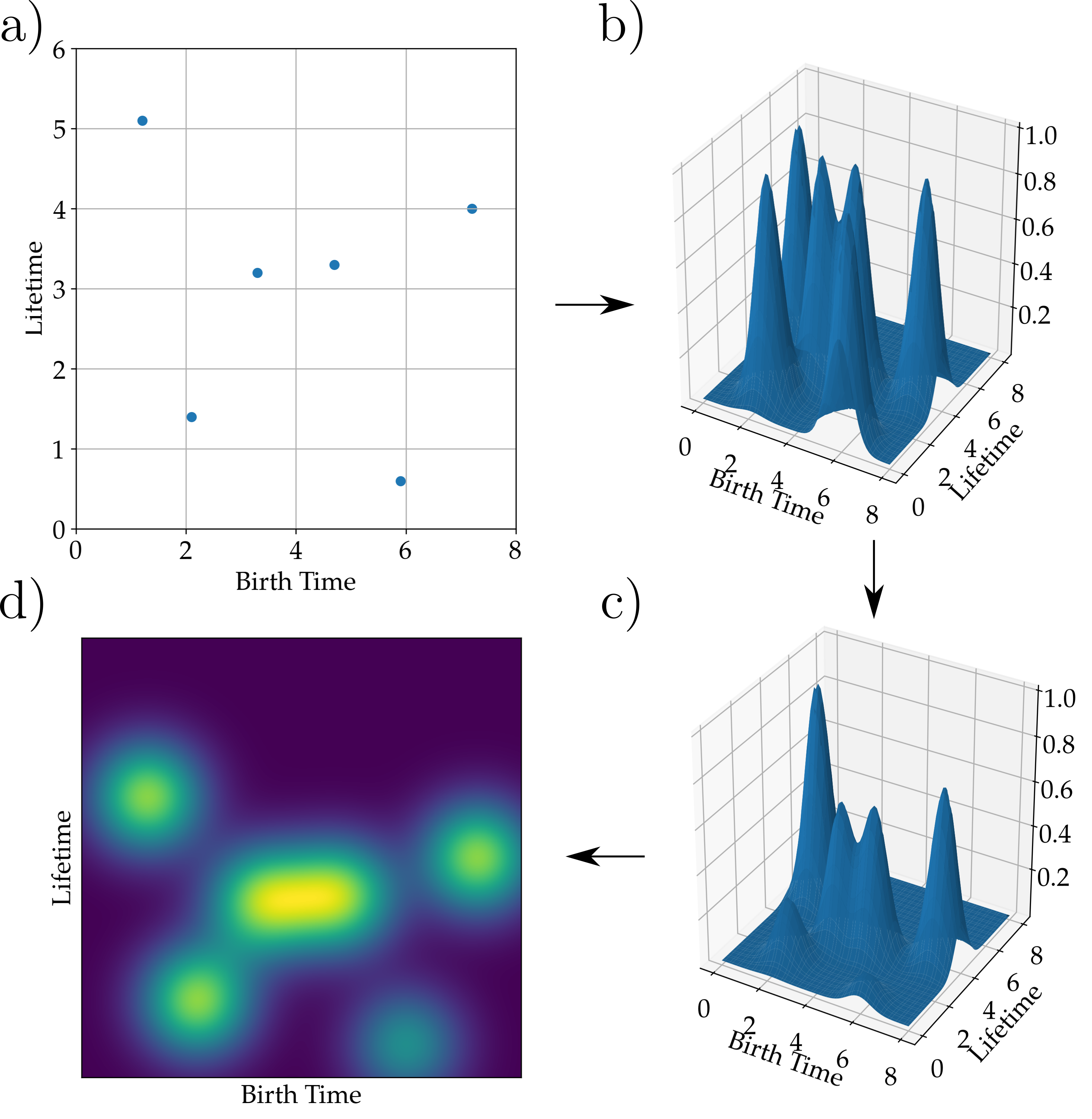}}
\caption{The illustration that shows the steps to obtain persistence images from the persistence diagram. a)Persistence (death-birth) vs. birth time diagram. b) Gaussians are placed on each point on the persistence vs. birth time diagram. c) Gaussians are weighted with respect to their persistence value. d) 3D surface is obtained using the 3D surface of Gaussians.}
\label{fig:PImages_example}
\end{center}
\end{figure}

Persistence diagrams can be converted into persistence images. 
The first step is to define a linear transformation such that \cite{Adams2017}
\begin{equation*}
T(b_{i},d_{i}) = (b_{i},d_{i}-b_{i})= (b_{i},l_{i}).
\end{equation*}
This transformation converts a birth time versus death time diagram into a birth time versus lifetime diagram (see Fig.~\ref{fig:PImages_example}a). 
Then, we place a Gaussian to each point on the transformed diagram as shown in Fig.~\ref{fig:PImages_example}b and the Gaussian distribution  is defined as
\begin{equation*}
D_k(x,y)=\frac{1}{2\pi\sigma^2}e^{-\lbrack(x-b_k)^2+(y-l_k)^2\rbrack/2\sigma^2},
\end{equation*}
where $\sigma$ is the standard deviation and we choose it as $0.1$ in this implementation.
Each Gaussian is weighted with respect to weighting function defined as $W(k) = W(b_{k},l_{k}): (b_{k},l_{k})\in T(D)\to \mathbb{R}$. The definition of the weighting function is given as  \cite{Adams2017}
\begin{equation*}
W(k)=W(b_k, l_k) = \begin{cases}
0 & \text{ if } l_k \leq 0; \\
\frac{l_k}{b} & \text{ if } 0 < l_k < b; \\
1 & \text{ if } l_k \geq b.
\end{cases}
\end{equation*} 
According to this function, the points with larger persistence values have more weight and the weighted Gaussians are shown in Fig.~\ref{fig:PImages_example}c.
The next step was to define the persistence surface using the weighted Gaussian such that
\begin{equation*}
S(x,y) = \sum_{k\in T(D)} W(k)D_{k}(x,y).
\end{equation*}
Persistence image is computed on the grids defined on the domain of the persistence surface.
The value of each grid or pixel is found by the integration given as
\begin{equation*}
PI_{i,j} = \iint Sdxdy.
\end{equation*}
After computing all pixel values, a persistence image is formed. An example is given in Fig.~\ref{fig:PImages_example}d.

In this implementation, we use the \href{https://gitlab.com/csu-tda/PersistenceImages}{PersistenceImages} package to compute the images.
The resulting image is a matrix that contains the pixel values and we vectorize this matrix to generate the feature matrix.
The rows of the output matrix are concatenated and the resulting vector is used as features for the corresponding persistence diagram.
The number of the features depends on the selected range of the birth time and lifetime and the size of the pixel.
In this implementation, we only test the features obtained using a pixel size of 0.1.
This means that if we have a range from 0 to 1 for both birth time and lifetime, we will end up with a $10\times 10$ matrix for the image.
Therefore, the resulting feature vector for the image will be 100 in length.
Features of all persistence diagrams are computed and then they are sent to supervised classification algorithms to predict chatter.

% ------------------------------------- Persistence Landscapes  --------------------------------%
\subsubsection{Persistence Landscapes}
\begin{figure}
\begin{center}
\centerline{\includegraphics[width=1\columnwidth,keepaspectratio]{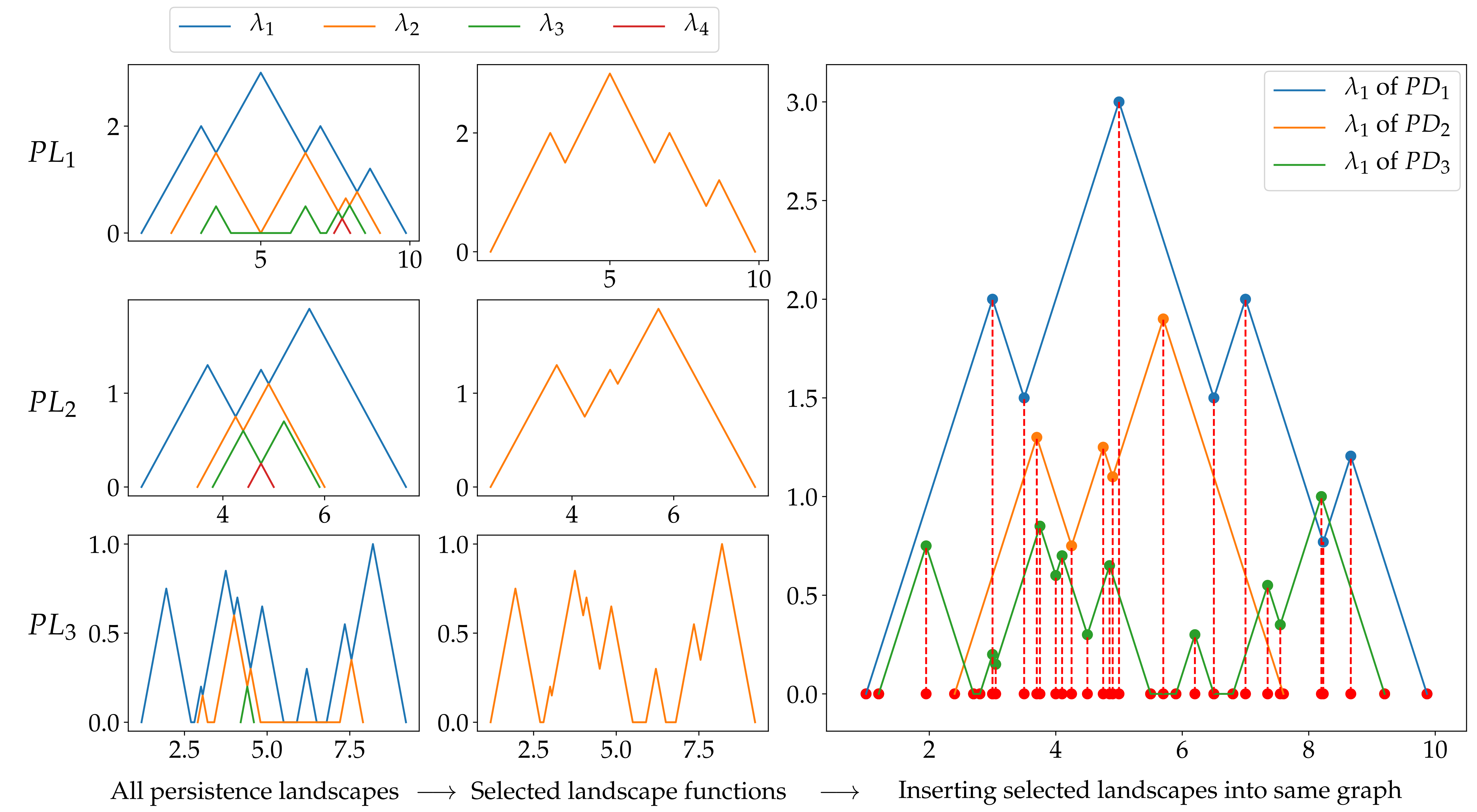}}
\caption{The illustration that shows the mesh generation using first persistence landscapes ($\lambda_{1}$) for three different persistence diagrams. $PL_{i}$ where $i=1,2,3$ represents the persistence landscape sets of three different persistence diagrams. Red dots are the projection of the nodes of landscape functions.}
\label{fig:mesh_generation}
\end{center}
\end{figure}

Another functional summary of persistence diagrams is persistence landscapes introduced by Bubenik and D\l{}otko \cite{Bubenik2017}.
Persistence landscapes are piecewise functions that are defined in the domain of the persistence diagrams. 
A persistence diagram is rotated 45 degrees in the clockwise direction and then the isosceles triangles are drawn for each point on the diagram \cite{Berry2020}.
The expression for the persistence landscapes is described as \cite{Bubenik2017}
\begin{equation*}
  g_{(b,d)}(x) =
  \begin{cases}
    0 & \text{if $x \not\in (b,d) $} \\
  x-b & \text{if $x \in (b, \frac{b+d}{2}]$} \\
 -x+d & \text{if $x \in (\frac{b+d}{2}, d)$}.
  \end{cases}
\end{equation*}
An example for the persistence landscape is provided in Fig.~\ref{fig:PL_example}, and the persistence landscapes are denoted as $\lambda_{k}$.
\begin{figure}
\begin{center}
\centerline{\includegraphics[width=0.75\columnwidth,keepaspectratio]{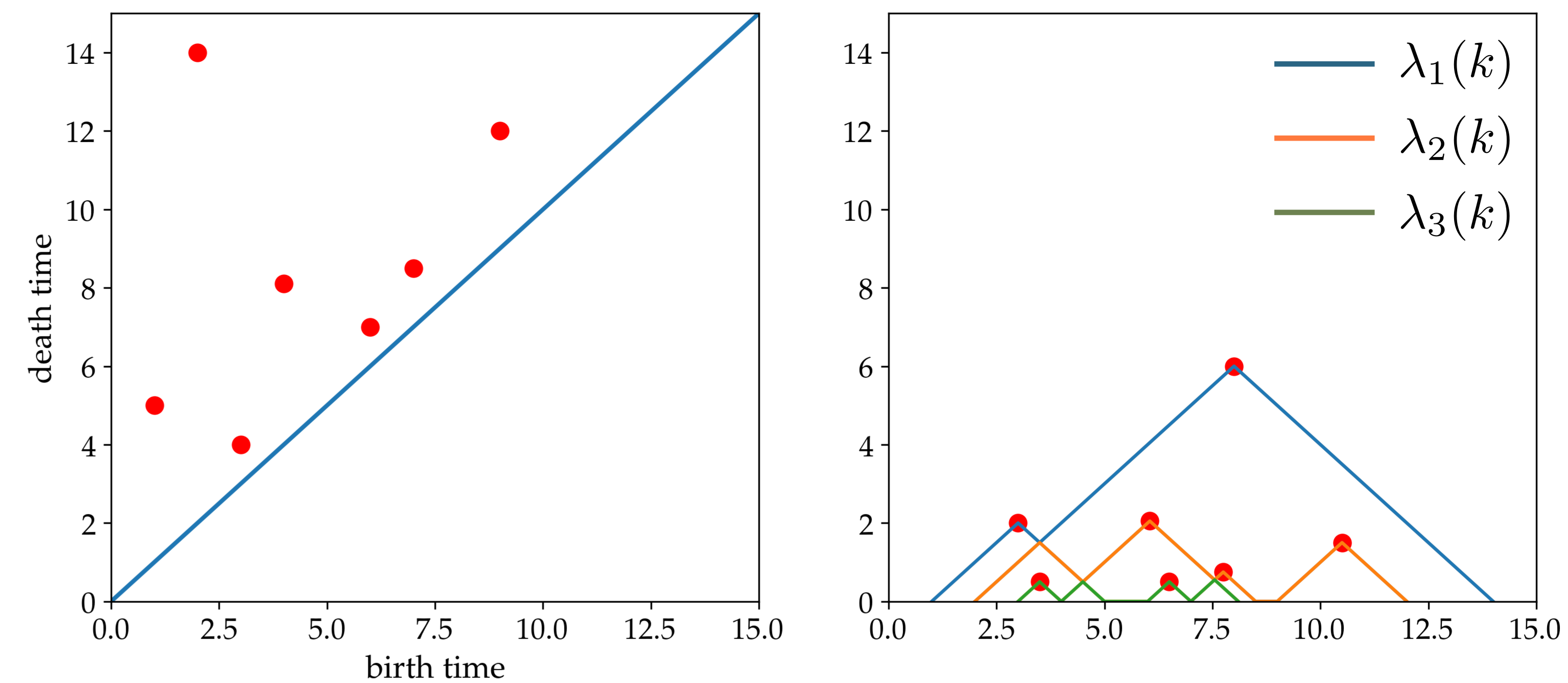}}
\caption{A simple example for persistence diagram and its persistence landscapes. (left) Persistence diagram. (right) Corresponding persistence landscapes for the diagram shown in the left.}
\label{fig:PL_example}
\end{center}
\end{figure}
\begin{figure}
\begin{center}
\centerline{\includegraphics[width=1\columnwidth,keepaspectratio]{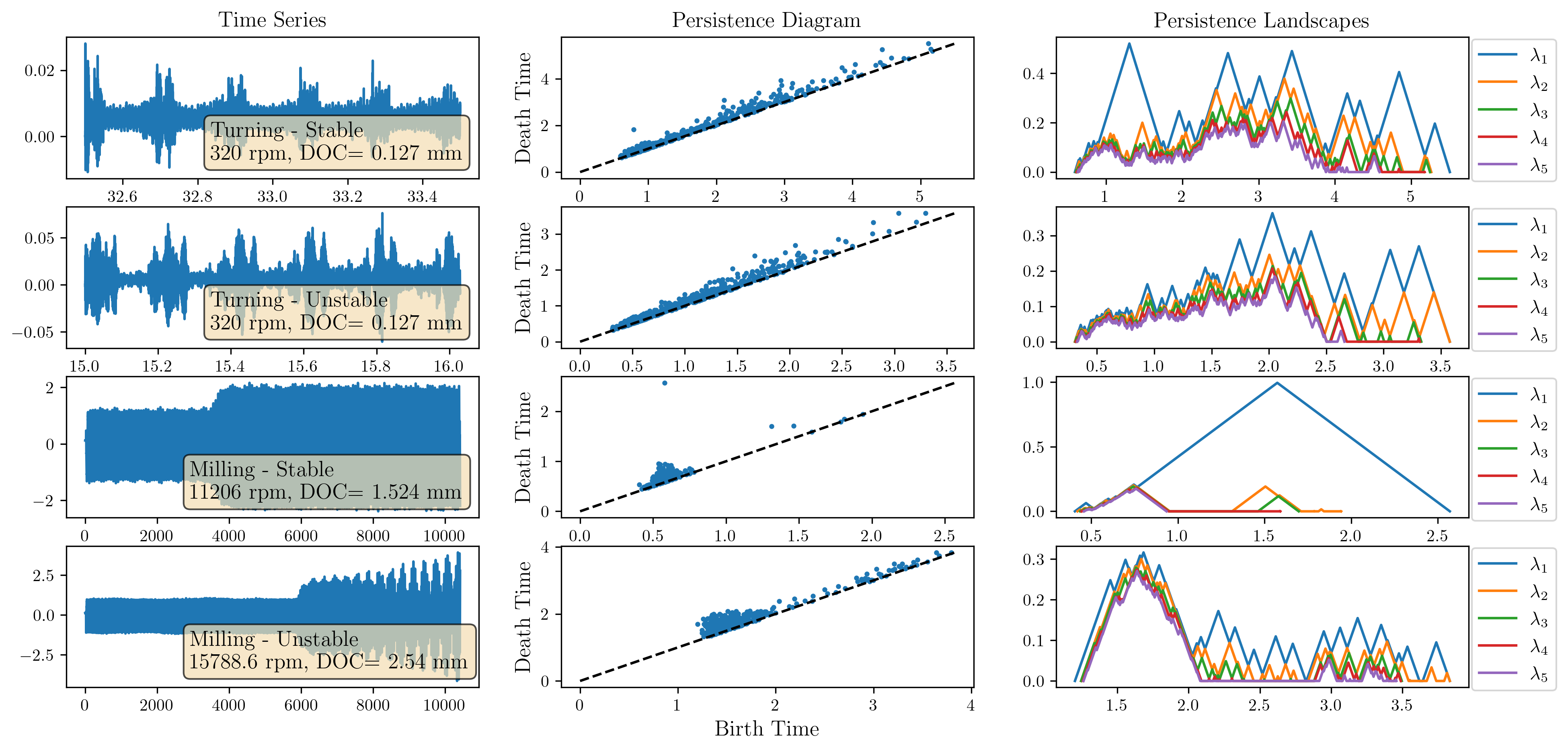}}
\caption{Persistence diagrams and persistence landscapes of four different time series from turning and milling experiments. (First row) Stable turning signal with RPM=320 and depth of cut (DOC) of 0.127 mm. (Second Row) Unstable turning signal with RPM=320 and depth of cut (DOC) of 0.127 mm.
(Third Row) Stable milling signal with RPM=11206 and depth of cut (DOC) of 1.524 mm. (Fourth Row) Stable milling signal with RPM=15788 and depth of cut (DOC) of 2.54 mm.}
\label{fig:PL_example}
\end{center}
\end{figure}
The next step was to convert these landscape functions into features.
We should first generate a mesh that we can use to define features from persistence landscapes functions for each diagram.
The mesh generation process is summarized in Fig.~\ref{fig:mesh_generation}.
The first column of the plots provides all persistence landscapes of three different persistence diagrams.
Then, we select a landscape number and corresponding landscape functions for each diagram.
Selected landscape functions are plotted on the same figure (see the last figure in Fig.~\ref{fig:mesh_generation}).
We take projections of the nodes of the landscape functions, and we show them with red dots in Fig.~\ref{fig:mesh_generation}.
Then, these points (red dots) are sorted in increasing order and the duplicates are removed.
The resulting points are called the mesh.
This mesh is then used to calculate the corresponding function value on each selected persistence landscape function using linear interpolation.
These function values are used as features for the persistence diagrams.
Then, we implement the supervised classification algorithm to detect chatter.
%%

% ------------------------------------- Template Functions  --------------------------------%
\subsubsection{Template Functions}
Template functions are introduced in Ref.~\cite{Perea2019a}.
As we do in persistence images, we represent the persistence diagrams in birth time and lifetime coordinates, so we represent the points on the diagram in the upper half-plane, $\mathbb{W} := \mathbb{R}\times \mathbb{R}_{>0}$.
Then, we define \textit{template function} on $\mathbb{R}^{2}$.
A function that has compact support in $\mathbb{W}$ could be a template function.
This function is evaluated on each point of the diagram to define a function such that
\[
v_f(D) = \sum_{x\in D} f(x) \quad f:\mathbb{W}\to \mathbb{R}.
\]

\textit{Template system} is defined as a collection of the template functions $\mathcal{T}$.
The function in the template system has distinct values for each persistence diagram.
A true template system has an infinite number of template functions, however, it is proved that a true template system can be approximated with a finite set of template functions in Ref.~\cite{Perea2019a}.
In this work, we implement the interpolating polynomials as template functions to vectorize the persistence diagrams.

We define two meshes denoted as $\mathcal{A}\subset \mathbb{R}$ and $\mathcal{B} \subset \mathbb{R}_{>0}$ in birth and lifetime coordinates, respectively.
The Lagrange polynomial is computed on a mesh such that
\[
\ell_j^\mathcal{A} (x) = \prod_{i\neq j} \frac{x - a_i}{a_j - a_i},
\]
where $a_{j} \in \mathcal{A} \subset \mathbb{R}$. The Lagrange polynomial also satisfies 
\[
\ell_j^{\mathcal{A}} (a_k) =
\begin{cases}
1 & j=k \\
0 & \text{otherwise}.
\end{cases}
\]
Template function with respect to chosen two meshes is defined as
\[
f(x,y) = \beta(x,y) \cdot | \ell_{i}^\mathcal{A}(x) \ell_{j}^\mathcal{B}(y) |,
\]
where $\beta$ is the bump function that makes the template function have compact support. 
However, it does not have to be defined, since we choose the meshes which contain all points in the persistence diagram.

\section{Results}
\label{sec:Results}
This section describes the classification approach and the transfer learning details.  As we mentioned in Sec.~\ref{sec:E_P}, the turning data set contains four different cases and the milling data does not have categorization.
Therefore, the total number of combinations between the cases of turning data and the milling data is 20.
We performed classification for all 20 combinations. 
In addition to transfer learning results, we also provide the results obtained from traditional machine learning in Section~\ref{sec:traditional_ml_results}.
Section~\ref{sec:TL_Turning} provides the results for the combination between cases of the turning data set, while Sec.~\ref{sec:TL_Turning_Milling} discusses the results for the combinations between turning and milling data set.
In addition, we take into account the mild chatter cases in turning data set as unstable while performing the classification. This is performed since the turning data is labeled in three classes (see Sec.~\ref{sec:E_P}). All results provided in this section belong to two-class classification for both milling and turning data sets.

%%%%%%%%%%%%%%%%%%%%%% Traditional ML %%%%%%%%%%%%%%%%%%%%%%%%%%%

\subsection{Results of Traditional Machine Learning Approach for Turning Data Set}
\label{sec:traditional_ml_results}
\begin{figure}[h]
\begin{center}
\centerline{\includegraphics[width=1\textwidth,keepaspectratio]{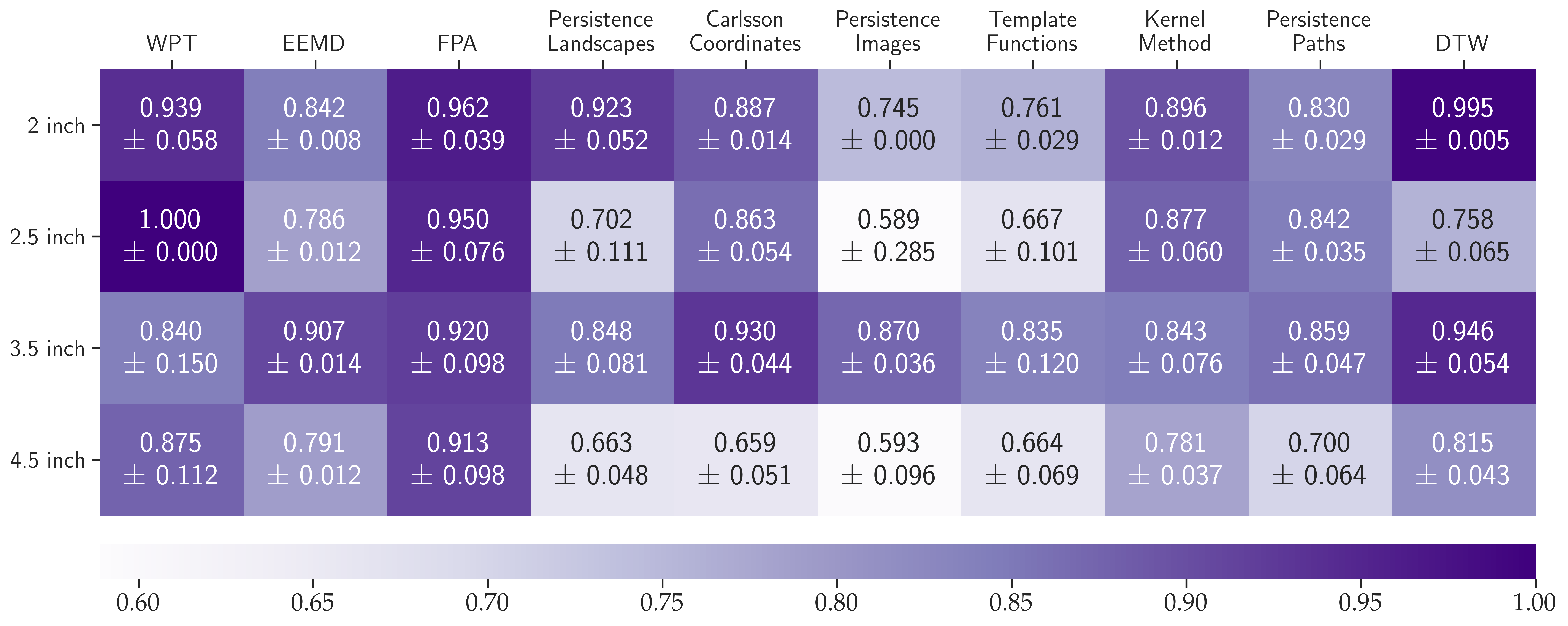}}
\caption{The highest accuracy out of four different classifiers (or out of selected numbers of nearest neighbor for DTW) for each approach used in traditional machine learning applications between overhang distance cases of turning experiments. WPT: Wavelet Packet Transform, EEMD: Ensemble Empirical Mode Decomposition, FPA: FFT/PSD/ACF, DTW: Dynamic Time Warping. These results are retrieved from Ref.~\cite{Yesilli2019a,Yesilli2019b,Yesilli2020}.}
\label{fig:Traditional_ML_Turning}
\end{center}
\end{figure}

In this section, we compare the results obtained with featurization approaches defined in Sec.~\ref{sec:Methods}.
We only provide the results for the turning data set in this section.
Traditional machine learning is the application where we train and test a classifier with data obtained from the same cutting configuration. 
Since we have four overhang distance cases in the turning experiment, Fig.~\ref{fig:Traditional_ML_Turning} provides the results for each overhang distance case in a heat map. 
The colors in the heat map correspond to the accuracy shown in each box, and the deviations are not taken into account for coloring the heat map.

Fig.~\ref{fig:Traditional_ML_Turning} shows us that the DTW approach provides the highest accuracy with a small amount of deviation compared to other approaches for 5.08 cm (2 inch) overhang distances. For the same case, persistence landscapes from the TDA-based approach can provide 92\% accuracy, while the highest accuracy from the widely adopted approach is provided by FPA with 96\%.
In the 6.35 cm (2.5 inch) overhang distance case, the highest accuracy is obtained with WPT with 100\%. It is worth noting that the test set size for the 6.35 cm case is smaller compared to the other overhang distance cases. This may explain 100\% accuracy in the test set. 
On the other hand, Carlson Coordinates from the TDA-based approach provides 86\% for the same case.
For the 8.89 cm (3.5 inch) case, the highest accuracy is obtained with the DTW approach and it is followed by Carlsson Coordinates. The deviations for these two results are smaller compared to traditional approaches.
FPA provides the largest accuracy for the last case with a larger deviation compared to novel approaches. 
Overall, it can be said that the novel approaches can match the accuracy obtained with traditional and widely adopted approaches. 
However, they provide better automation potential for chatter diagnosis since they eliminate manual preprocessing during feature extraction. 
Therefore, we also investigate their transfer learning performance and compare them to the widely adopted and traditional approaches in the next sections.

%%%%%%%%%%%%%%%%%%%%%% TURNING-TURNING  %%%%%%%%%%%%%%%%%%%%%%%
\subsection{Results of Transfer Learning Applications Between the Overhang Distance Cases of Turning Data Set}
\label{sec:TL_Turning}
The classification was performed with 10 realizations of training and test set for each method.
$67\%$ of the training set and $70\%$ of the test set were used to train and test the classifier, respectively.
To be fair in the comparison of methods, the same training and test sets were used; they were generated with a set of predefined random state parameters. 
Support vector machine (SVM) with \textit{rbf} kernel, logistic regression, random forest classifier with 100 estimators and a maximum depth of two for the trees, and gradient boosting algorithm were used to train and test a classifier.
In addition to these classifiers, similarity measure based approach, DTW can only be used with the K-Nearest Neighbor (KNN) classifier, since it uses a pairwise distance matrix between the time series (see Fig.~\ref{fig:heatmaps}).
KNN is only used with the similarity-based approach.  
Predicted labels were used to compute the average and standard deviation of the accuracy and F1-score for training and test set separately.
In addition to this, we categorize the methods in three groups: 1) Time-Frequency based approaches (WPT, EEMD and FFT/PSD/ACF (FPA)), 2) TDA-based approach and 3) Similarity measure (DTW). 
We compare the results of these groups. Features and classifiers used for each approach are given in Tab.~\ref{tab:features_classifiers}. For more details about the features, one can refer to Refs.~\cite{Yesilli2019a,Yesilli2019b, Yesilli2020}.

\begin{table}[!htbp]
\centering
\caption{Features and classifiers used for three main category of approaches. SVM: Support Vector Machine, LR: Logistic Regression, RF: Random Forest, GB: Gradient Boosting. }
\label{tab:features_classifiers}
\resizebox{1\textwidth}{!} {
\begin{tabular}{c|c|c}
Category&\multicolumn{1}{c|}{Features} &  \multicolumn{1}{c}{Classifiers} \\
\hline
\makecell{Time-Frequency-based}& \makecell{\textbf{WPT}: Mean, Standard Deviation, Root Mean Square (RMS), Peak, \\ Skewness, Kurtosis, Crest Factor, Clearance Factor, Shape Factor, \\ Impulse Factor, Mean Square Frequency, Standard Frequency, \\One Step Auto-Correlation Function, Frequency Center \\ \textbf{EEMD:} Energy Ratio, Peak to Peak, Standard Deviation, \\ RMS, Crest Factor, Skewness, Kurtosis \\ \textbf{FF/PSD/ACF (FPA):} The coordinates of the peaks}& SVM, LR, RF, GB\\
\makecell{TDA-Based}& \makecell{Carlsson Coordinates, Persistence Landscapes, Persistence Images\\ Template Functions} &  SVM, LR, RF, GB\\
\makecell{Similarity measure (DTW)}& Pairwise Distance Matrix& K-Nearest Neighbor\\ 
\end{tabular}
}
\end{table} 

\begin{figure}
\begin{center}
\centerline{\includegraphics[width=1\textwidth,keepaspectratio]{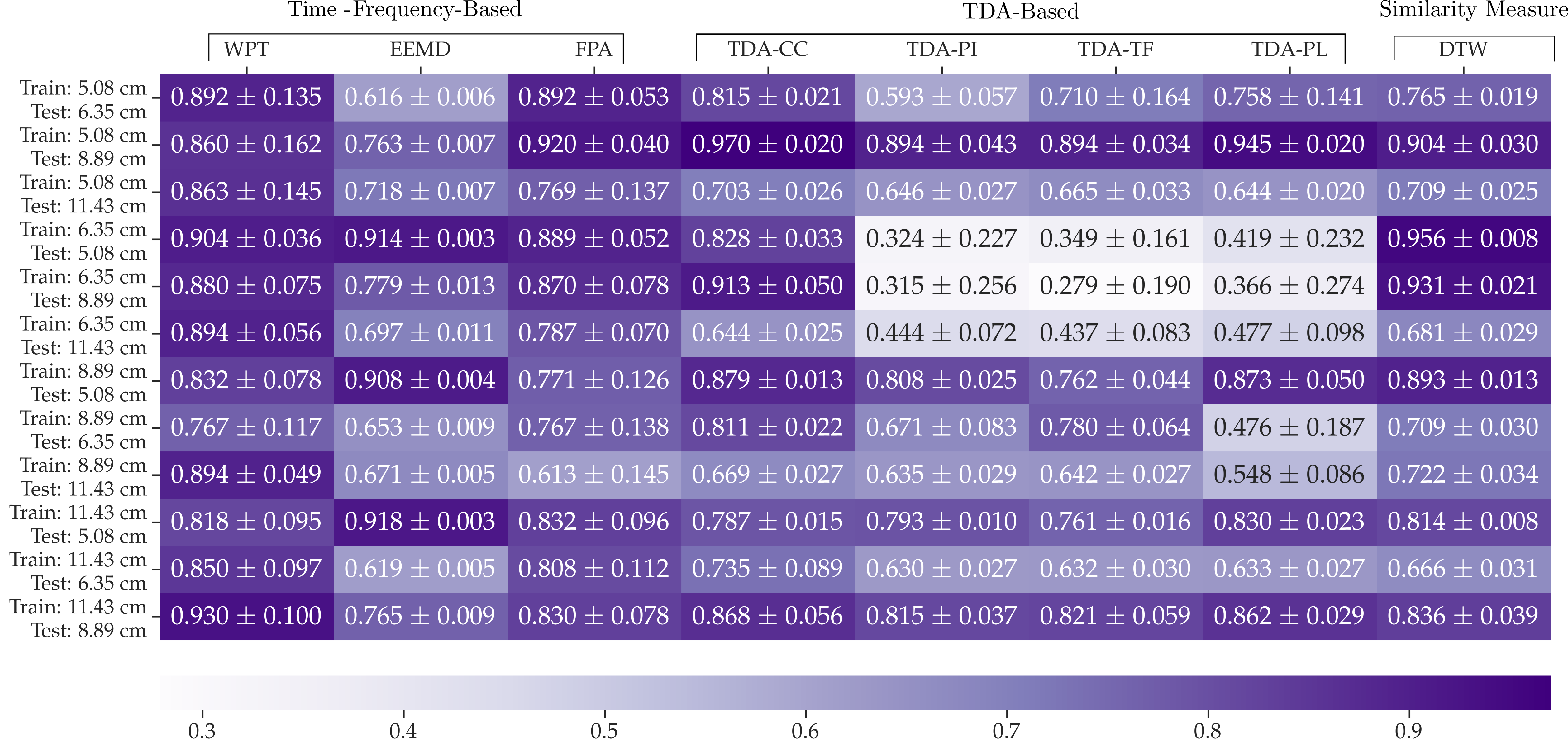}}
\caption{The highest accuracy out of four different classifiers (or out of selected numbers of nearest neighbor for DTW) for each approach used in transfer learning applications between overhang distance cases of turning experiments.}
\label{fig:Turning_TL_WPT_CC_DTW_Summary_Accuracy}
\end{center}
\end{figure}

\begin{figure}
\begin{center}
\centerline{\includegraphics[width=0.85\textwidth,keepaspectratio]{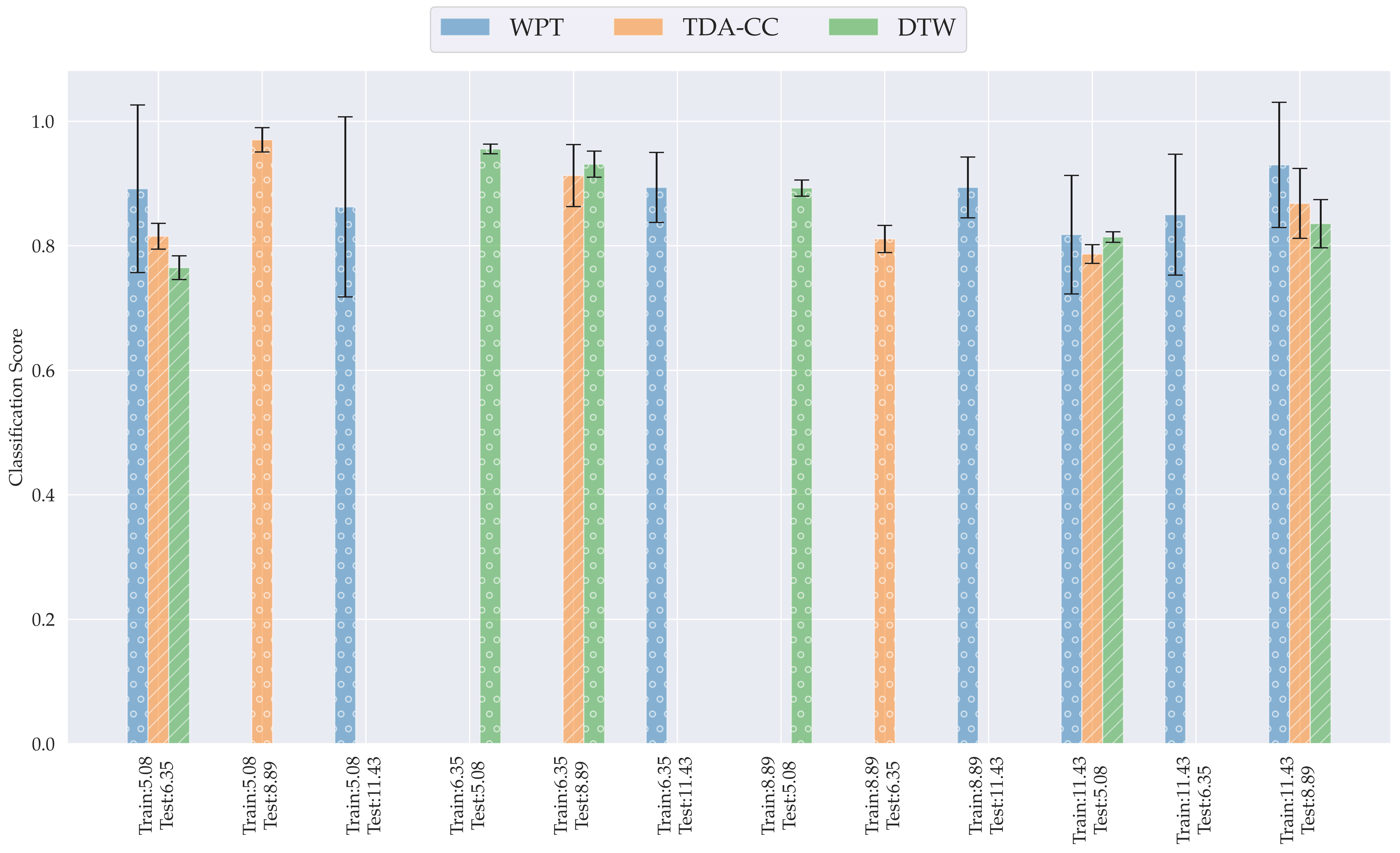}}
\caption{The classification results are obtained from the selected methods when we train and test between the overhang distance cases of the turning data set. The selected methods that give the highest accuracy are represented with the '$\circ$' bar hatch and the ones that are in the error band of the highest accuracy are shown with `$/$' bar hatch. One approach is selected from each category of the methods, and these are Wavelet Packet Transform (WPT), Carlsson Coordinates (TDA-CC), and Dynamic Time Warping (DTW).}
\label{fig:Turning_Selected_Best_Methods}
\end{center}
\end{figure}

For each combination of the cases between turning data sets, we have provided figures which show the accuracy of each feature extraction method for the classifiers mentioned above.
These plots are provided in Figs.~\ref{fig:training_overhang_2_test_turning}-{\ref{fig:Turning_DTW}} in the Appendix.
However, these plots can only compare the methods within the same application of transfer learning.
Therefore, we have provided a summary plot in Fig.~\ref{fig:Turning_TL_WPT_CC_DTW_Summary_Accuracy}. 
It provides the highest accuracies obtained out of four classifiers for all methods except DTW, while it shows the highest score out of all KNN, where K=(1,$\ldots$,5), for DTW.
It can be seen that the time-frequency-based methods, such as WPT, EEMD, and FPA, are the methods that give the highest score when we train and test between the overhang cases of the turning data set. 
However, WPT outperforms other approaches in most of the applications in the group of time-frequency-based approaches. 
On the other hand, the TDA-based approach and DTW have the highest accuracy in a few applications.
For TDA-based approaches, Carlsson Coordinates (TDA-CC) performs better than other featurization techniques within the group.
It is not easy to distinguish each result in Figure~\ref{fig:Turning_TL_WPT_CC_DTW_Summary_Accuracy}.
Therefore, we select WPT, TDA-CC, and DTW to summarize their results for different applications of transfer learning between turning data set cases and presented the results in Fig.~\ref{fig:Turning_Selected_Best_Methods}.

Fig.~\ref{fig:Turning_Selected_Best_Methods} contains only the highest scores of the selected approaches and the ones that are in the error band of the highest score. 
Each color represents a method; the best results and the ones in the error band are represented with two different hatches on the bar plots.
The bars with `$\circ$' hatch are the methods with the highest accuracy and the `$/$' hatch shows the methods that are in the error band.
% It is worth noting that the same highest score can be obtained with multiple methods for an application.
% In this case, we compare the deviations of each of these methods and pick the method with the smallest deviation as the best method for the corresponding application. 
% The remaining methods are still shown on the plot, but they have `$/$' type hatch on their bar plot.
Using Fig.~\ref{fig:Turning_Selected_Best_Methods}, we can define how many times a group of methods is the best (BM) or the method in the error band (MIEB), and these numbers are given in Table~\ref{tab:distribution_of_numbers}.
It is seen that the frequency-based approach (WPT) has the highest score in 7 out of 12 transfer learning applications between the cases of turning data set and it is not in the error band when the highest score is provided by the TDA-based approach and DTW.
On the other hand, the TDA based-approach and DTW provide the highest score two times and three times, respectively.
The TDA-based method is in the error band of the highest accuracy in 4 out 12 applications, while this number is three for DTW.
It is also worth noting that WPT results provided in Fig.\ref{fig:Turning_Selected_Best_Methods} have a larger deviation compared to DTW and TDA-based approaches, even though WPT provides the highest accuracies in most of the applications.

\begin{table}[!htbp]
\centering
\caption{The number of times when a selected method gives the highest accuracy out of 12 different applications between the cases of turning data set is denoted with BM. The number of times when a method is in the error band of the highest accuracy is denoted with MIEB. These two numbers are provided for accuracy and F1-score.}
\label{tab:distribution_of_numbers}
\resizebox{0.5\columnwidth}{!} {
\begin{tabular}{c|c|c|c|c}
&\multicolumn{2}{c|}{Accuracy} &  \multicolumn{2}{c}{F1-Score} \\
\hline
Method& \makecell{BM} &  MIEB &BM & MIEB\\
\hline
\makecell{Time - Frequency-based (WPT)}&7 & 0 &9 &1\\
\makecell{TDA-based (TDA-CC)}&2& 4 &3&0\\
\makecell{Similarity Measure (DTW)}&3&3&0&0
\end{tabular}}
\end{table} 

% When a method from a group provides the highest accuracy for a transfer learning application and another method from the same group are in the error band of that highest accuracy, we only add a count to BM column in Tab.~\ref{tab:distribution_of_numbers}.
% For example, WPT has the highest score when we train on 5.08 cm overhang case and test on 11.43 cm case of turning data set, while results of EEMD and FPA are in the error band of this highest accuracy.
% Since all of these methods are in the same group defined above, we only add one to number of times when a method provides the highest accuracy (BM), and the number that shows the how many times when a method is in the error band is still zero.
\begin{figure}
\begin{center}
\centerline{\includegraphics[width=1\textwidth,keepaspectratio]{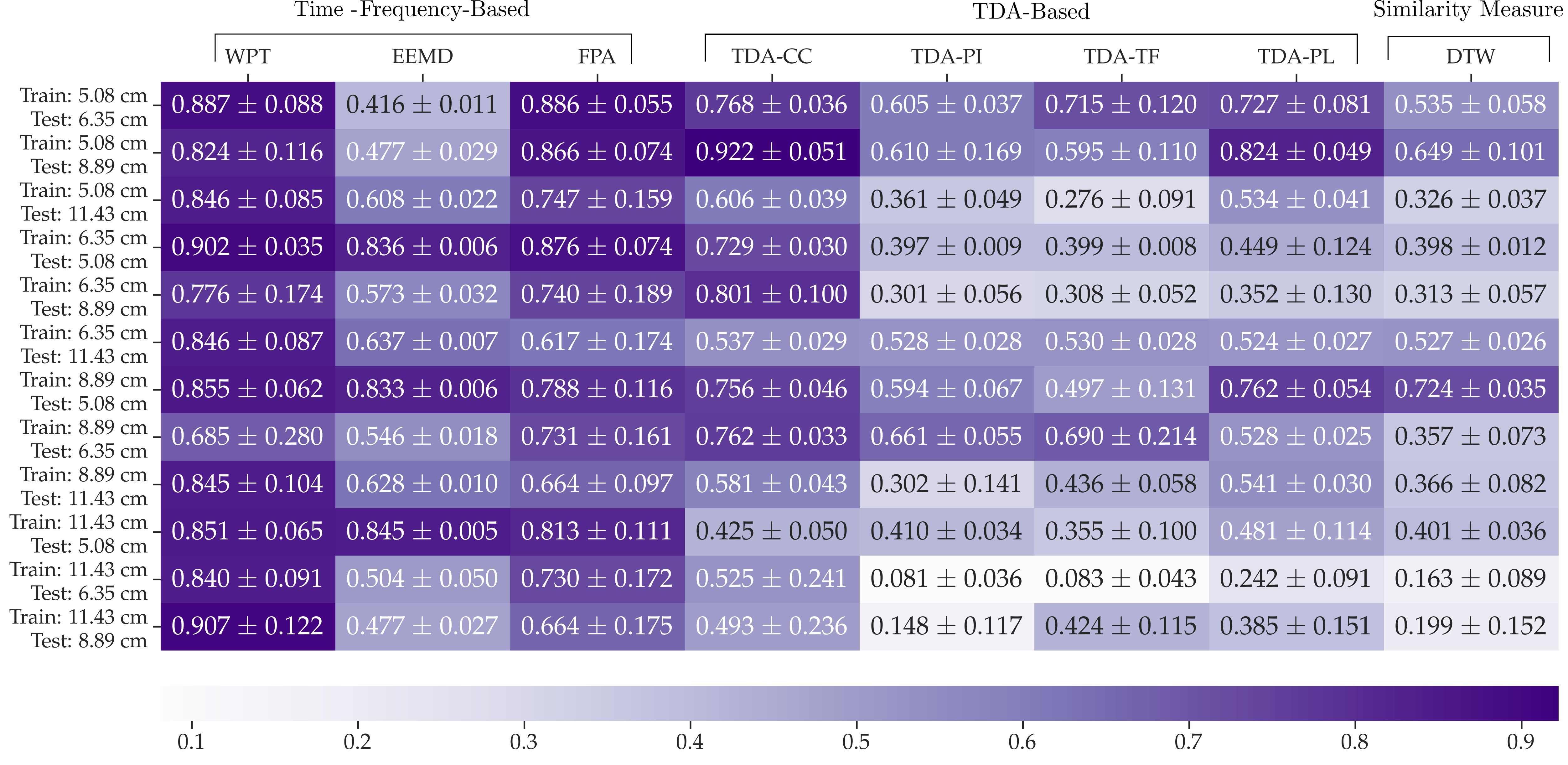}}
\caption{The highest F1-Score out of four different classifiers (or out of selected numbers of nearest neighbor for DTW) for each approach used in transfer learning applications between overhang distance cases of turning experiments.}
\label{fig:Turning_Summary_F1_score}
\end{center}
\end{figure}

\begin{figure}
\begin{center}
\centerline{\includegraphics[width=0.8\textwidth,keepaspectratio]{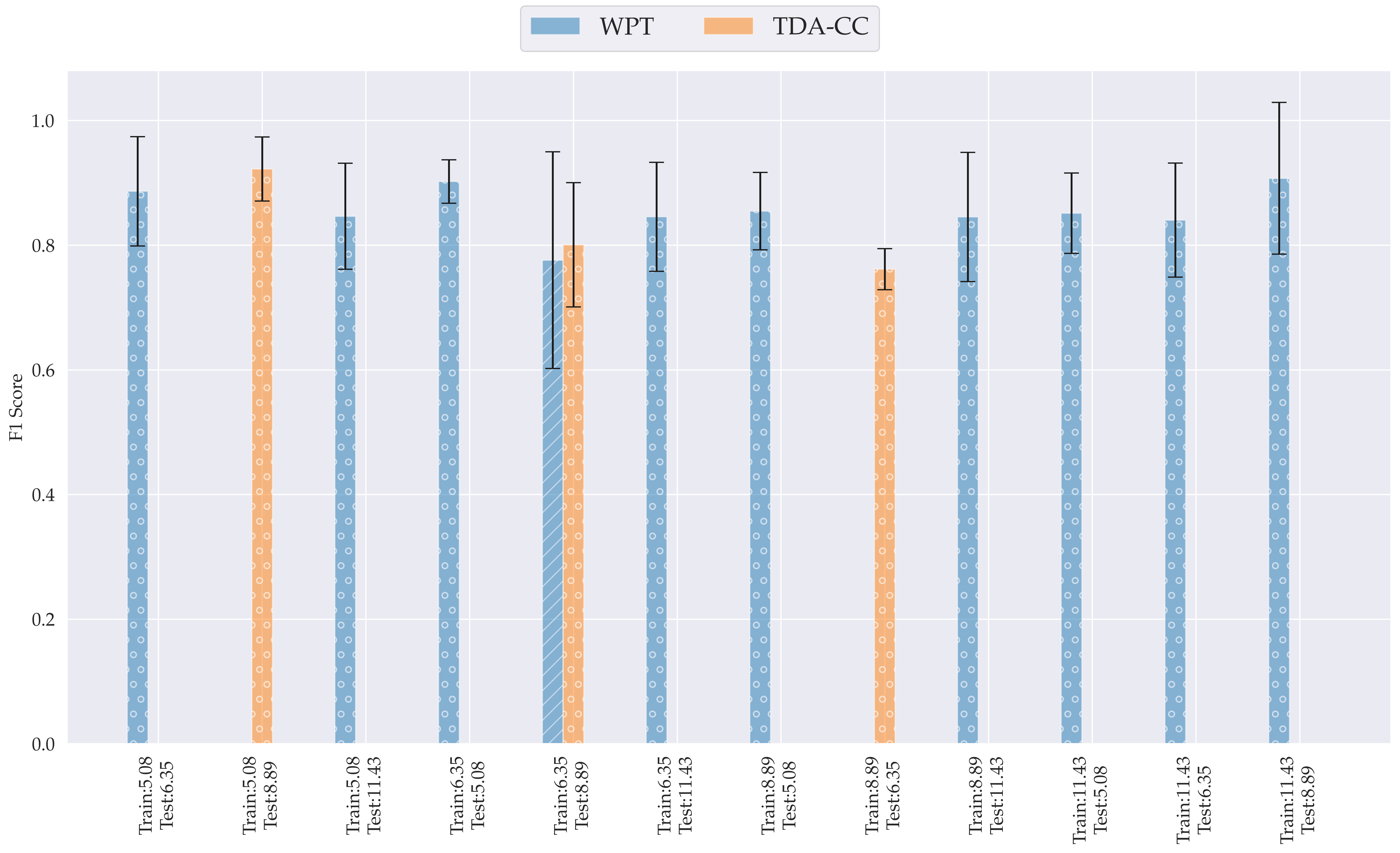}}
\caption{F1-Scores obtained from the selected methods when we train and test between the overhang distance cases of the turning data set. The selected methods that give the highest accuracy are represented with the '$\circ$' bar hatch and the ones that are in the error band of the highest accuracy are shown with the `$/$' bar hatch. One approach is selected from each category of the methods, and these are Wavelet Packet Transform (WPT), Carlsson Coordinates (TDA-CC), and Dynamic Time Warping (DTW).}
\label{fig:Turning_Best_Methods_F1_score}
\end{center}
\end{figure}

Another criterion to compare the performance of the methods is to check the F1-score. 
The F1-score is computed for all transfer learning applications and for each method.
Then, the highest F1-scores out of all classifiers were chosen, and the summary plots are given in Fig.~\ref{fig:Turning_Summary_F1_score} and~\ref{fig:Turning_Best_Methods_F1_score}.
In addition, we perform the counting the number of best methods and the methods that are in the error band as in the case of accuracy, and these numbers are reported in Tab.~\ref{tab:distribution_of_numbers}.
It is seen that the performance of the time-frequency-based approach is better since it has the highest F1-score in 9 out of 12 cases of the applications.
The TDA-based approach provides the best F1-score three times.
WPT again provides the best results with larger deviations compared to the TDA-based approach (see Fig.~\ref{fig:Turning_Best_Methods_F1_score}).

%%%%%%%%%%%%%%%%%%%%%% TURNING-MILLING  %%%%%%%%%%%%%%%%%%%%%%%%
\subsection{Results of Transfer Learning Applications Between Turning and Milling Data Sets}
\label{sec:TL_Turning_Milling}

\begin{figure}[h] 
\begin{center}
\centerline{\includegraphics[width=1\textwidth,keepaspectratio]{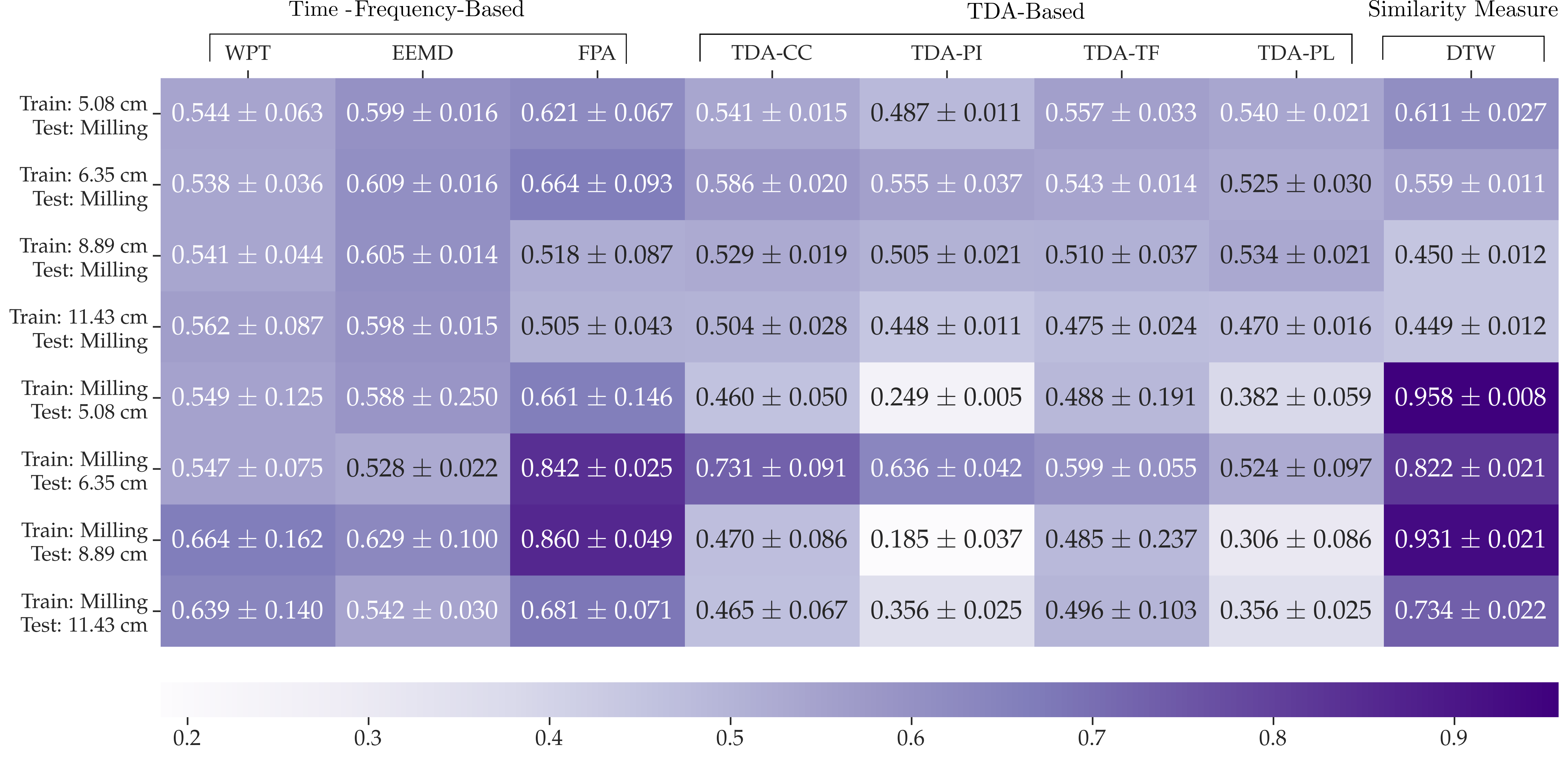}}
\caption{The highest accuracy out of four different classifiers (or out of selected numbers of nearest neighbor for DTW) for each approach used in transfer learning applications between overhang distance cases of turning and milling experiments.}
\label{fig:Milling_turning_TL_Best_Scores}
\end{center}
\end{figure}

\begin{figure}[h] 
\begin{center}
\centerline{\includegraphics[width=1\textwidth,keepaspectratio]{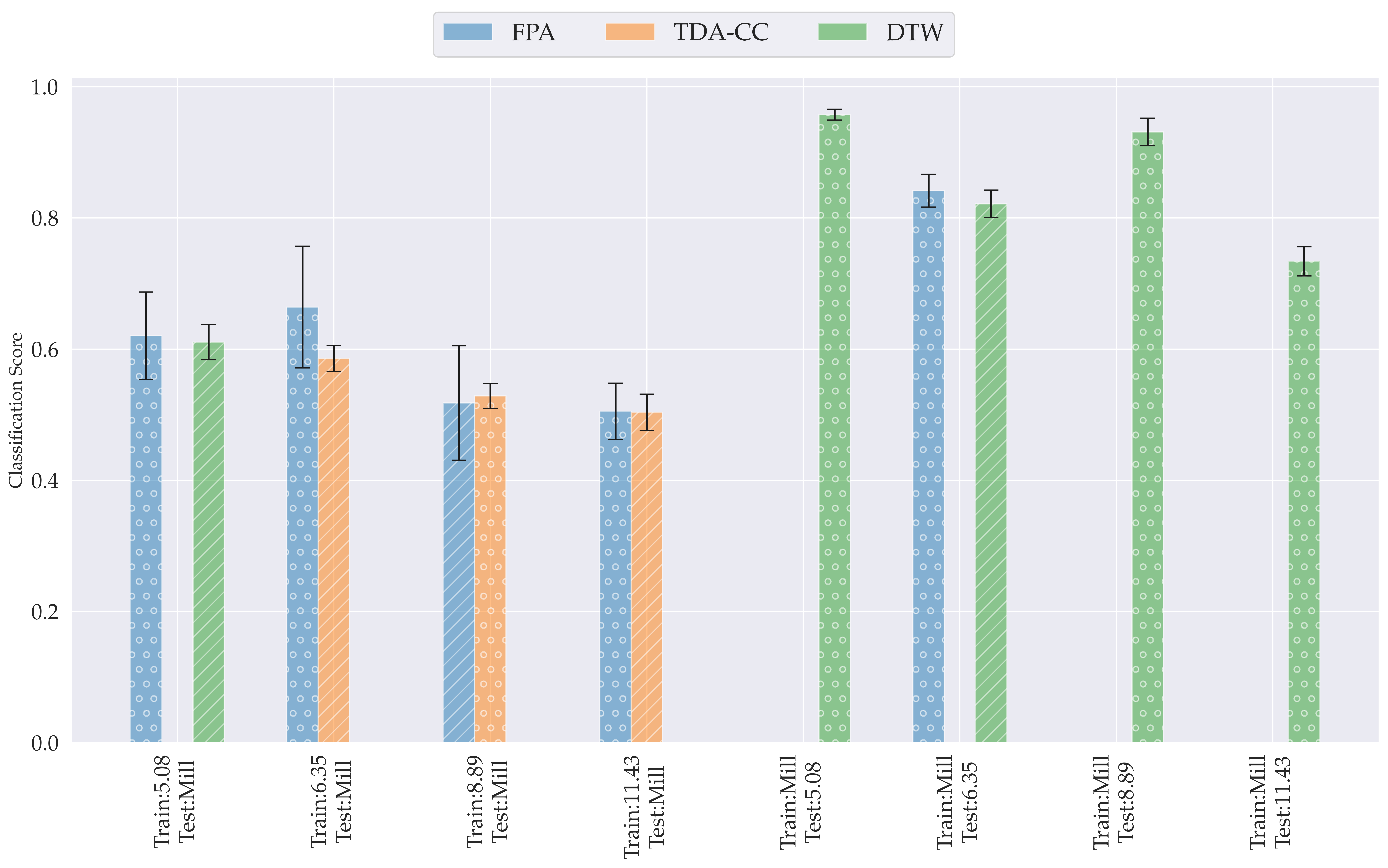}}
\caption{The classification results obtained from the selected methods when we train and test between the overhang distance cases of the turning data set and the milling data set. The selected methods that give the highest accuracy are represented with the '$\circ$' bar hatch and the ones that are in the error band of the highest accuracy are shown with the `$/$' bar hatch. One approach is selected from each category of the methods, and these are FPA, TDA-CC, and DTW.}
\label{fig:Turning_Milling_Selected_Best_Methods}
\end{center}
\end{figure}

There are 8 different permutations when we apply transfer learning between the turning and milling operations. 
The classification scores for four different classifiers are provided in Fig.~\ref{fig:training_milling_turning}-~\ref{fig:Turning_Milling_DTW}.
However, these plots include detailed results within the same application of transfer learning.
Therefore, we provide summary plots for these permutations in Fig.~\ref{fig:Milling_turning_TL_Best_Scores}. It is seen that FFT/PSD/ACF (FPA) gives the highest accuracies between time-frequency-based approaches in most of the transfer learning applications. The results obtained with TDA-based approaches are similar to each other, especially for Carlsson Coordinates (TDA-CC) and Template Functions (TDA-TF).
Since the TDA-CC provides the highest accuracy when we train on 6.35 cm overhang distance of turning data set and test on the milling data set, we choose TDA-CC to represent the TDA-based approaches. Accordingly, FPA, TDA-CC, and DTW are compared with respect to classification scores in Fig.~\ref{fig:Turning_Milling_Selected_Best_Methods}.
Similar figures for F1-Score can also be found in Fig.~\ref{fig:Turning_Milling_Summary_F1_score} and~\ref{fig:Turning_Milling_Best_Methods_F1_score}.

In Ref.~\cite{Yesilli2020,Yesilli2020a}, the authors mentioned several drawbacks of the time-frequency-based approaches.
One of the main drawbacks of these methods is that they require checking the frequency spectrum of each time series manually to decide informative decomposition for WPT and EEMD or the restriction parameters for FPA. 
In this study, we only performed this manual preprocessing for a couple of time series. 
This process gets cumbersome as the size of the data set increases.
In a real-time application, when a new time series is introduced to a classifier, the frequency spectrum of it and its reconstructed time series obtained from the wavelet packets need to be investigated to find the decomposition whose spectrum has the largest overlap with the signal's spectrum. 
On the other hand, the processes for the TDA-based approach and the DTW do not require any parameter selection, and all steps can be completed autonomously. 

Based on Fig.~\ref{fig:Turning_Milling_Selected_Best_Methods} and~\ref{fig:Turning_Milling_Best_Methods_F1_score}, we generated Tab.~\ref{tab:distribution_of_numbers_Milling_Turning} that shows the number of times when a selected method gives the highest accuracy (BM) or it is in the error band of the highest accuracy (MIEB). 
If we only consider the accuracy as the main criterion, the DTW method provides the highest accuracy in three out of eight applications, and it is in the error band of the highest accuracy in two applications.
In addition, the results of the TDA-based approach is in the error band in two out of eight applications.
Considering the drawbacks of the frequency-based approach and deviations of the results of the frequency-based approach, DTW and TDA-based approaches can be preferred when we apply transfer learning between different machining operations.

\begin{table}[!htbp]
\centering
\caption{The number of times when a selected method gives the highest accuracy out of 8 different applications between the cases of turning data set and the milling data set is denoted with BM. The number of times when a method is in the error band of the highest accuracy is denoted with MIEB. These two numbers are provided for accuracy and F1-score.}
\label{tab:distribution_of_numbers_Milling_Turning}
\resizebox{0.5\columnwidth}{!} {
\begin{tabular}{c|c|c|c|c}
&\multicolumn{2}{c|}{Accuracy} &  \multicolumn{2}{c}{F1-Score} \\
\hline
Method& \makecell{BM} &  MIEB &BM & MIEB\\
\hline
\makecell{Time - Frequency-based (FPA)}&4 & 1&6 &0\\
\makecell{TDA-based (TDA-CC)}&1& 2 &0&4\\
\makecell{Similarity Measure (DTW)}&3&2&2&1
\end{tabular}}
\end{table} 

\subsection{Transfer Learning Using Deep Learning}

In addition to traditional machine learning algorithms, we also utilized Artificial Neural Networks(ANNs) to test the performance of several approaches. 
Deep learning frameworks can learn from raw data set without the need of feature extraction. However, Zhoa et. al state that inadequate size of data set, noisy raw signal, and complex machining operations make it necessary to preprocess data before feeding it into deep learning algoritms~\cite{Zhao2019}. 
Therefore, we use some of the features extracted from TDA-based approaches to apply deep learning in transfer learning. 
Some of the studies in the literature (see Refs.~\cite{Unver2021,Postel2020}) trained the deep learning algorithms using the simulation data set to eliminate the need for an extensive amount of experimental data set to train the classifier.
However, in this work, we only used the existing experimental data and the features extracted from them to train deep learning algorithms to compare the results to traditional machine learning algorithms. 
We are aware of the fact that we need more observation to have a fair comparison between deep learning-based transfer learning and traditional machine learning based transfer learning. 
Since we do not split the raw experimental signals into small pieces for Time-Frequency based approaches, we have fewer observations for these approaches.
Hence, we do not utilize the features extracted using Time-Frequency based approaches to train deep learning algorithms. 

\begin{figure}[h] 
\begin{center}
\centerline{\includegraphics[width=1\textwidth,keepaspectratio]{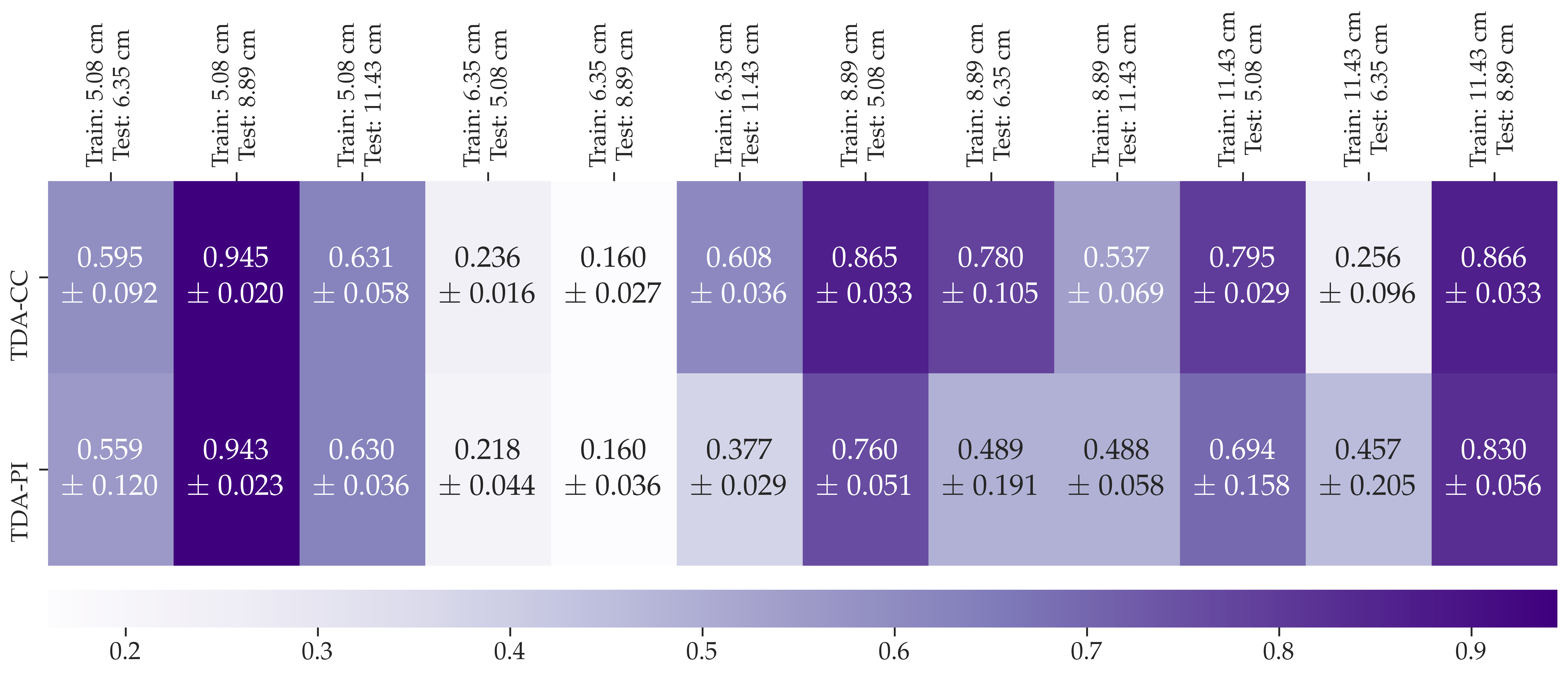}}
\caption{The classification accuracies obtained using Carlsson Coordinates and Persistence Images features with ANN algorithms for the transfer learning between the cases of turning experiments.}
\label{fig:turning_DL_results}
\end{center}
\end{figure}

\begin{figure}[h] 
\begin{center}
\centerline{\includegraphics[width=0.8\textwidth,keepaspectratio]{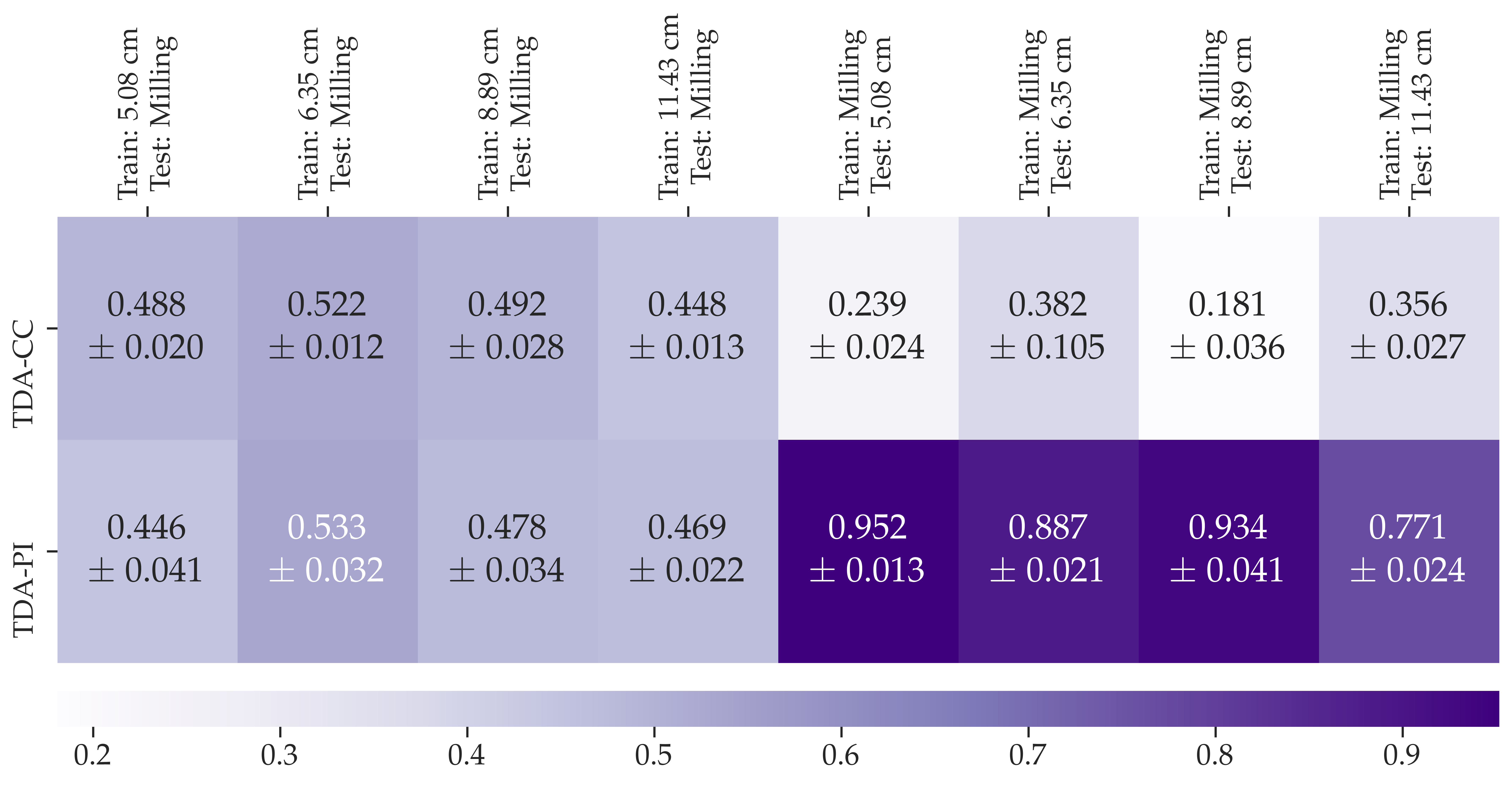}}
\caption{The classification accuracies obtained using Carlsson Coordinates and Persistence Images features with ANN algorithms for the transfer learning between the milling and turning experiments.}
\label{fig:turning_milling_DL_results}
\end{center}
\end{figure}

The ANN structure used in this work has one input, three hidden, and one output layer. The number of inputs fed into the input layer is based on the number of features extracted from TDA-based approaches. For instance, Carlsson Coordinates can provide five features for each persistence diagram, so the number of inputs will be five for this approach. The first and last hidden layers have 25 neurons, while the second hidden layer has 12 neurons. The hyperbolic tangent function is used as an activation function in all layers except the output layer. Since the classification output is binary, we have chosen the sigmoid function as the activation function in the output layer.
Adam optimization algorithm and binary cross-entropy loss functions are used to compile the ANN. Epoch number and the batch size to update the weights of the fully connected layers are selected as 100 and 5, respectively.
We have 12 permutations between the overhang distance cases of the turning data set, and 8 permutations between the turning and milling data set.
We used 67\% of the training data set as the training set, and 70\% of the test set data set to test the ANNs.
Train-test split is repeated for 10 different pre-defined random state numbers, and we computed the mean accuracy and standard deviation out of these 10 realizations.
The results for transfer learning applications between the overhang distance cases of turning data set is provided in Fig.~\ref{fig:turning_DL_results}, while Fig.~\ref{fig:turning_milling_DL_results} provides the accuracies with error bands for the transfer learning between the milling and turning data set.

From Fig.~\ref{fig:turning_DL_results} and~\ref{fig:Turning_TL_WPT_CC_DTW_Summary_Accuracy}, it is seen that traditional machine learning algorithms provide better accuracies compared to deep learning in 11 out of 12 different transfer learning applications for Carlsson Coordinates, while deep learning is outperformed in 9 out of 12 applications of transfer learning between the cases of turning data set using persistence images.
When we look into Fig.~\ref{fig:turning_milling_DL_results} and~\ref{fig:Milling_turning_TL_Best_Scores}, traditional machine learning algorithms outperform deep Learning in all applications of transfer learning for Carlsson Coordinates. However, deep learning outperforms traditional machine learning in 4 out of 8 applications of transfer learning between the milling and turning data set using persistence images.
Overall, it is obvious that the amount of experimental data set fed to deep learning is insufficient, and this leads to poor performance against the traditional machine learning algorithms.
In addition, we do not perform hyperparameter tuning for ANNs in this work, this could also be another reason for the poor performance of deep learning.

\section{Discussion and Conclusion}
\label{sec:Discussion_and_Conclusion}
The highest scores obtained from the transfer learning applications, between the cases of the turning data, was between 80\% and 100\%, while that drops to 60\% when we train on turning data and test on the milling data.
A period-doubling bifurcation was observed in 19 out of 318 time series of milling data and a Hopf bifurcation was observed in the rest of the unstable cases of milling data. On the other hand, the turning data only contains the Hopf bifurcation.
When we train on the turning data set, the classification model is not trained to recognize the descriptors of period-doubling bifurcation, so it performs poorly when it is tested on milling data. 
On the other hand, a classifier is trained with both features of Hopf and period-doubling when the milling data is used as the training set. This explains why training on milling data and testing on turning data set performs better. 
In addition, the mathematical model for the milling process has time-varying coefficients, while the turning process has an autonomous system.
Since the coefficients are constant in turning processes, this can lead to misclassification when the classifier is tested on milling processes with time-varying coefficients.

This study compared the performance of feature extraction methods from established methods alongside those recently proposed in the literature. 
Turning and milling data sets were used to evaluate the performance of each method.
The size of the training sets and the test sets were kept the same for each method. Since the training set data and test data are different from each other, we used 67\% and 70\% of the training set and test data to train and test a classifier, respectively.
Ten random state numbers were used to generate training and test splits, and these were used to train and test a classifier for each method. The 
average and standard deviation of the 10 realizations were computed and the final results were reported.
This has been repeated for all 20 combinations between the milling data and overhang cases of turning data. 

To compare the results, we provided two types of figures for each comparison criterion.
Fig.~\ref{fig:Turning_TL_WPT_CC_DTW_Summary_Accuracy},~\ref{fig:Turning_Selected_Best_Methods},~\ref{fig:Milling_turning_TL_Best_Scores}, and~\ref{fig:Turning_Milling_Selected_Best_Methods} were obtained when the criterion was accuracy, while  Fig.~\ref{fig:Turning_Summary_F1_score},~\ref{fig:Turning_Best_Methods_F1_score},~\ref{fig:Turning_Milling_Summary_F1_score}, and~\ref{fig:Turning_Milling_Best_Methods_F1_score} were given for the F1-score.
It can be seen that the time-frequency-based approaches give the highest accuracy in most of the applications of transfer learning with larger deviations in comparison to the TDA-based approach and DTW.
When we only consider the transfer learning between the milling and turning data sets, we see that the accuracies obtained from DTW can be as high as 96\% while the time-frequency-based approaches can be up to 86\% (see Fig.~\ref{fig:Milling_turning_TL_Best_Scores}).
For the same cases of transfer learning, the highest score obtained from TDA is  73\% (see Fig.~\ref{fig:Milling_turning_TL_Best_Scores}). 
For the transfer learning applications where we train and test between the cases of turning, the time-frequency-based approach has the highest accuracy of 93\%; the best score for the TDA approach and DTW are 97\% and 96\%, respectively (see Fig.~\ref{fig:Turning_TL_WPT_CC_DTW_Summary_Accuracy}). 
We also compared the results of traditional machine learning algorithms to the ones obtained from ANNs. It is seen that insufficient experimental data set leads to poor results against traditional machine learning approaches. The small size of the experimental data set also avoid us to compare different techniques to each other using deep learning algorithms. In this work, we were only able to compare several TDA-based approaches to each other. Using synthetic data sets generated using the analytical model of milling and turning operations can allow us to further extend the comparison of more approaches using deep learning frameworks in the future.

In summary, the TDA-based and DTW approaches can provide accuracies and F1-scores as high as the time-frequency-based methods.  
DTW outperforms all other methods when training on the milling data set and testing on the turning data set.
In addition, the TDA-based approach and DTW can be applied without needing manual preprocessing, all of the steps in their pipeline can be completed automatically.
Therefore, these approaches may be preferred over the time-frequency-based approaches in either real-time or in fully automated chatter detection schemes.
It is worth noting that we have not performed any optimization on hyperparameters of the traditional machine learning and deep learning algorithms. Thus future studies should also consider the effect of hyperparameter tuning.

\section*{Acknowledgement}
This material is based upon work supported by the National Science Foundation under Grant Nos.~CMMI-1759823 and DMS-1759824 with PI FAK.

\appendix

\newpage
\appendixpage
\label{sec:appendix}
%\section{Transfer learning results of turning data set}
% Train 2 - Test 2p5,3p5,4p5 ---------------------------------------------

\begin{figure}[h]
\centering
\includegraphics[width=1\textwidth,keepaspectratio]{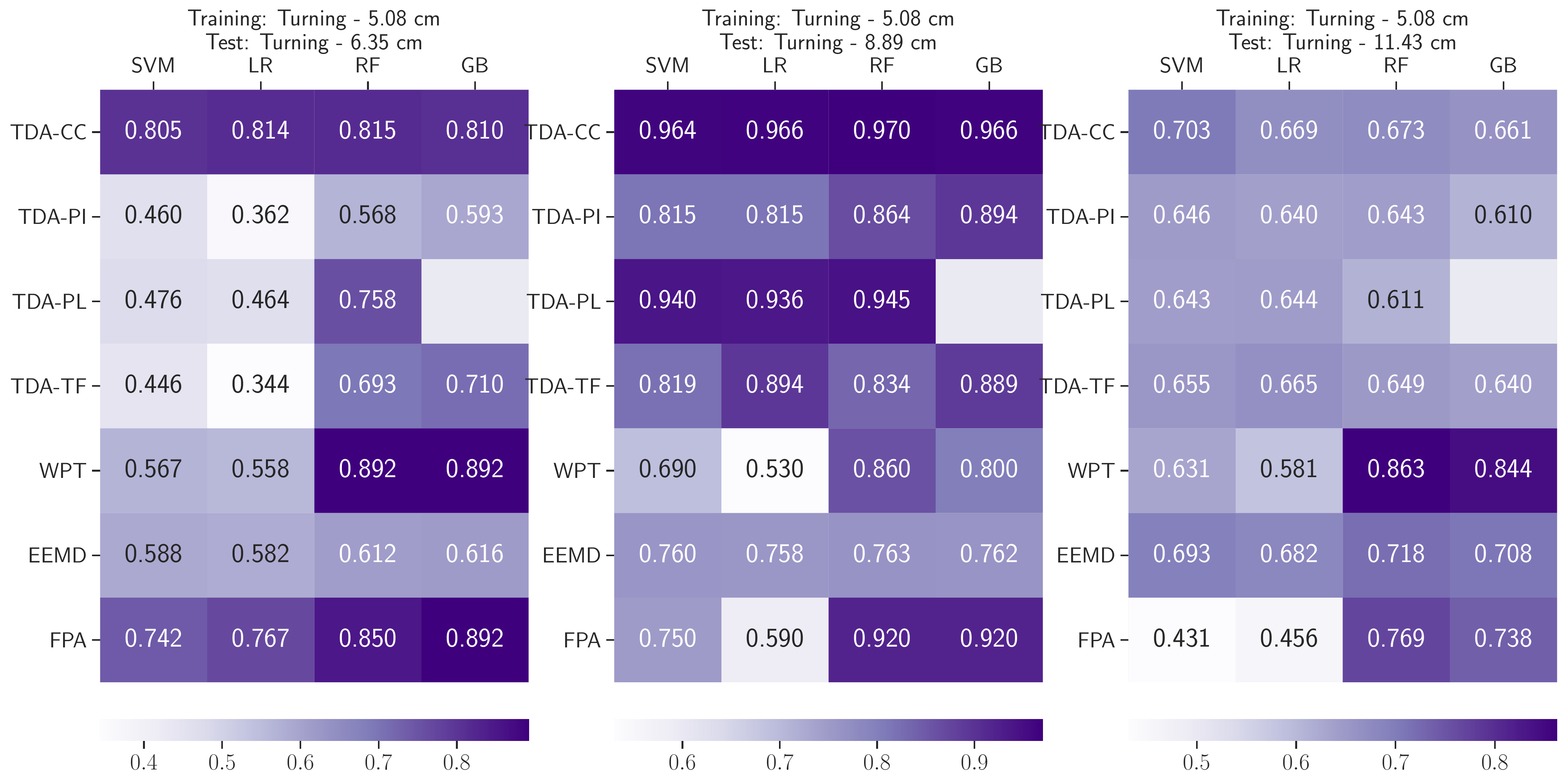}
\caption{Classification accuracies obtained from transfer learning applications for turning experiment case with 5.08 cm overhang distance . (left) Training: 5.08 cm Test: 6.35 cm (middle) Training: 5.08 cm Test: 8.89 cm, (right) Training: 5.08 cm Test: 11.43 cm. CC: Carlsson Coordinates, PI: Persistence Images, PL: Persistence Landscapes, TF: Template Functions, WPT: Wavelet Packet Transform, EEMD: Ensemble Empirical Mode Decomposition, FPA: FFT/PSD/ACF, SVM: Support Vector Machine, RF: Random Forest, GB: Gradient Boosting. The results for TDA-PL implementation with Gradient Boosting classifier (GB) is not available due to large amount of time training and testing. Therefore, it represents an empty box in the figure.}
\label{fig:training_overhang_2_test_turning}
\end{figure}

% Train 2p5 - Test 2,3p5,4p5 ---------------------------------------------

\begin{figure}[h]
\centering
\includegraphics[width=1\textwidth,keepaspectratio]{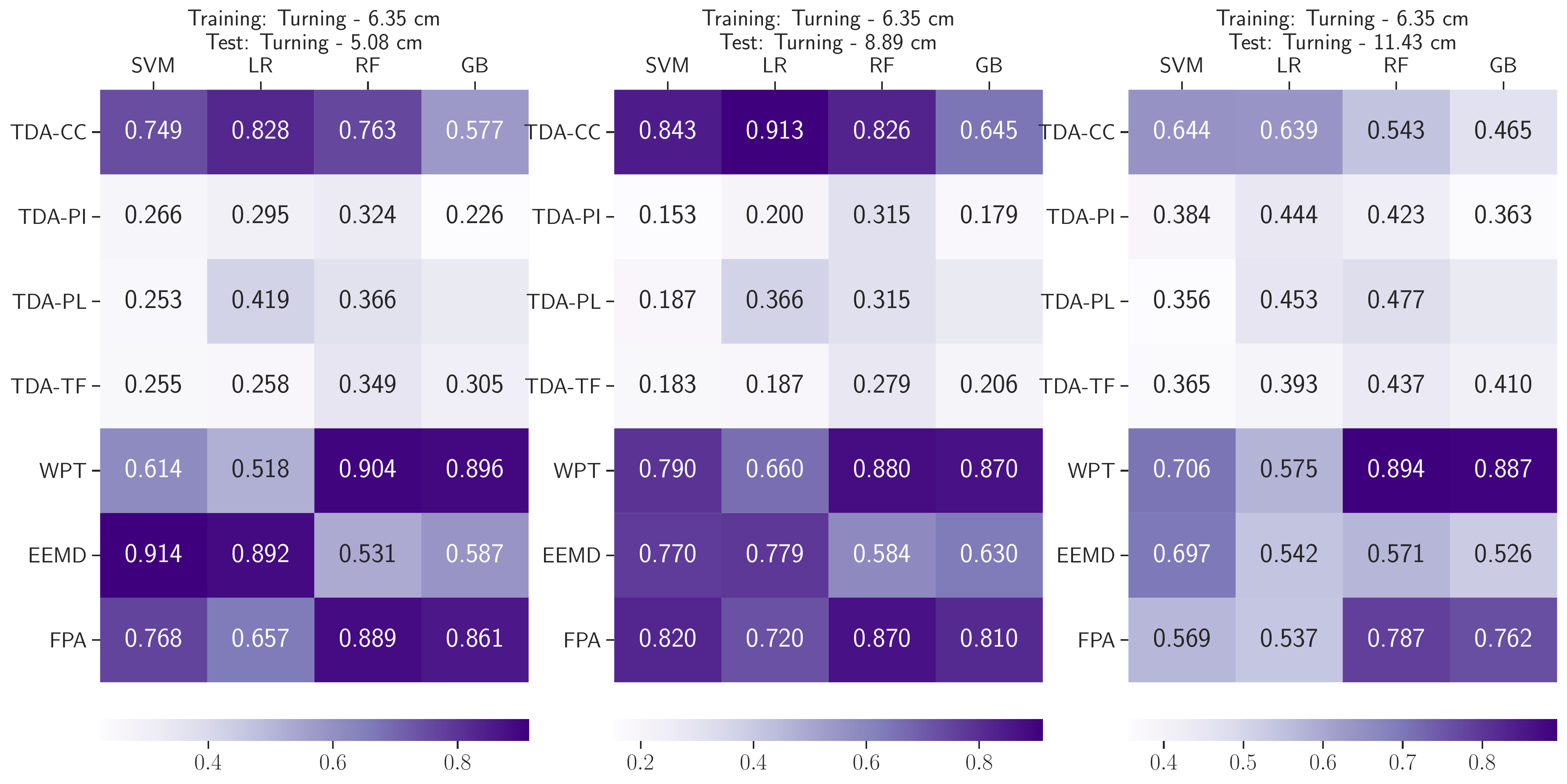}
\caption{Classification accuracies obtained from transfer learning applications for turning experiment case with 6.35 cm overhang distance . (left) Training: 6.35 cm Test: 5.08 cm (middle) Training: 6.35 cm Test: 8.89 cm, (right) Training: 6.35 cm Test: 11.43 cm. CC: Carlsson Coordinates, PI: Persistence Images, PL: Persistence Landscapes, TF: Template Functions, WPT: Wavelet Packet Transform, EEMD: Ensemble Empirical Mode Decomposition, FPA: FFT/PSD/ACF, SVM: Support Vector Machine, RF: Random Forest, GB: Gradient Boosting. The results for TDA-PL implementation with Gradient Boosting classifier (GB) is not available due to large amount of time training and testing. Therefore, it represents an empty box in the figure.}
\label{fig:training_overhang_2p5_test_turning}
\end{figure}

% Train 3p5 - Test 2,2p5,4p5 ---------------------------------------------

\begin{figure}[h]
\centering
\includegraphics[width=1\textwidth,keepaspectratio]{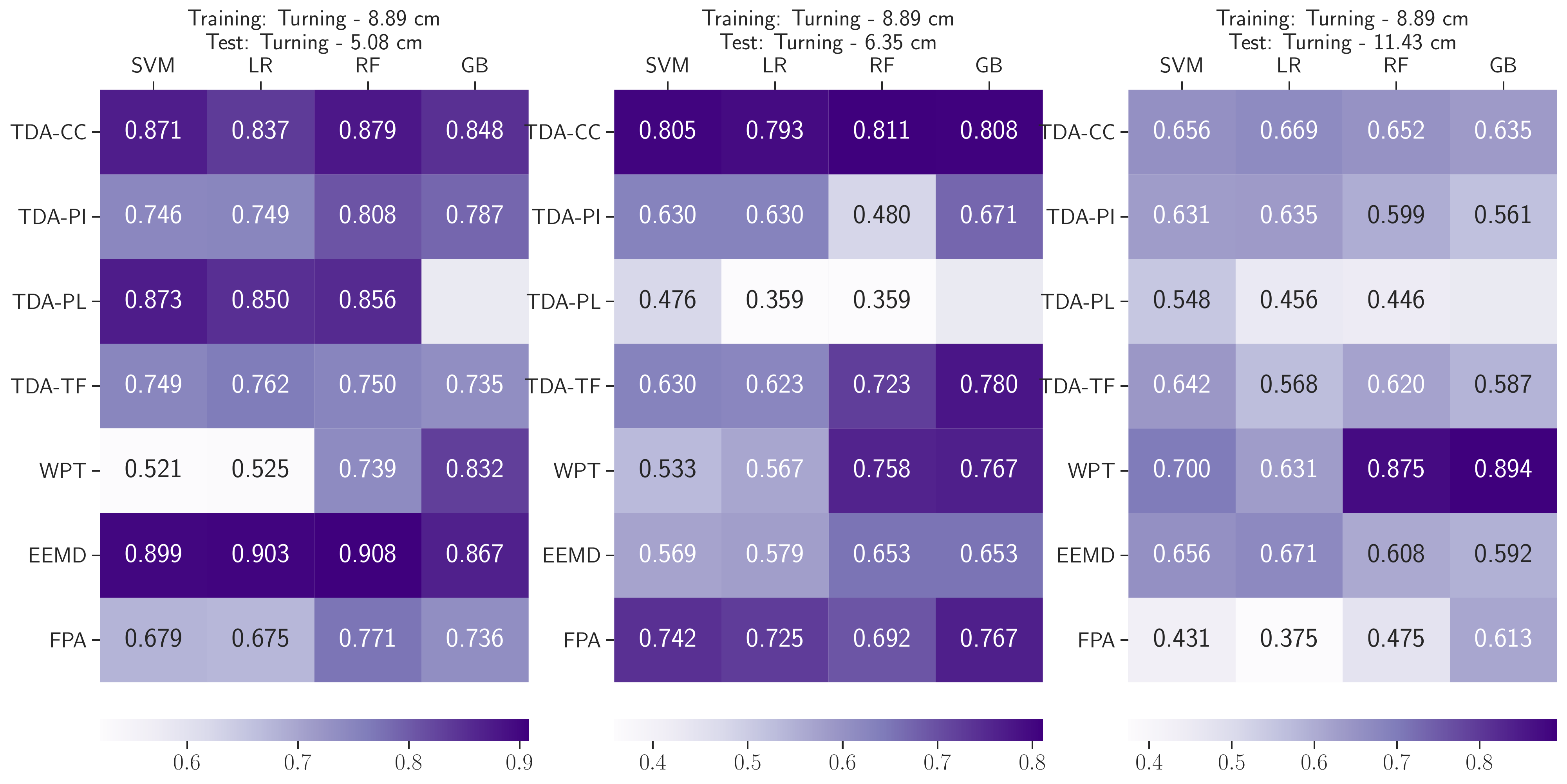}
\caption{Classification accuracies obtained from transfer learning applications for turning experiment case with 8.89 cm overhang distance . (left) Training: 8.89 cm Test: 5.08 cm (middle) Training: 8.89 cm Test: 6.35 cm, (right) Training: 8.89 cm Test: 11.43 cm. CC: Carlsson Coordinates, PI: Persistence Images, PL: Persistence Landscapes, TF: Template Functions, WPT: Wavelet Packet Transform, EEMD: Ensemble Empirical Mode Decomposition, FPA: FFT/PSD/ACF, SVM: Support Vector Machine, RF: Random Forest, GB: Gradient Boosting. The results for TDA-PL implementation with Gradient Boosting classifier (GB) is not available due to large amount of time training and testing. Therefore, it represents an empty box in the figure.}
\label{fig:training_overhang_3p5_test_turning}
\end{figure}

% Train 4p5 - Test 2,2p5,3p5 ---------------------------------------------

\begin{figure}
\centering
\includegraphics[width=1\textwidth,keepaspectratio]{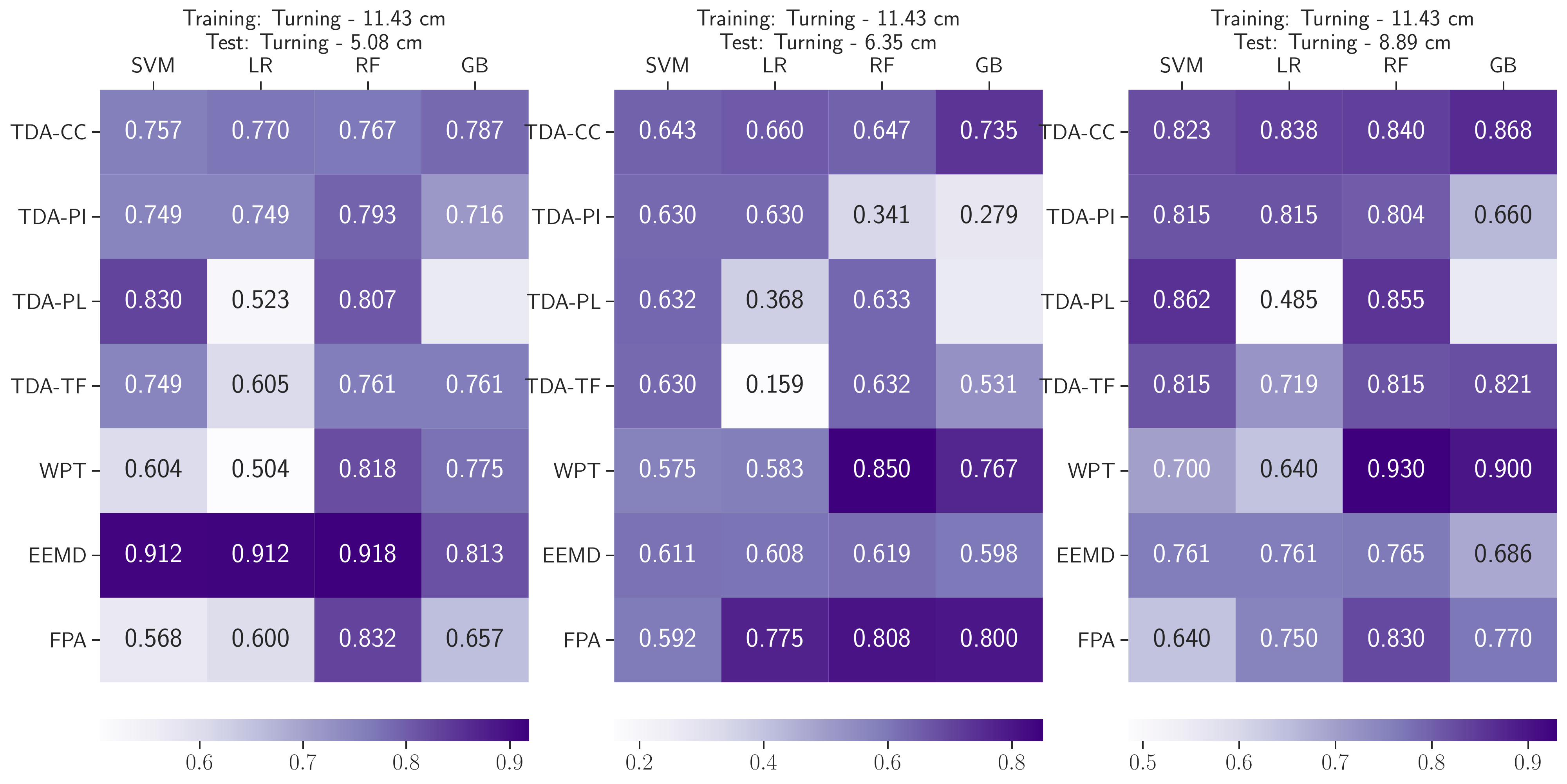}
\caption{Classification accuracies obtained from transfer learning applications for turning experiment case with 11.43 cm overhang distance . (left) Training: 11.43 cm Test: 5.08 cm (middle) Training: 11.43 cm Test: 6.35 cm, (right) Training: 11.43 cm Test: 8.89 cm. CC: Carlsson Coordinates, PI: Persistence Images, PL: Persistence Landscapes, TF: Template Functions, WPT: Wavelet Packet Transform, EEMD: Ensemble Empirical Mode Decomposition, FPA: FFT/PSD/ACF, SVM: Support Vector Machine, RF: Random Forest, GB: Gradient Boosting. The results for TDA-PL implementation with Gradient Boosting classifier (GB) is not available due to large amount of time training and testing. Therefore, it represents an empty box in the figure.}
\label{fig:training_overhang_4p5_test_turning}
\end{figure}

% DTW Results ---------------------------------------------
\begin{figure}
\centering
\includegraphics[width=1\textwidth,keepaspectratio]{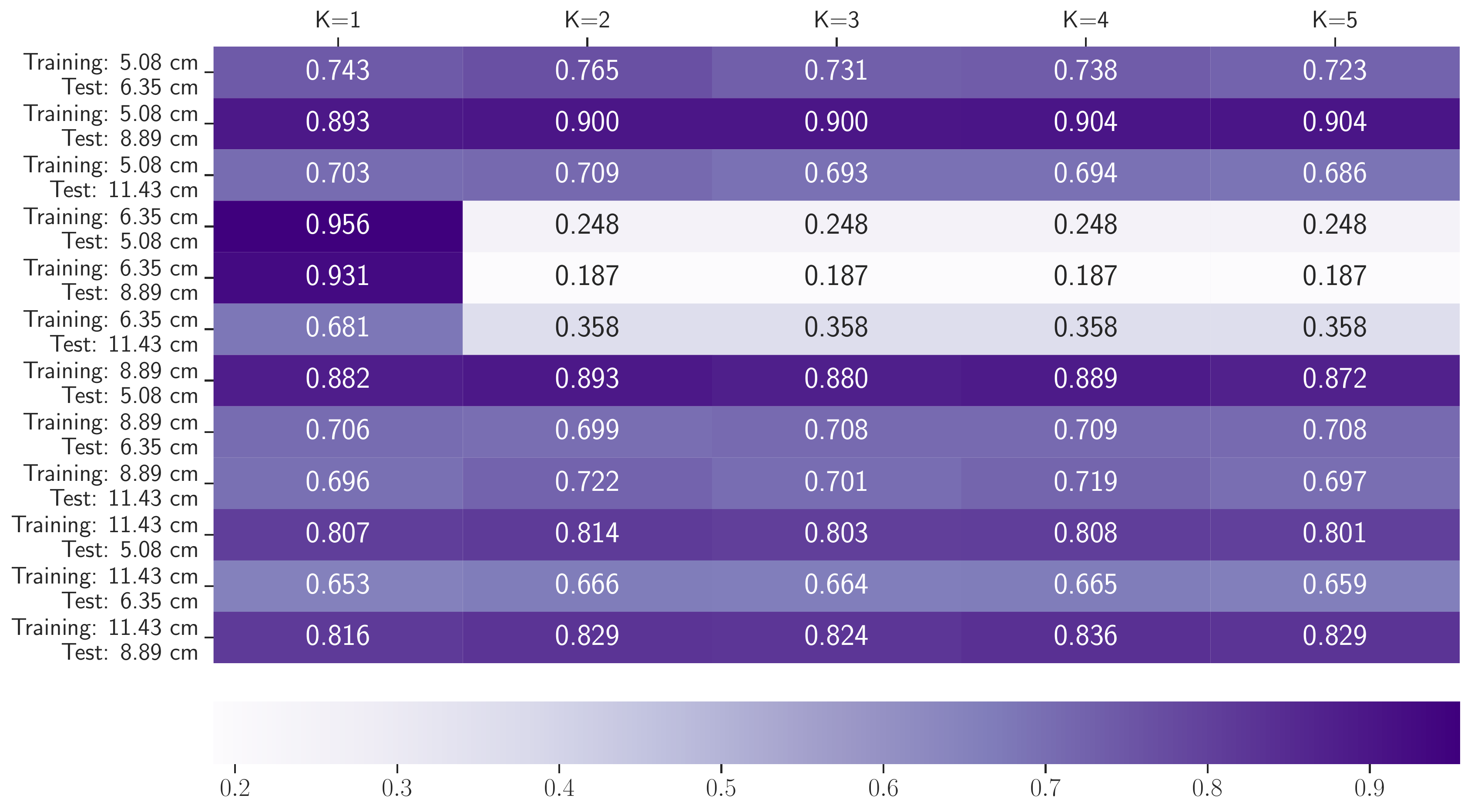}
\caption{Classification accuracies obtained from transfer learning applications for turning data set using DTW approach with $K=1,2,3,4,5$, where $K$ represent the nearest neighbor number. Overhang distances used as training and testing data set are shown in $y$-axis. }
\label{fig:Turning_DTW}
\end{figure}

%\section{Transfer learning results of milling data set}
% Train: 2-2p5-3p5-4p5 - Test:Milling ---------------------------------------------

\begin{figure}
\centering
\includegraphics[width=1\textwidth,keepaspectratio]{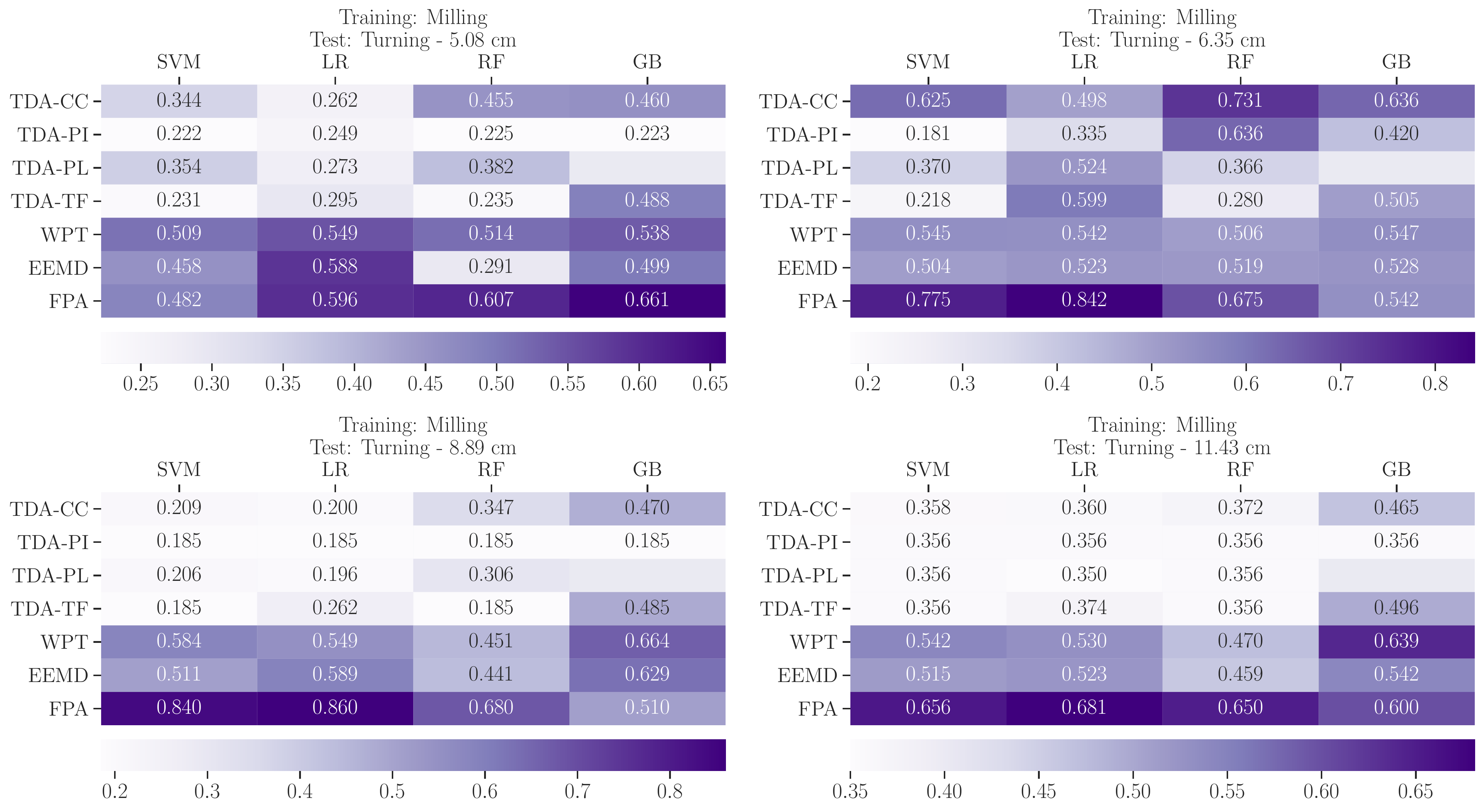}
\caption{Classification accuracies obtained from transfer learning applications when milling data set is used as training set and turning data set is used as test set. CC: Carlsson Coordinates, PI: Persistence Images, PL: Persistence Landscapes, TF: Template Functions, WPT: Wavelet Packet Transform, EEMD: Ensemble Empirical Mode Decomposition, FPA: FFT/PSD/ACF, SVM: Support Vector Machine, RF: Random Forest, GB: Gradient Boosting. The results for TDA-PL implementation with Gradient Boosting classifier (GB) is not available due to large amount of time training and testing. Therefore, it represents an empty box in the figure.}
\label{fig:training_milling_turning}
\end{figure}

\begin{figure}
\centering
\includegraphics[width=1\textwidth,keepaspectratio]{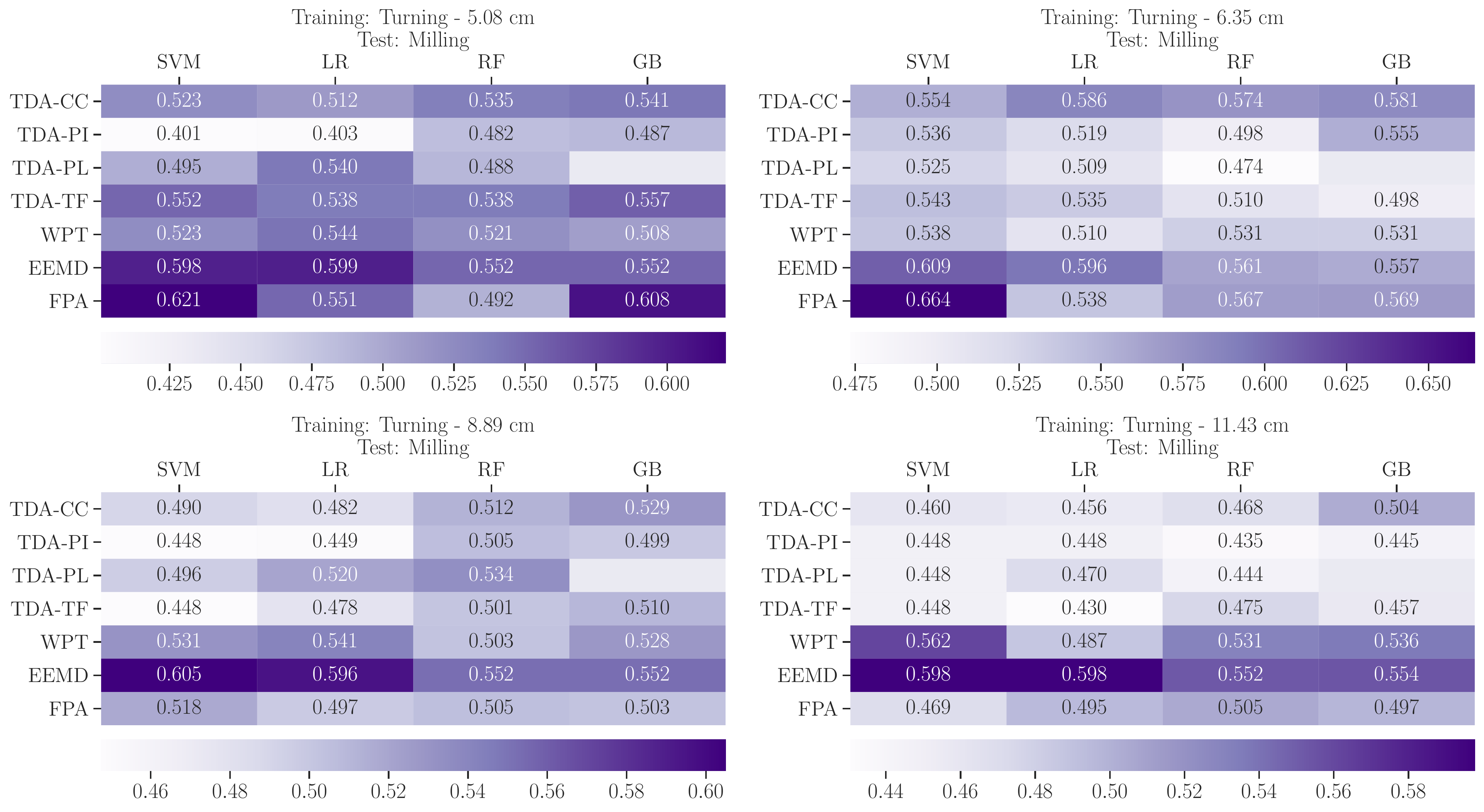}
\caption{Classification accuracies obtained from transfer learning applications when milling data set is used as test set and turning data set is used as training set. CC: Carlsson Coordinates, PI: Persistence Images, PL: Persistence Landscapes, TF: Template Functions, WPT: Wavelet Packet Transform, EEMD: Ensemble Empirical Mode Decomposition, FPA: FFT/PSD/ACF, SVM: Support Vector Machine, RF: Random Forest, GB: Gradient Boosting. The results for TDA-PL implementation with Gradient Boosting classifier (GB) is not available due to large amount of time training and testing. Therefore, it represents an empty box in the figure.}
\label{fig:training_turning_milling}
\end{figure}

% Milling-Turning DTW Results ---------------------------------------------
\begin{figure}
\centering
\includegraphics[width=1\textwidth,keepaspectratio]{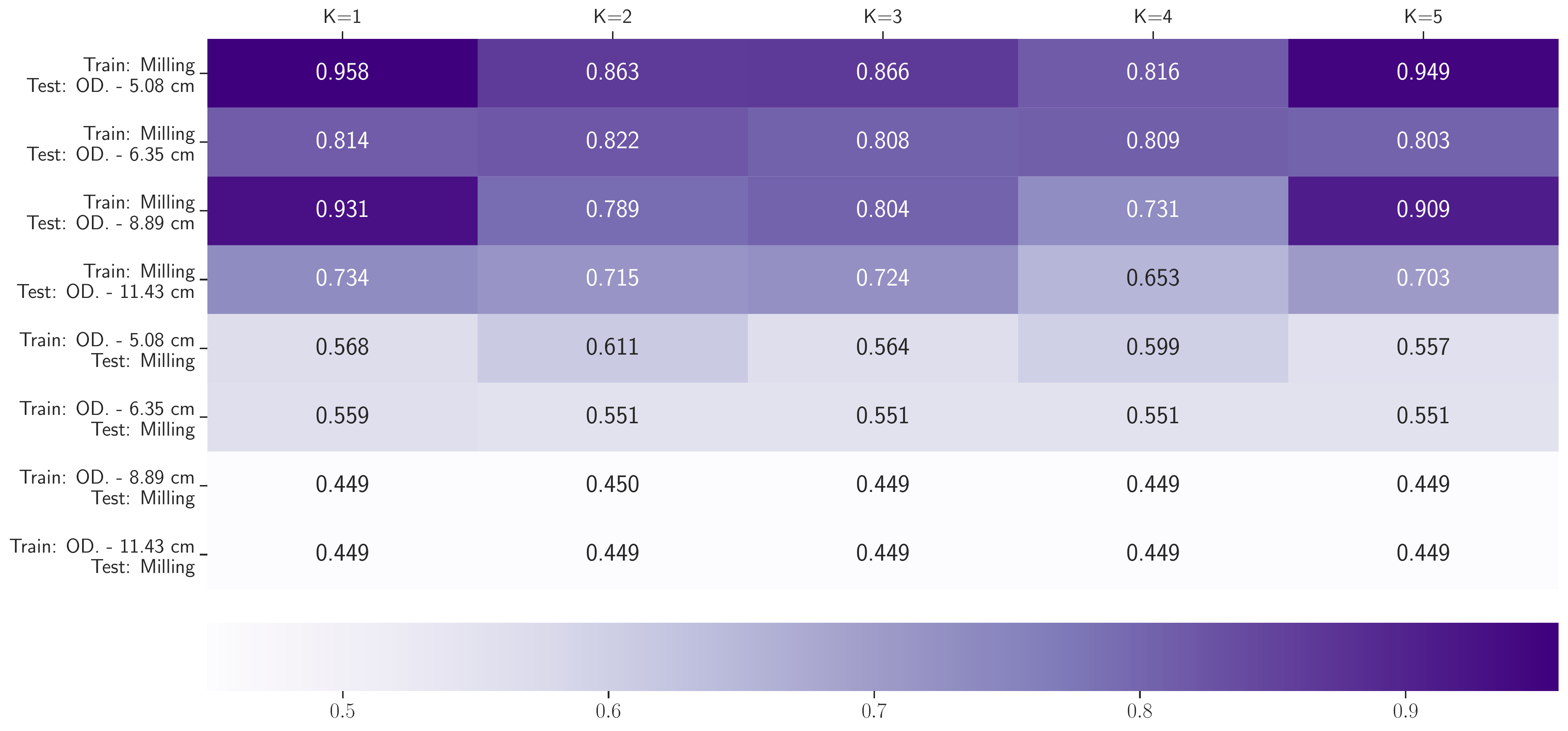}
\caption{Classification accuracies obtained from transfer learning applications between turning and milling data sets using DTW approach with $K=1,2,3,4,5$, where $K$ represent the nearest neighbor number. Overhang distances (OD.) used as training or testing data set are shown in $y$-axis.}
\label{fig:Turning_Milling_DTW}
\end{figure}

% F1 Score between milling-turning applications
\begin{figure}
\begin{center}
\centerline{\includegraphics[width=1\textwidth,keepaspectratio]{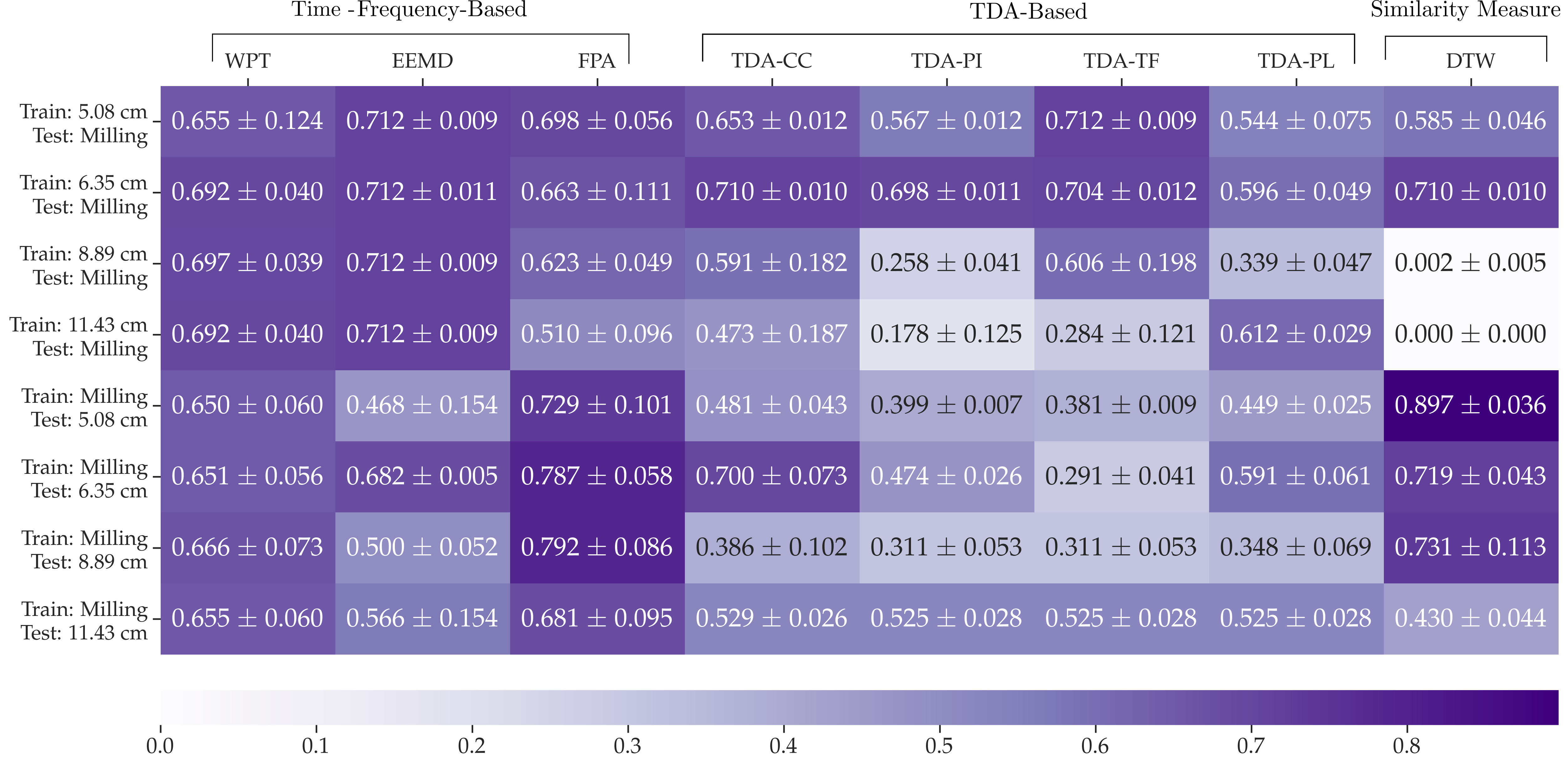}}
\caption{The highest F1-score out of four different classifier (or out of selected numbers of nearest neighbor for DTW) for each approach used in transfer learning applications between overhang distance cases of the turning and milling experiments.}
\label{fig:Turning_Milling_Summary_F1_score}
\end{center}
\end{figure}

\begin{figure}[h] 
\begin{center}
\centerline{\includegraphics[width=0.8\textwidth,keepaspectratio]{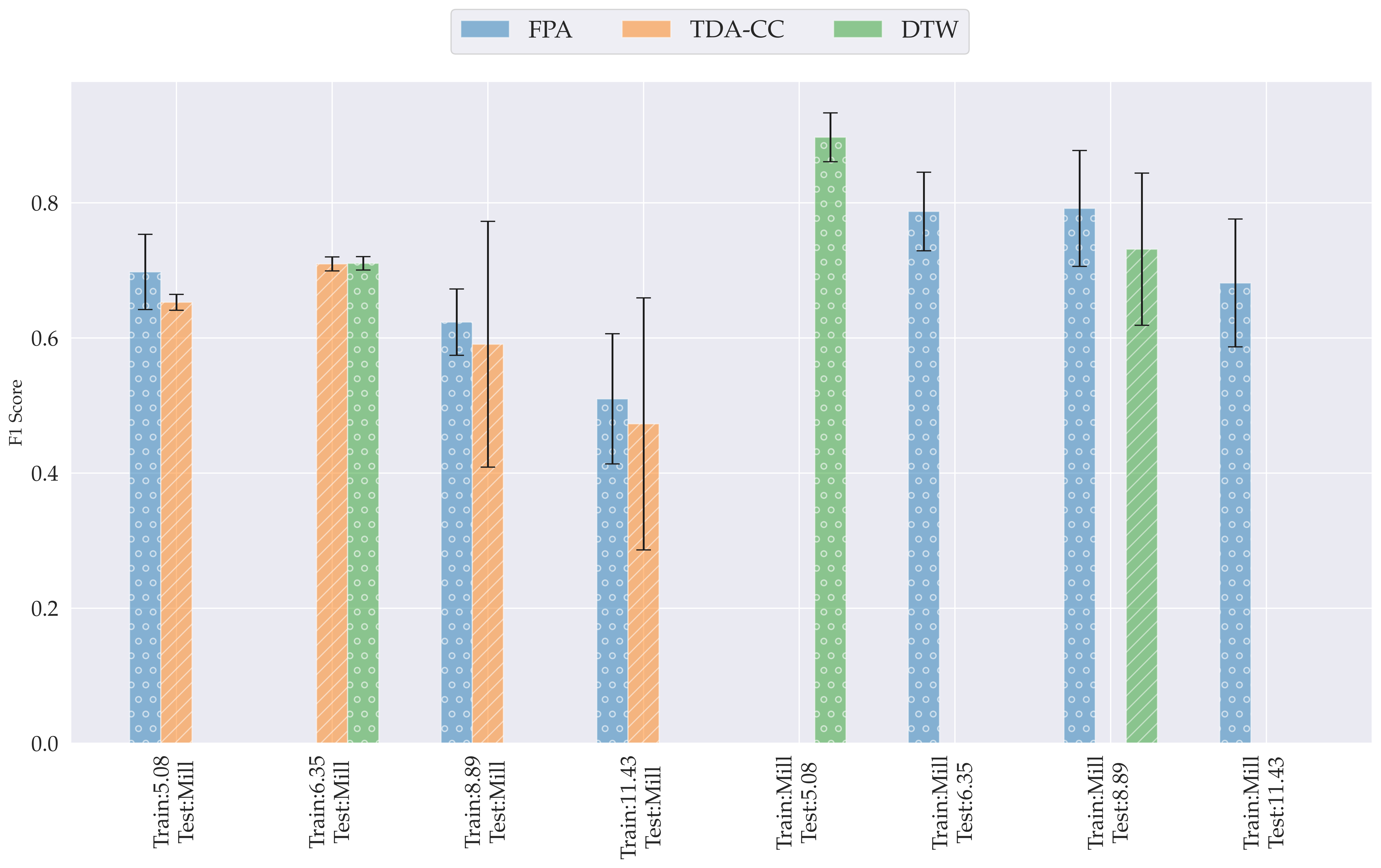}}
\caption{F1-Scores obtained from the selected methods when we train and test between the overhang distance cases of the turning data set and the milling data set. The selected methods that give the highest accuracy are represented with '$\circ$' bar hatch and the ones which are in the error band of the highest accuracy are shown with `$/$' bar hatch. One approach is selected from each category of the methods, and these are FPA, TDA-CC and DTW.}
\label{fig:Turning_Milling_Best_Methods_F1_score}
\end{center}
\end{figure}

\clearpage
\newpage
\bibliography{TF_Learning}
% \clearpage

\end{document}